\documentclass[a4paper,11pt]{article}
\pdfoutput=1 

\usepackage{jcappub}
\usepackage{amssymb,amsmath,verbatim,mathtools,needspace,enumitem,etoolbox,graphicx,physics,microtype,afterpage,xspace,tabularx,lmodern,multirow}
\usepackage{gensymb}
\usepackage[dvipsnames, usenames]{xcolor}
\colorlet{MAGENTA}{Magenta}
\definecolor{linkcolor}{rgb}{0.0,0.3,0.5}
\usepackage{hyperref}
\usepackage[all]{hypcap}
\usepackage[T1]{fontenc}
\usepackage[utf8]{inputenc}
\usepackage[usenames,dvipsnames]{xcolor}
\usepackage{soul}
\usepackage{float}
\usepackage{comment}
\usepackage{tipa}
\usepackage{appendix}
\usepackage{stackengine}
\hypersetup{colorlinks=true,citecolor=romared,linkcolor=romared,urlcolor=romared}
\usepackage{threeparttable}
\graphicspath{{NewPlotStyle/}}
\usepackage{subfig}

\usepackage[capitalise]{cleveref}
\crefformat{equation}{Eq.~#2#1#3}
\crefformat{table}{Tab.~#2#1#3}
\crefformat{figure}{Fig.~#2#1#3}
\crefformat{appendix}{App.~#2#1#3}
\crefformat{section}{Sec.~#2#1#3}
\crefformat{subsection}{Subsec.~#2#1#3}

\definecolor{romared}{RGB}{142,0,28}

\newcommand{\be}{\begin{equation}}
\newcommand{\ee}{\end{equation}}
\newcommand{\bea}{\begin{eqnarray}}
\newcommand{\eea}{\end{eqnarray}}

\begin{document}

\title{Disentangling photon rings beyond General Relativity with future radio-telescope arrays}
\author[a]{Raúl Carballo-Rubio,}
\author[a]{Héloïse Delaporte,}
\author[a]{Astrid Eichhorn}
\author[b]{and Aaron Held} 

\affiliation[a]{CP3-Origins, University of Southern Denmark, Campusvej 55, DK-5230 Odense M, Denmark}
\affiliation[b]{
Institut de Physique Théorique Philippe Meyer, Laboratoire de Physique de l’\'Ecole normale sup\'erieure (ENS), Universit\'e PSL, CNRS, Sorbonne Universit\'e, Universit\'e Paris Cité, F-75005 Paris, France
}

\emailAdd{raulc@sdu.dk}
\emailAdd{hdel@sdu.dk}
\emailAdd{eichhorn@cp3.sdu.dk}
\emailAdd{aaron.held@uni-jena.de}

\abstract{New physics beyond General Relativity can modify image features of black holes and horizonless spacetimes and increase the separation between photon rings. 
This motivates us to explore synthetic images consisting of two thin rings. 
Our synthetic images are parameterized by the separation as well as the relative flux density of the two rings.
We perform fits to the visibility amplitude and analyze closure quantities. The current Event Horizon Telescope array cannot detect the presence of a second ring in the region of parameters motivated by particular new-physics cases. We show that this can be improved in three ways: first, if the array is upgraded with Earth-based telescopes with sufficiently high sensitivity, second, if the array is upgraded with a space-based station and third, if super-resolution techniques are used for the data obtained by the array.
}

\maketitle

\section{Introduction and Motivation: Two rings as a smoking gun of new physics}

Black holes are among the most promising gravitational systems to discover new physics, because the theoretical understanding of these objects based on General Relativity (GR) and quantum mechanics is incomplete. Where exactly in a black-hole spacetime our understanding breaks down, is unknown. It was long assumed that GR holds up to very high curvature scales, right up to the core of the black hole. Thus, our incomplete understanding was considered as a mere theoretical conundrum with no impact on observations. However, new theoretical developments \cite{Chapline:2000en,Mazur:2001fv,Mathur:2005zp,Almheiri:2012rt,Barcelo:2014npa,Rovelli:2014cta,Haggard:2014rza,Barcelo:2014cla,Barcelo:2015noa,Haggard:2016ibp,Bacchini:2021fig,Eichhorn:2022bbn,Bena:2022rna,Eichhorn:2023xxx} are starting to challenge 
this perspective, while
recent observations are pushing the boundary of what is observable~\cite{LIGOScientific:2016aoc,EventHorizonTelescope:2019dse,NANOGrav:2023gor}. This trend will continue with the next-generation Event Horizon Telescope (ngEHT) \cite{Roelofs:2022lux,Johnson:2023ynn,Ayzenberg:2023hfw}.

We focus on very-long-baseline interferometry (VLBI) observations of supermassive black holes. The groundbreaking observations of M87$^*$~\cite{EventHorizonTelescope:2019dse, EventHorizonTelescope:2019uob, EventHorizonTelescope:2019jan, EventHorizonTelescope:2019ths, EventHorizonTelescope:2019pgp, EventHorizonTelescope:2019ggy, EventHorizonTelescope:2021bee, EventHorizonTelescope:2021srq} and Sgr A$^*$~\cite{EventHorizonTelescope:2022wkp, EventHorizonTelescope:2022apq, EventHorizonTelescope:2022wok, EventHorizonTelescope:2022exc, EventHorizonTelescope:2022urf, EventHorizonTelescope:2022xqj} by the Event Horizon Telescope (EHT) collaboration are consistent with GR, but may also be explained by alternatives to GR black holes. 
Possible alternatives that have been studied in this context are regular black holes~\cite{Lamy:2018zvj,Vincent:2020dij,Kumar:2020ltt,Eichhorn:2021etc,Eichhorn:2021iwq,Eichhorn:2022oma,KumarWalia:2022ddq,Ling:2022vrv,Islam:2022wck}, horizonless (ultra) compact objects~\cite{Cunha:2017wao,Eichhorn:2022oma,Eichhorn:2022fcl,Eichhorn:2022bbn,Guerrero:2022msp,Carballo-Rubio:2022bgh}, wormholes~\cite{Guerrero:2022qkh,Olmo:2023lil,Neto:2022pmu},
as well as black holes in theories beyond GR \cite{Gyulchev:2021dvt, Sengo:2022jif}, see \cite{Ayzenberg:2023hfw} for an overview.  

Because the image of a black hole depends both on the spacetime geometry and on the emission properties and structure of the accretion disk, a clean distinction of Kerr black holes from alternatives is not simple~\cite{Lara:2021zth,Kocherlakota:2022jnz,Younsi:2021dxe}. Photon rings \cite{darwin1959gravity, Bardeen:1972fi, Luminet:1979nyg, Gralla:2019xty, Johnson:2019ljv,Cardenas-Avendano:2023dzo} are increasingly coming into focus as the cleanest probe
of black-hole spacetimes in the context of images. Other VLBI image features, most importantly the direct emission, are highly dependent on the profile of the direct emission, which is in turn determined by the astrophysics of the disk. Different properties of accretion disks (e.g., MAD~\cite{bisnovatyi1974accretion, Igumenshchev:2003rt, Narayan:2003by} vs SANE~\cite{DeVilliers:2003gr, Gammie:2003rj, Narayan:2012yp} accretion flows) lead to distinct properties of the $n=0$ emission. At increasing order $n$ of the photon rings, the physics of the disk has a decreasing impact. This makes photon rings important targets for the EHT and ngEHT \cite{Broderick:2021ohx,Ayzenberg:2023hfw,Johnson:2023ynn}.

VLBI observations are not the only way to probe black holes beyond GR; gravitational waves are also a powerful probe \cite{LIGOScientific:2018dkp, LIGOScientific:2019fpa, LIGOScientific:2020tif, LIGOScientific:2021sio, Jiang:2023kpx}. However, we focus on VLBI observations for two reasons, a theoretical and a pragmatic one.

The theoretical reason is that we do not expect that black-hole uniqueness theorems generically hold beyond GR. Simple examples supporting this expectation include quadratic gravity \cite{Lu:2015cqa}, semi-classical gravity \cite{Fernandes:2023vux} or scalar-Gauss-Bonnet theory with scalarized black-hole solutions in addition to the Kerr solution \cite{Doneva:2017bvd,Dima:2020yac, Doneva:2022ewd}. 
Thus, it is not excluded that supermassive black holes, (so far) only accessed by VLBI techniques, and solar-mass black holes, accessed by the LIGO/Virgo collaboration, correspond to different branches of solutions of a theory beyond GR. A specific example in which supermassive black holes correspond to a different, non-Kerr branch of solutions, while solar-mass black holes are described by the Kerr solution, can be found in~\cite{Eichhorn:2023xxx}.
Thus, pursuing tests of the Kerr hypothesis across all mass ranges is crucial, even if LIGO/Virgo effectively accesses higher curvature scales than the EHT - it may be the case that even though the curvature at their horizon is lower, supermassive black holes exhibit a larger deviation from the Kerr spacetime, because their different formation history accesses a different branch of solutions of the theory beyond GR.

The pragmatic reason is that VLBI techniques only require knowledge of a spacetime metric to interpret the data. In contrast, gravitational-wave observations require knowledge of the full dynamics beyond GR and numerical simulations of binary black-hole mergers. This requires a well-posed initial-value formulation of the equations of motion that is amenable to numerical simulations. Finding such a formulation is a significant challenge in many settings beyond GR, that has only partially been met in a small subset of theories \cite{Witek:2020uzz,Silva:2020omi,Corman:2022xqg, Held:2023aap}. In contrast, to compare a spacetime beyond Kerr to VLBI data, one does not even need to know which dynamics that spacetime is a solution to. It is, therefore, a promising strategy to first use VLBI data to constrain possible deviations from the Kerr spacetime. If a deviation is found, the much more challenging steps of constructing a possible underlying theory, formulating the dynamics in such a way that it is amenable to numerical simulations~\cite{Pretorius:2005gq,Bishop:2016lgv} and then comparing to gravitational wave data (including from observatories at frequencies relevant for mergers of supermassive black holes), can be undertaken.

\subsection{Photon rings in spacetime geometries beyond GR
}\label{sec:models}

Black holes (in and beyond GR) and exotic compact objects produce strong gravitational lensing. They generically give rise to photon rings, which are higher-order (lensed) images of the objects' surroundings that can be labelled by a set of integers $n$. In GR, the separation between successively higher-order photon rings and their total, integrated flux density decrease exponentially~\cite{darwin1959gravity, Bardeen:1972fi, Luminet:1979nyg, Gralla:2019xty, Johnson:2019ljv}.\footnote{This is true at high enough order. For low orders, there is a strong dependence on the astrophysics of the disk and higher-order rings may even appear outside of lower-order ones in the image plane \cite{Broderick:2021ohx}, if the emission region is located close enough to the horizon. The latter may not be the physically most relevant situation, because emission typically comes from the accretion disk, which is expected to have a cutoff at the ISCO.} This makes higher-order photon rings challenging to access in GR. For instance, if M87* and Sgr A* are described by the Kerr geometry\footnote{In practise, the presence of the accretion disk introduces (essentially negligible) deviations from the Kerr spacetime even in GR.}, then their images, reconstructed from the EHT observations, are dominated by the $n=0$ direct emission. For the EHT array, already the $n=1$ ring is inaccessible without super-resolution techniques \cite{Broderick:2022tfu}. The $n=2$ ring requires a baseline that can only be achieved with space-VLBI \cite{Vincent:2022fwj}.

\begin{figure*}[h!]
\captionsetup[subfigure]{labelformat=empty}
\centering
\subfloat[]{\includegraphics[width=0.78\linewidth]{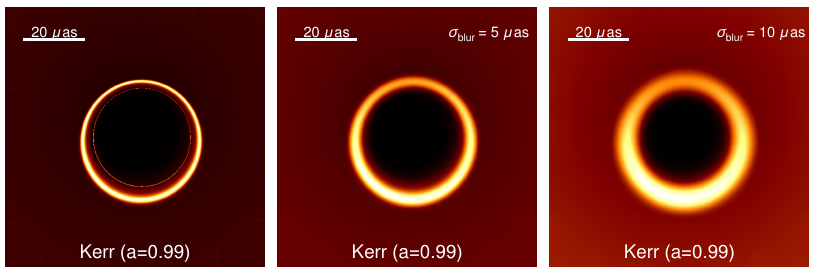}}\label{fig:multiring-Kerr}\\[-17pt]
\subfloat[]{\includegraphics[width=0.78\linewidth]{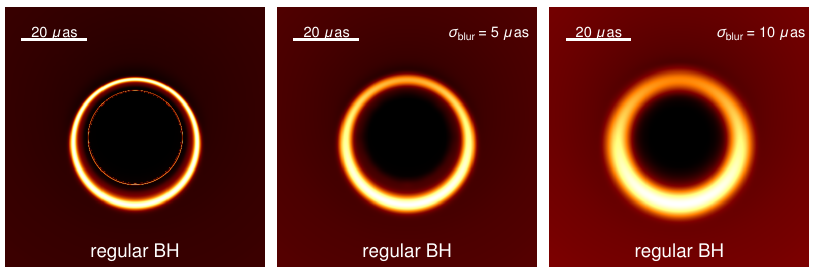}}\label{fig:multiring-regular}\\[-17pt]
\subfloat[]{\includegraphics[width=0.78\linewidth]{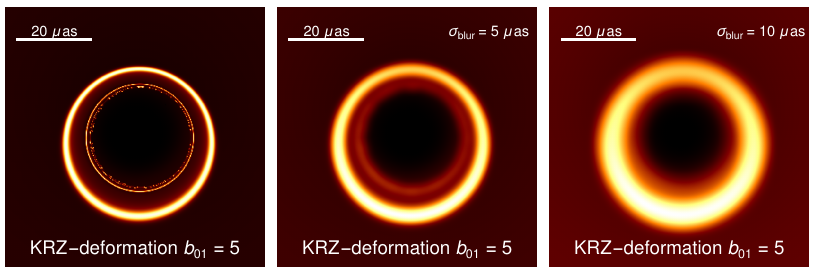}}\label{fig:multiring-KRZ}\\[-17pt]
\subfloat[]{\includegraphics[width=0.78\linewidth]{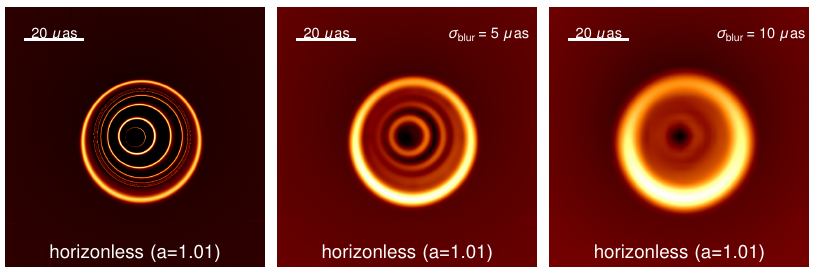}}\label{fig:multiring-horizonless}

\caption{For each spacetime example, we show three shadow images from the ideal image (left column) to the ideal image along with a Gaussian blurring of variance $\sigma_{\rm blur} = 5\, \mu{\rm as}$ (middle column) and finally with a Gaussian blurring of variance $\sigma_{\rm blur} = 10\, \mu{\rm as}$. As ${\rm FWHM} = 2 \sqrt{2 \ln{2}}\, \sigma_{\rm blur}$, the variances of the Gaussian blurrings correspond to FWHMs of $\sim 12\, \mu{\rm as}$ and $\sim 24\, \mu{\rm as}$ (within the current nominal EHT resolution), respectively.
 Top: Kerr black hole with spin $a=0.99\,M$. The image is generated with a disk model as in \cite[slow falloff model in Tab.~1]{Eichhorn:2022bbn}. Second row: regular black hole with exponential falloff function, see e.g.~\cite[Eq.~(3)]{Eichhorn:2021etc}. Third row: circular~\cite{Papapetrou:1966zz} deformation in the KRZ parameterization~\cite{Konoplya:2016jvv} of a Kerr black hole with spin $a=0.9\,M$ and a single deformation parameter $b_{01}=5$. Bottom: marginally overspun (with $a=1.01\,M$) and thus horizonless regular spacetime, cf.~\cite{Eichhorn:2022fcl}.}
\label{fig:motivation_images}
\end{figure*}

Many alternatives to Kerr black holes are characterized by a photon-ring structure that is distinct from the predictions of GR. In particular, photon rings may be more separated, calling for a broadening of dedicated studies in GR~\cite{Johnson:2019ljv, Himwich:2020msm, Broderick:2021ohx, Paugnat:2022qzy, Vincent:2022fwj, Tiede:2022grp, Lockhart:2022rui} to settings beyond GR \cite{Wielgus:2021peu,Staelens:2023jgr,Ayzenberg:2022twz,Cardenas-Avendano:2023obg}. We discuss three examples to make this point: regular black holes, horizonless spacetimes and parametrized black-hole spacetimes beyond GR. We will compare these examples with a Kerr black hole, depicted in the top row of Fig.~\ref{fig:motivation_images}.

As a first example, for regular black holes, the photon rings are more widely separated under the conditions spelled out in \cite{Eichhorn:2022oma}.  Regular black holes arise: (i) in approaches to quantum gravity, e.g., in asymptotically safe quantum gravity \cite{Bonanno:2000ep,Falls:2010he,Platania:2019kyx}, (see~\cite{Eichhorn:2022bgu,Platania:2023srt} for recent reviews) and Loop Quantum Gravity \cite{Ashtekar:2005qt,Modesto:2005zm,Campiglia:2007pb} (see \cite{Ashtekar:2023cod} for a recent review), (ii) in GR coupled to particular matter theories \cite{Ayon-Beato:1998hmi,Ayon-Beato:1999kuh,Bronnikov:2000vy} (see~\cite{Bronnikov:2022ofk} for a recent review), and (iii) in degenerate-higher-order-scalar-tensor (DHOST) theories \cite{Babichev:2020qpr, Baake:2021jzv}. They have been proposed as phenomenological models for black-hole spacetimes beyond GR \cite{Hayward:2005gi,Simpson:2018tsi,Carballo-Rubio:2019fnb,Mazza:2021rgq,Simpson:2021dyo,Carballo-Rubio:2023mvr}, see also \cite{Bambibook} for a recent review of various aspects of regular black-hole spacetimes\footnote{Regular black holes are usually understood as phenomenological models, but not yet as the ultimate and correct description of a fully consistent black-hole spacetime~\cite{Carballo-Rubio:2018pmi}, due to the following reason: regular black holes contain inner horizons, which become Cauchy horizons if not disappearing in finite time due to evaporation or some other process. The spacetime region around inner horizons generically displays an exponential focusing of null rays unless the inner surface gravity vanishes~\cite{Carballo-Rubio:2022kad,Franzin:2022wai}, which results in an exponential mass inflation phase in which curvature invariants grow exponentially~\cite{Carballo-Rubio:2021bpr,Bonanno:2020fgp} (see also~\cite{Barcelo:2022gii,DiFilippo:2022qkl,Bonanno:2022jjp,Carballo-Rubio:2022twq}). The endpoint of this dynamical evolution is unknown, and is a question to be addressed in specific dynamical frameworks leading to regular black hole solutions.}. To tame the curvature singularity of a Kerr black hole, a regular black hole is based on a \emph{mass function} which approaches the ADM mass in Kerr at asymptotic infinity and goes to zero in the core of the black hole. This mathematical description models an expected physical effect, namely a weakening of gravity through quantum fluctuations or appropriate matter fields. Because gravity is weakened compared to a Kerr black hole, the event horizon and the photon sphere are both located closer to the center of the black hole, with a larger shift for the event horizon than for the photon sphere. Similarly, null geodesics that approach the photon sphere are pulled further inwards more strongly, if they orbit closer to the photon sphere. Thus, higher-order photon rings end up further away from the low-order photon rings than they do in a Kerr black hole \cite{Eichhorn:2022oma}, cf. second row in Fig.~\ref{fig:motivation_images}.

As a second example, photon-ring separations in parametrized spacetimes beyond Kerr can also be significantly larger than in the Kerr spacetime.
To describe black-hole spacetimes beyond GR, deviations from the Kerr geometry are encoded in general parameterizations. Parameterizations differ in the degree of symmetry they impose: a completely general axisymmetric and stationary parameterization would include non-circular spacetimes~\cite{Delaporte:2022acp}. Circular spacetimes are parameterized in~\cite{Papapetrou:1966zz, Weyl:1917gp, Lewis:1932zz, Konoplya:2016jvv} and spacetimes which additionally feature a Carter-like constant of motion are parameterized in \cite{Benenti:1979erw, Johannsen:2013szh, Vigeland:2011ji}. Within the parameter space covered by, e.g., the circular KRZ parameterization \cite{Konoplya:2016jvv}, there are cases with a photon-ring separation that is significantly larger than that of a Kerr black hole with the same mass and embedded in the same accretion disk, cf. third row in Fig.~\ref{fig:motivation_images}. The particular deformation that we have included does not introduce deviations of the first parametrized post-Newtonian (PPN) parameters~\cite{1993tegp.book.....W} and, therefore, evades all solar-system constraints.

Horizonless spacetimes are our third example, which have been proposed in several frameworks, including semiclassical gravity~\cite{Carballo-Rubio:2017tlh,Arrechea:2021xkp,Carballo-Rubio:2022nuj}, string theory~\cite{Mathur:2005zp,Bena:2022rna} and asymptotic safety~
\cite{Bonanno:2000ep,Falls:2010he,Bonanno:2019ilz,Eichhorn:2022bbn}. They can feature a secondary set of (inner) photon rings which can be labelled by a second set of (positive) integers $m>0$. These can arise from trajectories that approach the photon sphere from the inside, e.g., for regular black-hole spacetimes at supercritical spin parameter or supercritical new-physics scale~\cite{Guerrero:2022qkh,Eichhorn:2022oma,Eichhorn:2022fcl,Eichhorn:2022bbn}. Even in the absence of a photon sphere, a finite number of such inner photon rings can exist \cite{Eichhorn:2022oma}. If the horizonless spacetime contains an ultracompact object, reflection off the surface can also result in a secondary ring structure due to strong lensing \cite{Carballo-Rubio:2022aed}. Finally, even naked singularities can result in such a secondary set of photon rings~\cite{Shaikh:2018lcc,Guerrero:2022msp,Gyulchev:2020cvo}\footnote{It is widely expected that singularities should be absent from physical solutions of an everywhere viable theory of gravity. Nevertheless, investigating the images of spacetimes with naked singularities is important, because it lays the basis to test this expectation observationally.}.
The $n=1$ and $m=1$ rings can be well separated and have similar flux densities, e.g., for a reflective surface with reflection coefficient close to unity~\cite{Carballo-Rubio:2022bgh}, or for an overspun regular black hole, cf. bottom row in Fig.~\ref{fig:motivation_images}.

Thus, beyond GR, images containing two rings with larger separation can occur. Similarly, the difference between the flux densities can be much less pronounced. This motivates a study of the capabilities of current and future VLBI arrays for the detection of two such rings, irrespective of theoretical assumptions from GR. We will take a step in this direction by considering a geometric image-plane model of two thin rings.

This paper is organized as follows: in Sec.~\ref{sec:syntheticdata}, we define the geometric two-ring model. In Sec.~\ref{sec:Fourier plane-analysis}, we work in the Fourier plane, as is appropriate for image reconstruction from VLBI data. We discuss the visibility amplitude associated to the 2-ring model and perform fits to the synthetic data obtained when images of the same model are observed with current and future VLBI arrays. We find that a part of the parameter space, which we consider to be physically relevant, is in fact detectable by the (ng)EHT. 
In Sec.~\ref{sec:closure}, we perform an analysis using closure quantities, which provide a cleaner measurement removing some of the errors associated to the VLBI array, while offering a different perspective on the comparison between 1-ring and 2-ring models. 
We conclude and provide an outlook in Sec.~\ref{sec:Conclusions and outlook}.

\section{The model: synthetic data}
\label{sec:syntheticdata}

Images of black holes are reconstructed from complex visibilities, which are related to the Fourier transform of the flux density. The complex visibilities associated to a pair of telescopes in a VLBI array are the data products coming out of VLBI observations.  
We define our model in the image plane and translate it into the Fourier plane.

Our synthetic data depends on six parameters, two of which are held fixed at values inspired by the EHT observations of M87*~\cite{EventHorizonTelescope:2019dse}.

The first parameter is $F_{\rm tot}$, the total flux density, which we keep fixed to $F_{\rm tot}=0.7\,\text{Jy}$, motivated by the EHT observation of M87*~\cite{EventHorizonTelescope:2019ths}. The total flux density enters our analysis in relation to the sensitivity of the respective telescopes in current and future VLBI arrays.

The second parameter is $\Delta F$, the relative flux density between the two rings, i.e.,
\be
\Delta F = \frac{F_2}{F_1},
\ee
where $F_i$ is the flux density in the $i$th-ring. Thus, the respective flux densities are given by
\be
F_1 = \frac{F_{\rm tot}}{1+\Delta F},\quad F_2 = \frac{F_{\rm tot}}{1+1/\Delta F}.
\label{eq:FluxInEachRing}
\ee

Further, we specify the geometry of the rings. The outer ring is kept at fixed diameter, $d_1=42\,\text{$\mu$as}$, again inspired by the EHT observation of M87* \cite{EventHorizonTelescope:2019dse}. 
We do not vary the widths $\omega_{1,2}$ of both rings. Finally, the diameter of the second ring, $d_2$, is changed implicitly, by varying the separation $s$ between the two rings. The relation between separation and diameter is given by $d_2=d_1-2s$.

\begin{figure}[t]
    \centering
    \includegraphics[width=0.5\linewidth]{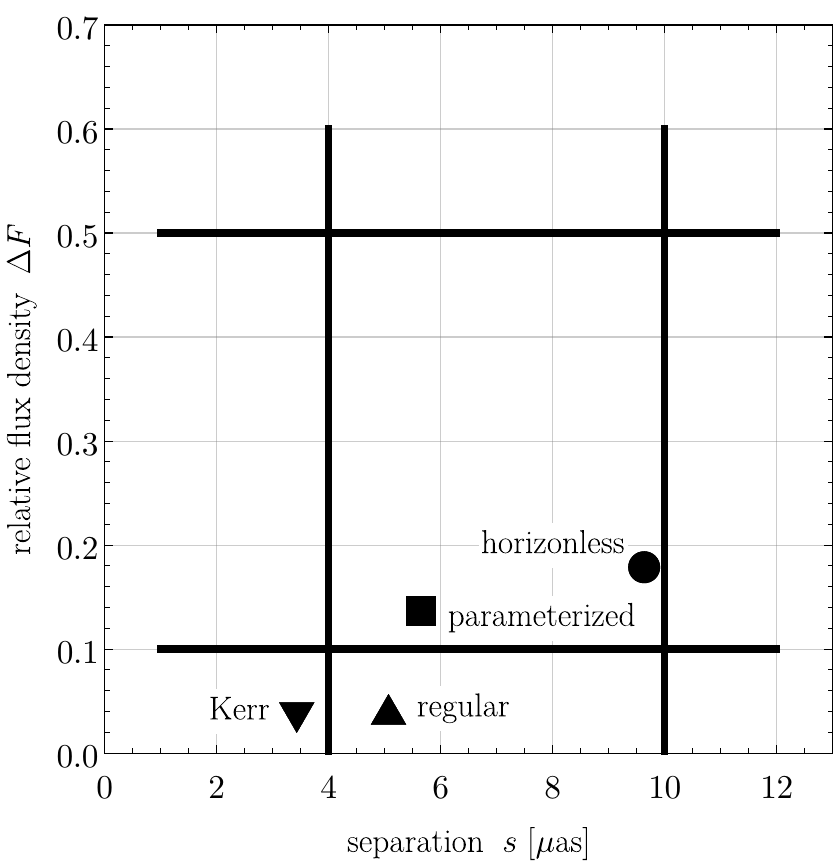}
    \caption{
    We visualize various estimates obtained from theoretically expected parameter ranges of possible 2-ring structures in the $(s,\,\Delta F)$ projection of the model parameter space: the different markers correspond to the expected parameter values of the four spacetimes in Fig.~\ref{fig:motivation_images}, i.e. Kerr, regular black holes, parametric deformations and horizonless objects, respectively. The points are obtained by minimizing the divergence between a Gaussian two-ring
    model, see Eq.~\eqref{eq:GaussianProfile}, and the obtained synthetic image.
    The black lines indicate lines along which we perform our analysis in Sec.~\ref{sec:Fourier plane-analysis}.
    \label{fig:s-DeltaF-plane}}
\end{figure}
From a simple Gaussian blurring of the images (cf. third and bottom rows in Fig.~\ref{fig:motivation_images}), we expect that the separation $s$ and the relative flux density $\Delta F$ are the most relevant parameters to determine the detectability of a second ring, at least in the limit of relatively thin rings which is realized in our examples. We thus visualize in Fig.~\ref{fig:s-DeltaF-plane} where the specific examples, which motivate our study, are located in the $(s,\,\Delta F)$ plane.\\

To generate synthetic data, we define a flat profile for each ring, based on the auxiliary function
\begin{equation}
\mu(r;d,\omega)=\frac{1}{\pi d \omega}\Theta\left(\frac{d\!+\! \omega}{2}\! - \! r\right)\Theta\left(r\! - \frac{d\! - \! \omega}{2}\right),
\label{eq:AuxiliaryFuncFlatProfile}
\end{equation}
which is normalized such that
\be
\int_0^{\infty}dr\, 2\pi r\int_0^{2\pi} d\theta\, \mu(r;d,\omega) = 1.
\label{eq:NormalisationAuxiliaryFunc}
\ee
The total resulting flux density profile combines two flat rings and is given by a function of the radius in the image plane as
\begin{equation}
F_{\rm Cresc}[r]= \frac{F_{\rm tot}}{1+\Delta F}\mu(r;d_1,\omega_1)+\frac{F_{\rm tot}}{1+1/\Delta F} \mu(r;d_1-2s,\omega_2).\nonumber
\label{eq:CrescentProfile}
\end{equation}
The prefactors appearing in front of the auxiliary function take this form, because we parameterize the flux density in terms of the total flux density in the image and the total flux density in each ring, instead of the peak flux- density. While the latter would be a simpler parameter to build a model with, the former is closer to the observations, because the EHT is sensitive to the total flux density in the image. The prefactors correspond to $F_{1,2}$, respectively, such that
\be
\int_0^{\infty}dr\, 2\pi r\int_0^{2\pi}d\theta F_{\rm Cresc}[r] = F_{\rm tot}, 
 \label{eq:NormalisationCrescentProfile}
\ee
and that each ring's total flux density is normalized to either $F_1$ or $F_2$. This flat profile $F_{\rm Cresc}$ corresponds to the addition of two crescent models introduced in \cite{Kamruddin:2013iea}.\\

Then, we produce a synthetic image based on a discretization of the crescent profile. In Cartesian coordinates $(x, y)$, we center both rings at $(x_0, y_0) = (0, 0)$ in our image, where $x$ refers to the relative right ascension (rRA, in $\mu \mathrm{as}$) on the horizontal axis, and $y$ to the relative declination (rDEC, in $\mu \mathrm{as}$) on the vertical axis. 

We construct the synthetic image data by discretizing the flux density profile in a two-dimensional array of pixels $\left(x_k, y_k, F(x_k, y_k)\right)$, where the index $k$ runs from $1$ to the square of the number of pixels $N_{\rm pix}^2$. Each pixel in our square array carries a flux density $F(x_k, y_k)$.\\
The spans of $x$- and $y-$axis, which are symmetric, equal and centered on $x_0 = 0$ ($y_0 = 0$ respectively), define a field of view (FOV, in $\mu \mathrm{as}$) as:
\begin{equation}
\rm{FOV} = (N_{\rm pix} - 1) \cdot \delta \theta,
    \label{eq:def_FOV}
\end{equation}
where $\delta \theta$ is the pixel ``length'' (in $\mu \mathrm{as}$).

The crescent profile in Eq.~\eqref{eq:CrescentProfile} makes up our synthetic data and has an analytically known form in the Fourier plane \cite{Kamruddin:2013iea}, which provides us with an analytical fitting function.
We do not vary the flux density profile within individual rings,
because we focus on the thin ring limit, for which
the widths of the two rings are small compared to their diameters $\omega_i \ll d_i$  
and hence the type of profile does not matter (see Fig.~\ref{fig:p-value-test_s-DeltaF-plane_width_profile} in~\cref{app:ComparisonProfiles} for a demonstration). In actual observations, in addition to thin rings\footnote{We expect that photon rings are thin compared to their diameter also in theories beyond GR and are not aware of a counterexample to this expectation.}, there is a much broader, diffuse emission from the accretion disk. We do not account for this broad image feature in our analysis and comment on the resulting limitations in the conclusions.

\section{Visibilities in the Fourier plane}\label{sec:Fourier plane-analysis}
The Fourier transform of any continuous intensity model in the image plane leads to the complex visibility
\begin{equation}
V(u,v) = \int \int dx\, dy\, I(x,y)\, e^{-\frac{2 \pi i (ux + vy)}{\lambda}},
 \label{eq:visibility}   
\end{equation}
where $\lambda$ is the the wavelength of observation, $(x,y)$ are angular coordinates on the sky, $(u,v)$ are
projected (on the plane orthogonal to the line of sight) baseline coordinates, and $I(x,y)$ in units of $10^{26} {\rm Jy} \cdot {\rm sr^{-1}}$ is the intensity model, related to the flux density profile from the previous section by a dimensionless factor of solid angle in steradian (${\rm sr}$), see below. It is standard to express $(u,v)$ in units of $10^9 \cdot \lambda$ (i.e. $\mathrm{G} \lambda$).

In practice, the discrete array of antennas does not sample $V(u,v)$ continuously. Thus we define its discrete counterpart
\begin{equation}
V_{ij}=V(u_{ij},v_{ij}),
\end{equation}
where $(u_{ij},v_{ij})$ is the vector associated with stations $i$ and $j$ in the array. 

\subsection{Complex visibilities for the crescent model}

Calculating $V_{ij}$ analytically is only possible for particular choices of synthetic data; one of those is the crescent model~\cite{Kamruddin:2013iea} on which our fitting profile $F_{\rm Cresc}$ is based. The resulting expression $V_{ij,\, \rm Cresc}$ is based on the Fourier transform of a single disk of radius $R$,
\be
V_d(k,
F_0, R) = \pi R^2 I_0\, \frac{2 J_1(k\, R)}{k\, R},
\ee
where $I_0$ is the constant intensity of the disk and $J_1$ is the Bessel function of the first kind of order 1. Because a disk provides a radially symmetric flux density in the image plane, the resulting Fourier transform does not depend on $u$ and $v$ separately, but only through the combination 
\be
k =\frac{2\pi}{\lambda} \sqrt{u^2+v^2}.
\ee 
Based on this expression,~\cite{Kamruddin:2013iea} provide the visibility amplitude for a 1-ring flux-profile, by subtracting the visibility amplitude of two disks with outer and inner radius:
\begin{align}
	V(k) &=
	\frac{2\pi I_0}{k} \Big[
		R_{\rm outer} J_1(k R_{\rm outer}) 
		- R_{{\rm inner}}J_1(k R_{{\rm inner}}
	)\Big] 
	\notag\\
	&= 
	\frac{2 F_0}{k \left(R_{\rm outer}^2 - R_{\rm inner}^2\right)} \Big[
		R_{{\rm outer}} J_1(k R_{{\rm outer}}) - R_{{\rm inner}} J_1(k R_{{\rm inner}})
	\Big]\;.
    \label{eq:visibility_1_ring}
\end{align}

To obtain the complex visibility for the 2-ring model, we add the visibilities for two rings, cf.~Eq.~\eqref{eq:visibility_1_ring}:
\begin{align}
	V(k) &= V_1(k) + V_2(k)\nonumber\\
	&= 
	\frac{2\pi I_1}{k} \Big[
		R_{{\rm outer},1} J_1(k R_{{\rm outer},1}) - R_{{\rm inner},1} J_1(k R_{{\rm inner},1})
	\Big]
	\nonumber\\
	&\quad +
	\frac{2\pi I_2}{k} \Big[
		R_{{\rm outer},2} J_1(k R_{{\rm outer},2}) - R_{{\rm inner},2} J_1(k R_{{\rm inner},2})
	\Big].
   \label{eq:visibility_2_rings}
\end{align}
where the intensities are related to the flux densities used in the previous section as $I_j = F_j / \pi (R_{{\rm outer,} j}^2 - R_{{\rm inner,} j}^2)$ and the radii are dimensionless quantities expressed in terms of the physical parameters as
\be
\begin{aligned}[c]
& R_{\rm outer,1} = \frac{d_1 + \omega_1}{2}, \\
& R_{\rm outer,2} = \frac{d_1 - 2s + \omega_2}{2},
\end{aligned}
\quad
\begin{aligned}[c]
& R_{\rm inner,1} = \frac{d_1 - \omega_1}{2}, \\
& R_{\rm inner,2} = \frac{d_1 - 2s - \omega_2}{2}.
\end{aligned}
\label{eq:correspondence_params}
\ee

We will explore the thin-width case, for which we gain some intuition from the strict limit of infinitely thin rings, in which the complex visibility is \cite{Johnson:2019ljv},
\be
V_{\rm thin\, rings}(k) = F_1\, J_0\left(\frac{k\, d_1}{2}\right)+ F_2\, J_0\left(\frac{k\, d_2}{2}\right).
\ee
For large real argument, $ k\, d_1 \gg 3/4$, the $m^{\rm th}$ Bessel function of the first kind can be expanded as
\begin{equation}
J_{m}(z) \simeq \sqrt{\frac{2}{\pi z}} \left[\cos\left(z - \frac{(1+2m)\pi}{4}\right) + \mathcal{O}(\vert z \vert^{-1})\right].
    \label{eq:Bessel1_large_arg}
\end{equation}
For one infinitely thin ring, the visibility amplitude is a damped oscillation with the period set by the inverse diameter, $1/d$. For two infinitely thin rings, where the separation $s$ is not much smaller than the diameter (cf.~Fig.~\ref{fig:s-DeltaF-plane}), two damped oscillations are superposed.

In the case of similar flux densities in the two infinitely thin rings, the zeros of the total visibility amplitude can lie at rather distinct locations from those of the two individual rings. 
However, for the cases we are most interested in, the outer ring carries the largest part of the flux density and thus dominates the visibility amplitude. In that case, the second ring only leads to a slight modulation of the overall visibility amplitude and shifts the locations of the zeros somewhat, but cannot fully remove them (at least in the range of $k$ that we consider and for ring diameters which are of the same order of magnitude, as we consider here).
An example of this is shown in \cref{fig:p-value-test_example}.

\subsection{Generating synthetic data for (ng)EHT observations}

The data from an actual observation with a VLBI array such as the EHT differs from the above idealized discussion in several ways.

First, the combination of limited baseline and observing frequency sets an effective cutoff on the sampling of the complex visibilities, i.e., it limits the maximal resolvable $uv$-distance. For any Earth-based VLBI campaign, the baseline is necessarily limited by Earth's diameter. The frequency is effectively limited by atmospheric scattering, see e.g.~\cite{EventHorizonTelescope:2019uob}. This limits us to $k \lesssim 8.5 \, \rm G\lambda$.
 
Second, the finite number of telescopes leads to a sparse sampling of the complex visibility. Due to Earth's rotation, the projected baselines change during an observation campaign. As a result, several data points in the $uv$-plane can be obtained from each baseline.

Third, the finite sensitivity of each telescope causes additional thermal and systematic errors in each observational data point. The thermal noise is dominated by the system equivalent flux density (SEFD) of the respective telescopes, see e.g.~\cite{EventHorizonTelescope:2019uob}. Overall, the lower the SEFD, the smaller the respective thermal noise. The systematic errors can be factorized as frequency- and time-dependent multiplicative station-based ``gains''.

To capture these effects and thereby gain first quantitative insight into the detectability of multi-ring features, we take the theoretical model defined in~\cref{sec:syntheticdata} and generate synthetic data, as expected from a given telescope array. To do so, we use the \texttt{ehtimaging} toolkit~\cite{chael_2023_8408352}, cf.~also~\cite{Chael:2018oym} for further details of the capabilities of~\texttt{ehtimaging}. In~\cref{tab:arrays}, we summarize the set of arrays used in our exploratory study. The detailed tables of telescope sites and SEFD values required to reproduce our results are provided in~\cref{app:arrays}.
\begin{table*}[h!]
\resizebox{\linewidth}{!}{%
\begin{threeparttable}[t]
\begin{tabular}{|c|c|c|c|}
\hline
{Arrays} & {Total number of sites} & {Frequencies $\nu_\text{obs}$ (GHz)} & {SEFD values (Jy) of the new sites\tnote{a}}
		\\\hline
		EHT 2022 & $11$ & 230 & See values in  Tab.~\ref{tab:array_EHT2022}
		\\\hline
		EHT 2022 & $11$ & 230 \& 345 & See values in  Tab.~\ref{tab:array_EHT2022}
		\\\hline
		ngEHT-230-low-SEFD & $19$ & 230 & 74 (low; ALMA value)\tnote{b}
		\\\hline
		ngEHT-230-high-SEFD & $19$ & 230 & 19300 (high; SPT value)\tnote{c}
		\\\hline
		ngEHT-dualfreq-low-SEFD & $19$ & 230 \& 345 & 250 (low; ALMA value)\tnote{b}
		\\\hline
		ngEHT-dualfreq-high-SEFD & $19$ & 230 \& 345 & 44970 (high; KPNO value)\tnote{c}
		\\\hline
		ngEHT-230-space & $19+1$\tnote{d} & 230 & 74 (low; ALMA value) \& 36600 (space)\tnote{e}
		\\\hline
		ngEHT-dualfreq-space & $19+1$\tnote{d} & 230 \& 345 & 250 (low; ALMA value) \& 56000 (space)\tnote{e}
		\\\hline
		\end{tabular}
\begin{tablenotes}
		\item[a] The new sites are defined as all the Earth-based sites added to the EHT 2022 array.
		\item[b] See the detail of all SEFD values in Tab.~\ref{tab:array_ngEHT_230_345_optimistic}.
		\item[c] See the detail of all SEFD values in Tab.~\ref{tab:array_ngEHT_230_345_pessimistic}.
		\item[d] The ``$+ 1$'' refers to the space-based site.
		\item[e] SEFD values for the space-based site at $230$ and $345$ GHz are estimated from Tab. 1 in \cite{Roelofs:2019nmh}.
\end{tablenotes}
\end{threeparttable}%
}
\caption{We tabulate the specifications of different VLBI arrays used in the complex-visibility analysis. The number of sites determines how sparse the sampling of the Fourier plane is. The frequency influences the maximum $u-v$ distance that is effectively resolved. The system equivalent flux density (SEFD) is a measure for the sensitivity of each telescope, hence the quality of single data points in the Fourier plane: high SEFD values thus correspond to worse data quality. The labels ``low'' and ``high'' in the arrays refer to the SEFD value of the new sites. More detail on the array specifications is given in \cref{app:arrays}.}
\label{tab:arrays}
\end{table*}

We start from the telescope array which is used in the 2022 observing campaign of the Event Horizon Telescope collaboration (EHT 2022). The EHT 2022 array includes 11 telescopes and operates at $230\,\text{GHz}$. Next, we add 8 telescope sites which are discussed as part of the next-generation Event Horizon Telscope (ngEHT) proposal, cf.~\cite{Johnson:2023ynn, 2023Galax..11..107D}. Finally, we also include a single space-based telescope (ngEHT-space) to quantify the potential gain in detectability as compared to purely Earth-based observation campaigns.

In addition to these variations in site selection, we also vary the observation frequency from the current $230\,\text{GHz}$ to $345\,\text{GHz}$, as proposed by the ngEHT collaboration~\cite{Johnson:2023ynn}. 

We work with known SEFD values for the EHT 2022 telescope sites~\cite{EventHorizonTelescope:2019uob,Broderick:2021aeg}, cf.~\cref{app:arrays} for details. The SEFD values for the remaining ngEHT sites depend on design choices which may be informed by science cases such as the one we investigate here. To investigate the effect of varying SEFD values in the additional ngEHT telescope sites, we distinguish between one set of lower and one set of higher SEFD values, low-SEFD and high-SEFD, respectively. For the low-SEFD case, we assume that all future ngEHT sites can reach the SEFD value of ALMA, the most sensitive site in the EHT 2022 array. For the high-SEFD case, we assume that the future ngEHT sites are limited to the SEFD value of SPT/KP, the least sensitive sites in the EHT 2022 array.

With this selection of reference arrays at hand, we use \texttt{ehtimaging} to generate synthetic data for each set of model parameters corresponding to the 2-ring model specified in~\cref{sec:syntheticdata}. 
In particular, we perform scans along each of the four rays indicated in~\cref{fig:s-DeltaF-plane}.

Before we proceed to discuss the results, we describe our procedure to quantify the detectability of multiple rings.

\subsection{A quantitative test of detectability}
\label{sec:p-value_setup}
To decide about the detectability of the second ring, we fit the synthetic dataset with both a 1-ring and a 2-ring model. To perform the fits, we use the \texttt{lmfit} python package~\cite{newville_matthew_2014_11813} which minimizes the least-square residuals between data points and fitting function, also taking into account the error budget at each data point.

With the two fits at hand, we perform a test to quantify whether the 2-ring fit is favored. The nonlinear nature of the problem makes using a reduced chi-squared test questionable~\cite{2010arXiv1012.3754A}. As an alternative, we determine the respective minimized residuals and perform a 2-sample Kolmogorov-Smirnov (KS) test. The latter returns a $p$-value which quantifies how confidently one can exclude the hypothesis that both sets of residuals are drawn from the same probability distribution. The $p$-value quantifies with how much confidence the `1-ring hypothesis' can be rejected. For instance, a $p\leqslant 0.01$ rejects the `1-ring hypothesis' with $99\%$ confidence. For two examples, the synthetic data (black crosses), together with the best-fit 1-ring and 2-ring model, is shown in~\cref{fig:p-value-test_example}. In the left panel, the $p$-value test shows that the two models cannot be distinguished. In the right panel, the data at the largest accessible $uv$-distances are sufficiently distinct to distinguish the two models.

\subsection{Results for simulated observations
}
\label{sec:p-value_s-DeltaF-plane}
Motivated by theoretical studies beyond GR, where photon rings are typically thin, just as in GR, see, e.g., \cite{Eichhorn:2022oma, Ayzenberg:2023hfw,Staelens:2023jgr}, we focus on synthetic data in the limit of relatively thin rings. Specifically, we set $\omega_1= 2\, \mu \rm as$ and $\omega_2= 1 \, \mu\rm as$ in everything that follows but expect that our results do not depend much on this exact choice and would remain
similar for other values of $\omega_{1,2}$, as long as $\omega_{1,2}\ll d$.\\

In this limit, changes of the flux density profile within a ring do not affect the (non-) detectability of the second ring, because these widths are sufficiently far below the resolution limit of the VLBI arrays we investigate. Thus, the simplest fitting profile, i.e., a flat flux density within the rings, described by the crescent model in Eq.~\eqref{eq:CrescentProfile}, suffices. One might worry that using the same fitting profile that is also used to generate the synthetic data could compromise the results. In~\cref{app:ComparisonProfiles}, 
we have explicitly checked that using, e.g., a Gaussian profile to generate synthetic data does not alter our conclusions regarding the (non-)detectability, as shown in Fig.~\ref{fig:p-value-test_s-DeltaF-plane_width_profile}.

\begin{figure*}[t]
    \centering
    \includegraphics[width=0.48\linewidth]{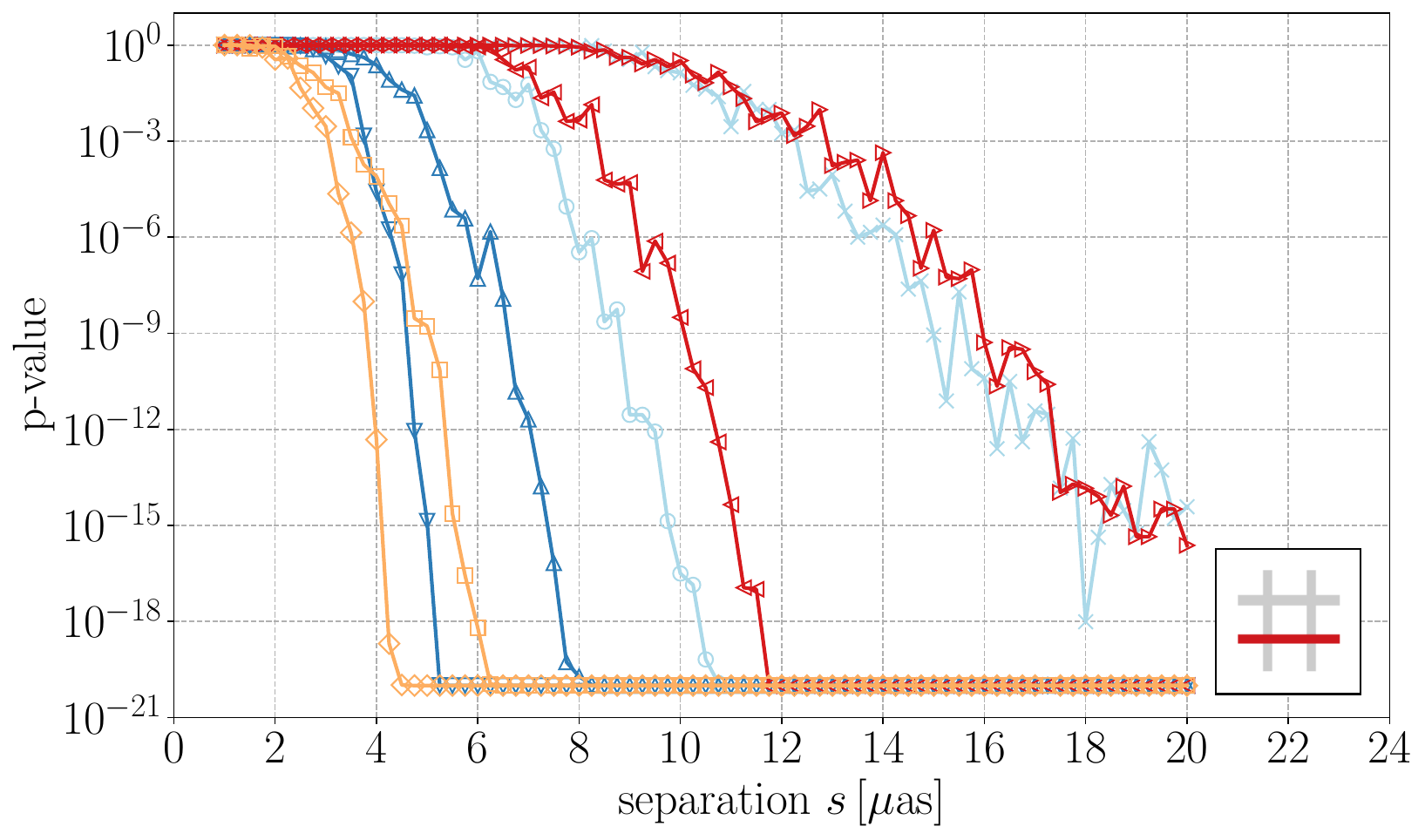}
    \includegraphics[width=0.48\linewidth]{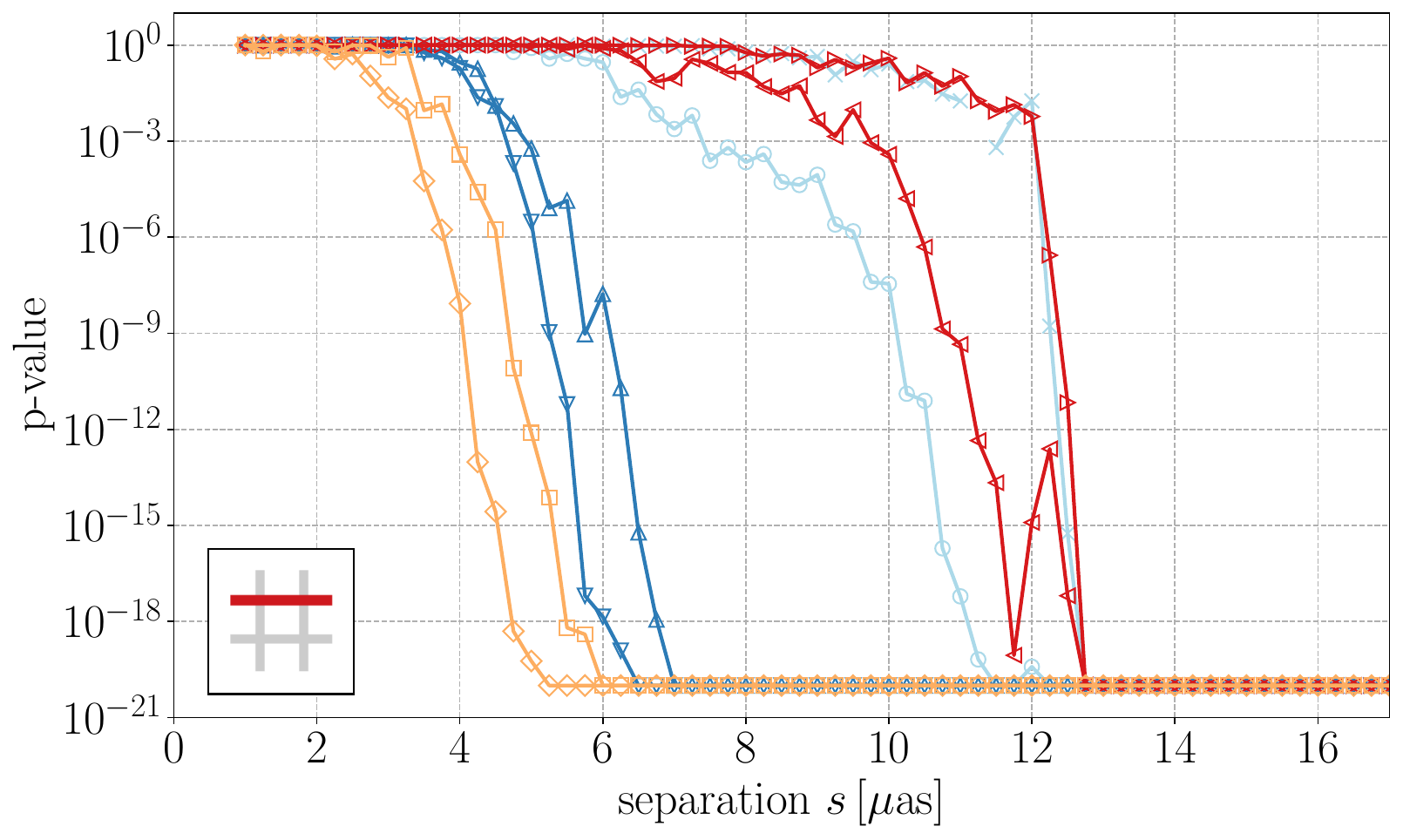}
    \\
    \includegraphics[width=0.48\linewidth]{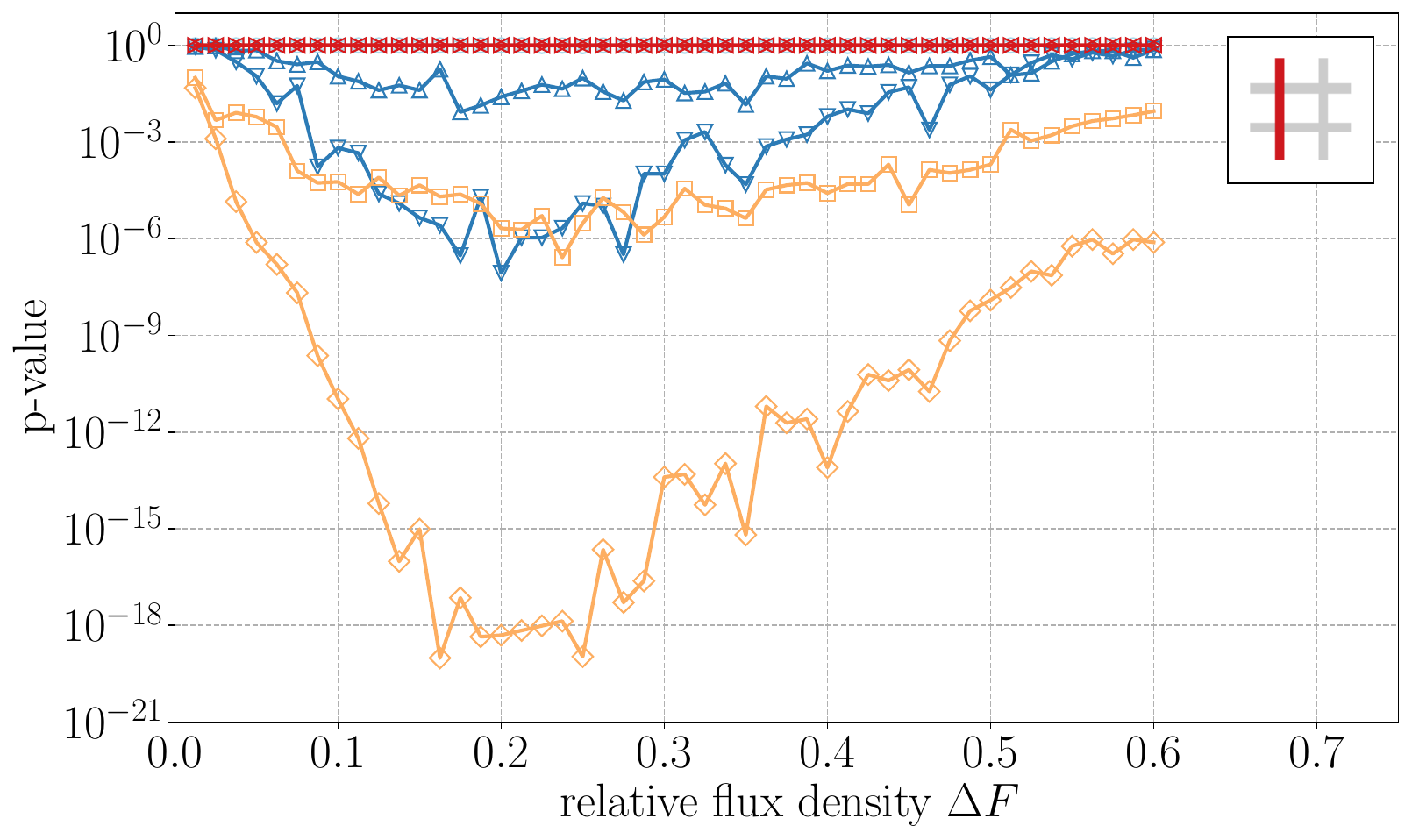}
    \includegraphics[width=0.48\linewidth]{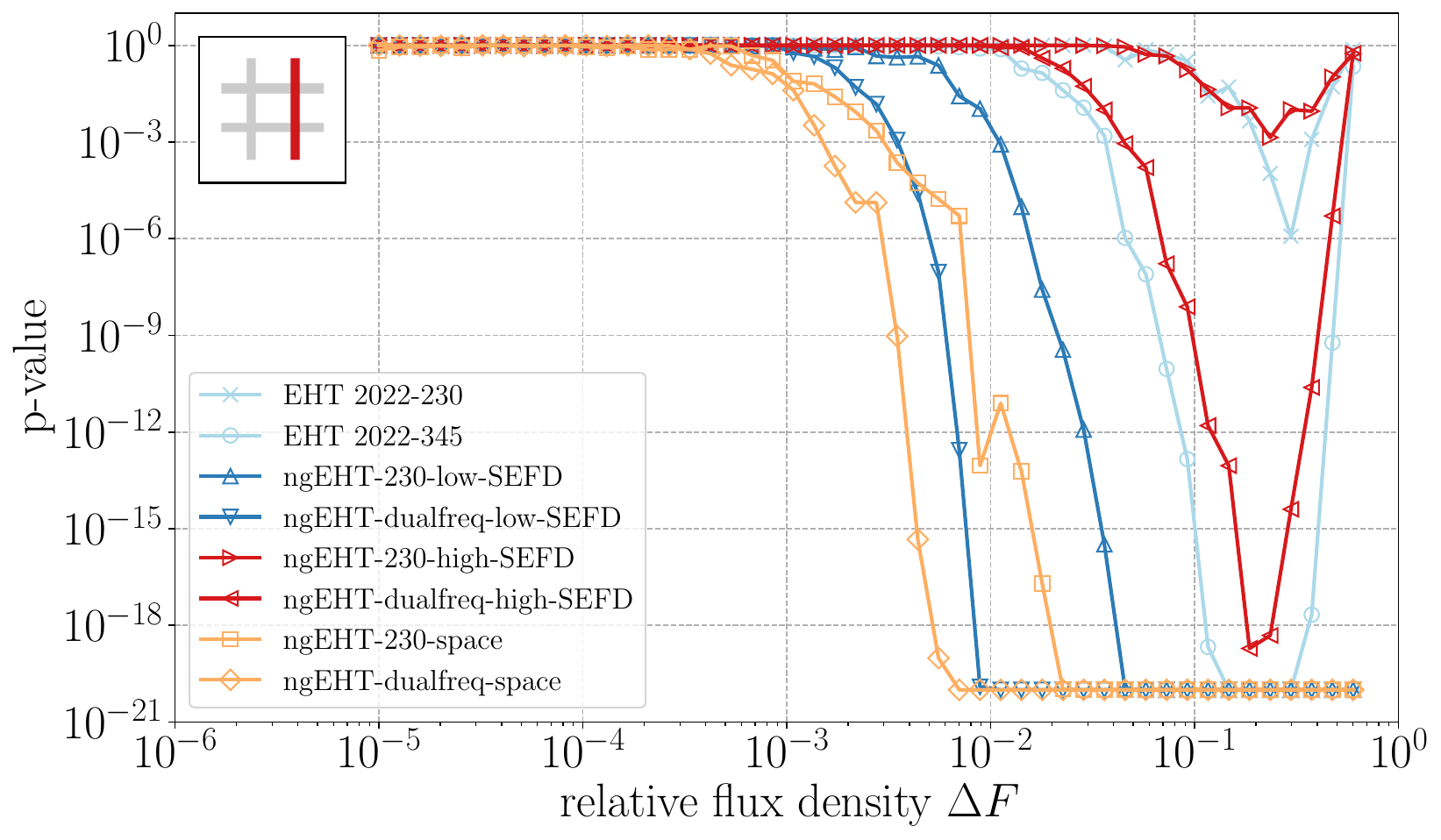}
    \caption{We show the 2-ring detectability (according to the $p$-value test, cf. main text) projected onto the four rays in the $(s,\,\Delta F)$ plane, cf.~Fig.~\ref{fig:s-DeltaF-plane}. A transition of the $p$-value from (close to) one to (close to) zero indicates the transition from non-detectable to detectable cases, see main text. For visual purposes, we have added a $p$-value floor of $10^{-20}$ to all data points.  
    The different lines therefore indicate the varying detectability thresholds that we find for various arrays as in Tab.~\ref{tab:arrays}. In all cases, we focus on the thin-ring limit, i.e., the remaining 2-ring parameters are chosen as $\omega_1 = 2\,\mu$as and $\omega_2 = 1\,\mu$as.
    Moreover, we generate and fit the data with a crescent profile, i.e., the conducted $p$-value test implicitly assumes perfect knowledge about the ring profile. No constraints, especially on the widths, have been added.
     \label{fig:p-value-test_s-DeltaF-plane_no_constraint}
    }
\end{figure*}

\begin{figure*}[t]
    \centering
    \includegraphics[width=0.48\linewidth]{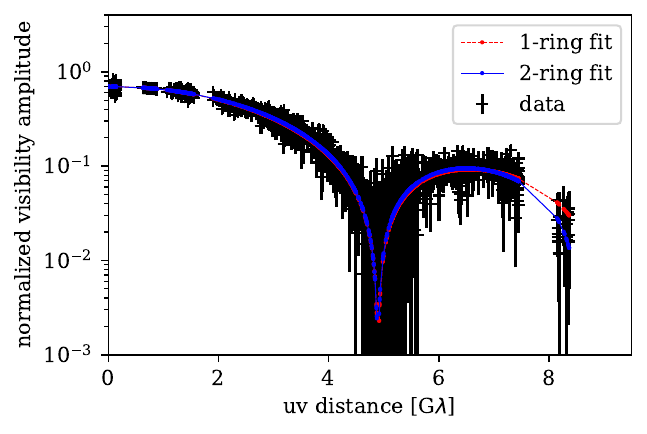}
    \hfill
    \includegraphics[width=0.48\linewidth]{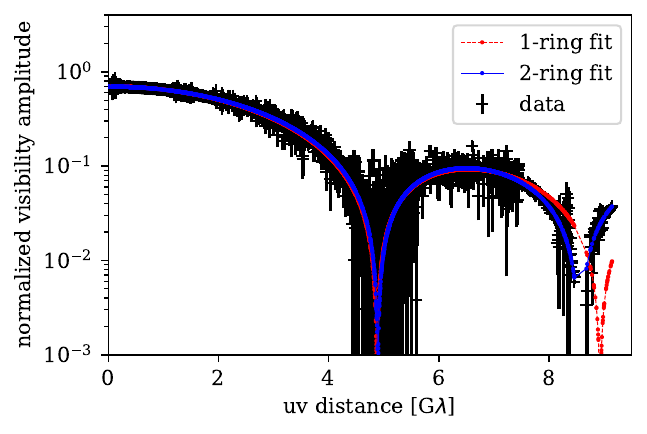}
    \caption{
    Two examples of fits corresponding to the $s=12\,\mu$as case in the right-hand upper panel in Fig.~\ref{fig:p-value-test_s-DeltaF-plane_no_constraint}. The left-hand panel shows simulated data taken with the EHT 2022-230 array and finds no detection. The right-hand panel shows simulated data taken with the ngEHT-230-low-SEFD array and finds a detection.
     \label{fig:p-value-test_example}
    }
\end{figure*}

Thus, the two remaining parameters which determine whether or not a 2-ring-model can be distinguished from a 1-ring model are the relative flux density $\Delta F$ and the separation between the rings, $s$; these two span our two-dimensional parameter space. We perform four scans through this parameter space, as indicated in Fig.~\ref{fig:s-DeltaF-plane}, which are motivated by the new-physics cases we have discussed. For each scan, we consider 8 different array configurations as specified in Tab.~\ref{tab:arrays}, cf.~Fig.~\ref{fig:p-value-test_s-DeltaF-plane_no_constraint}.

First, we observe that the 2022 EHT configuration is only sensitive to the presence of a second ring, if the separation between the two rings is larger than $\sim 12\, \mu \rm as$, which roughly corresponds to the expected resolution for this array configuration\footnote{The nominal resolution is given by $\theta = \lambda/B_{\rm max}$ with $B_{\rm max}$ the longest baseline. This works out to $25\, \mu$as at 230 GHz  and $16\, \mu$as at 345 GHZ. These can be reduced by a factor of roughly 2 by using regularized-maximum likelihood imaging methods, see section 2.1 in \cite{EventHorizonTelescope:2019uob}.}. Second, we find that a ring separation of $\sim 5\, \mu \rm as$ could be detectable with an Earth-based array; this however requires high sensitivity and is therefore only reachable with an optimistic ngEHT array design (ngEHT-low, in which the SEFD is very low, i.e., the sensitivity high, in all telescopes added beyond the 2022 array).

This is an important result in view of the fact that some of our new-physics cases have separations which are of the order of $\sim 5\, \mu \rm as$. It suggests that -- if the results from our idealized study extended to simulated observations of beyond-GR-spacetimes -- a ground-based array design with very low SEFD values could potentially probe spacetimes beyond GR. As an alternative, we achieve similar results with a space-based telescope (again with low SEFD).

Third, we find an interplay between separation and relative flux: at higher values of the relative flux density, the threshold in separation is lower, at least for the less advanced array configurations (cf.~left and right upper panels in Fig.~\ref{fig:p-value-test_s-DeltaF-plane_no_constraint}). This is as expected; for less sensitive arrays, even structures separated further than the nominal resolution cannot be resolved, if the total flux density in one of them is too low.

Fourth, we find that parameter scans at fixed separation and increasing relative flux density show a somewhat surprising result: at some low value of relative flux density, there is a detection threshold at which the $p$-value drops significantly below $10^{-3}$ or even $10^{-5}$. At higher values of the relative flux density, the $p$-value increases again, i.e., it becomes more difficult to distinguish the 1- and 2-ring models. The reason lies in the fact that at relative flux densities $\Delta F \simeq 1$, the visibility amplitude for low $k$ is nearly degenerate with that of a 1-ring model with a diameter that is roughly the average of the two diameters of the two rings. This degeneracy can be lifted once higher baselines are reached, which is achievable with a space-based array. The lower left panel in Fig.~\ref{fig:p-value-test_s-DeltaF-plane_no_constraint} highlights that only the space-based array confidently detects the presence of a second ring at the highest values of relative flux density that we consider.

This result motivates the use of super-resolution techniques, which have been pioneered for M87* in \cite{Broderick:2022tfu} and which we here implement as a constraint on the width of the rings.

\subsection{Detecting new physics at sub-resolution scales}
Both in and beyond GR, photon rings are typically thin compared to the shadow diameter~\cite{Johnson:2019ljv,Ayzenberg:2023hfw,Staelens:2023jgr}.  
Thus, we include a constraint on the width of the rings as a prior in our reconstruction to investigate, how strong such a prior has to be in order to significantly improve the detectability of a second ring feature.

\begin{figure*}[t!]
    \centering
    \includegraphics[width=0.48\linewidth]{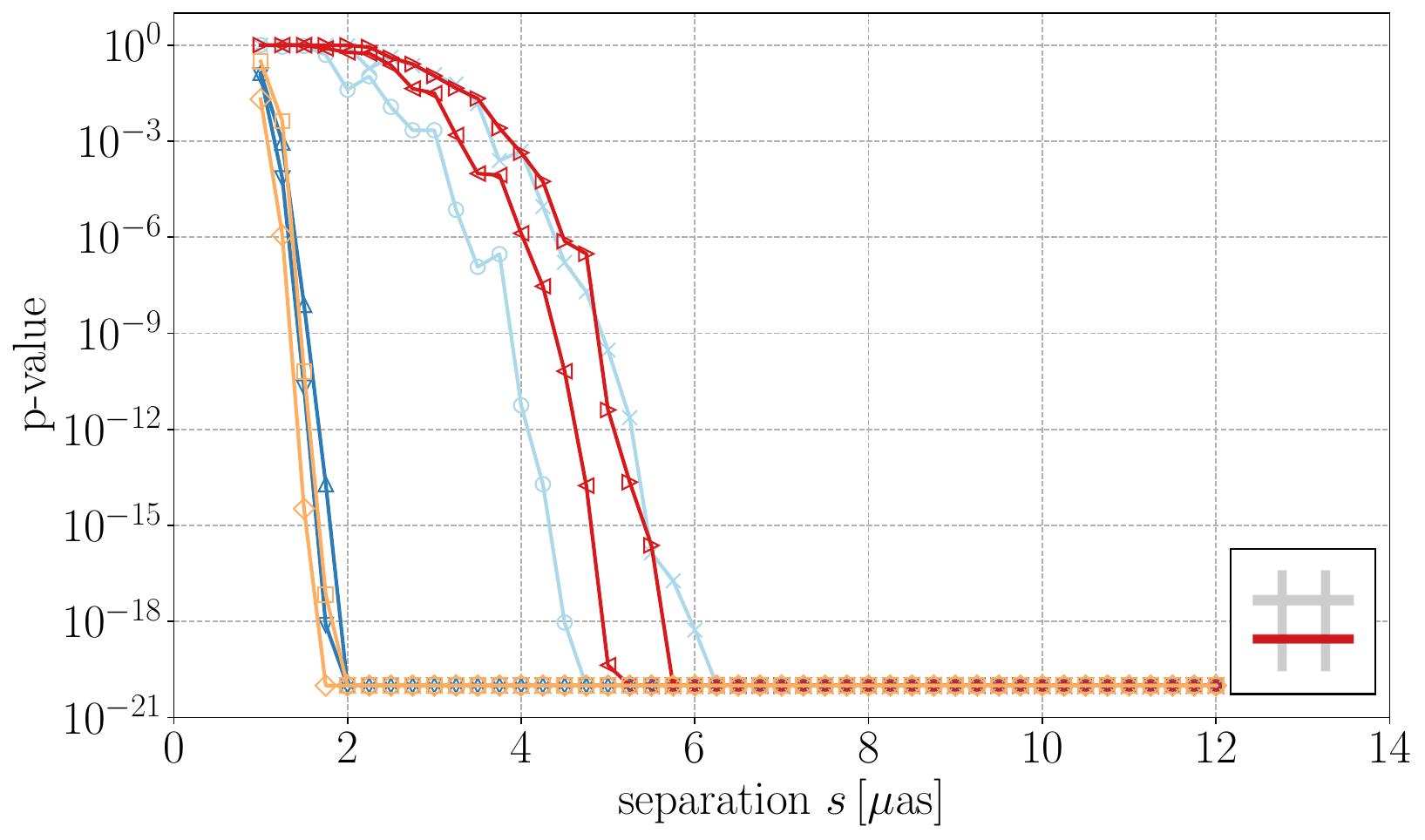}
    \includegraphics[width=0.48\linewidth]{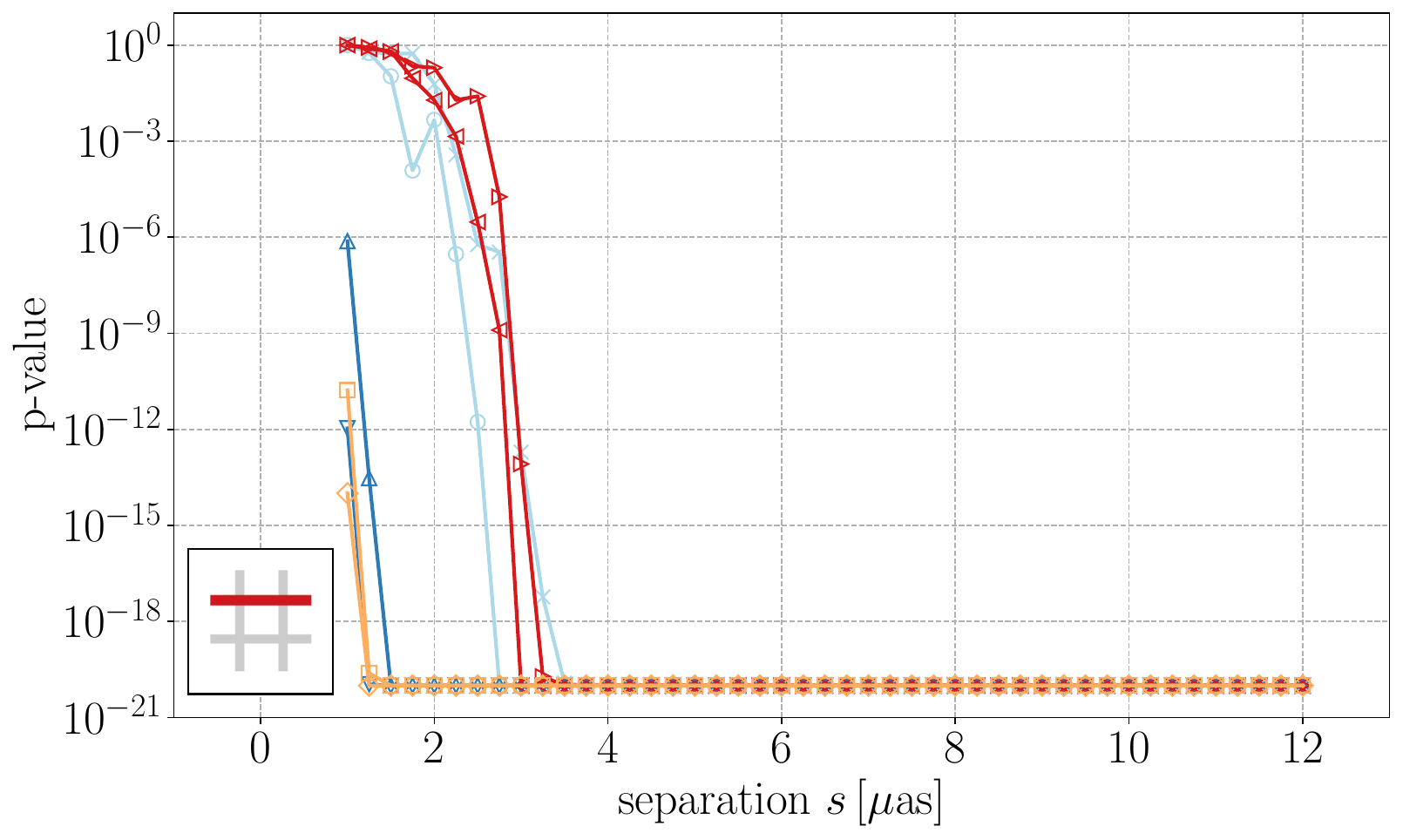}
    \\
    \includegraphics[width=0.48\linewidth]{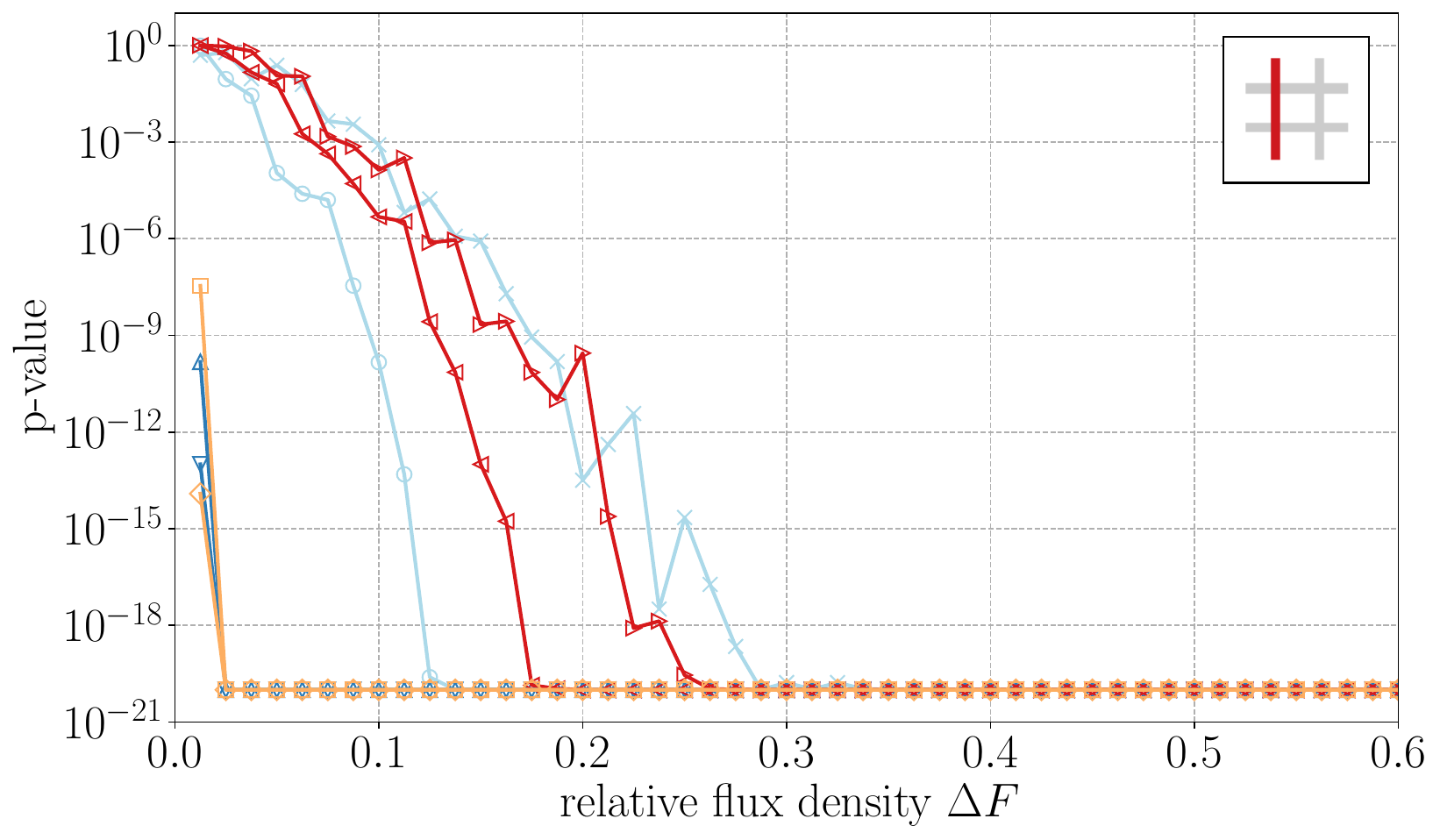}
    \includegraphics[width=0.48\linewidth]{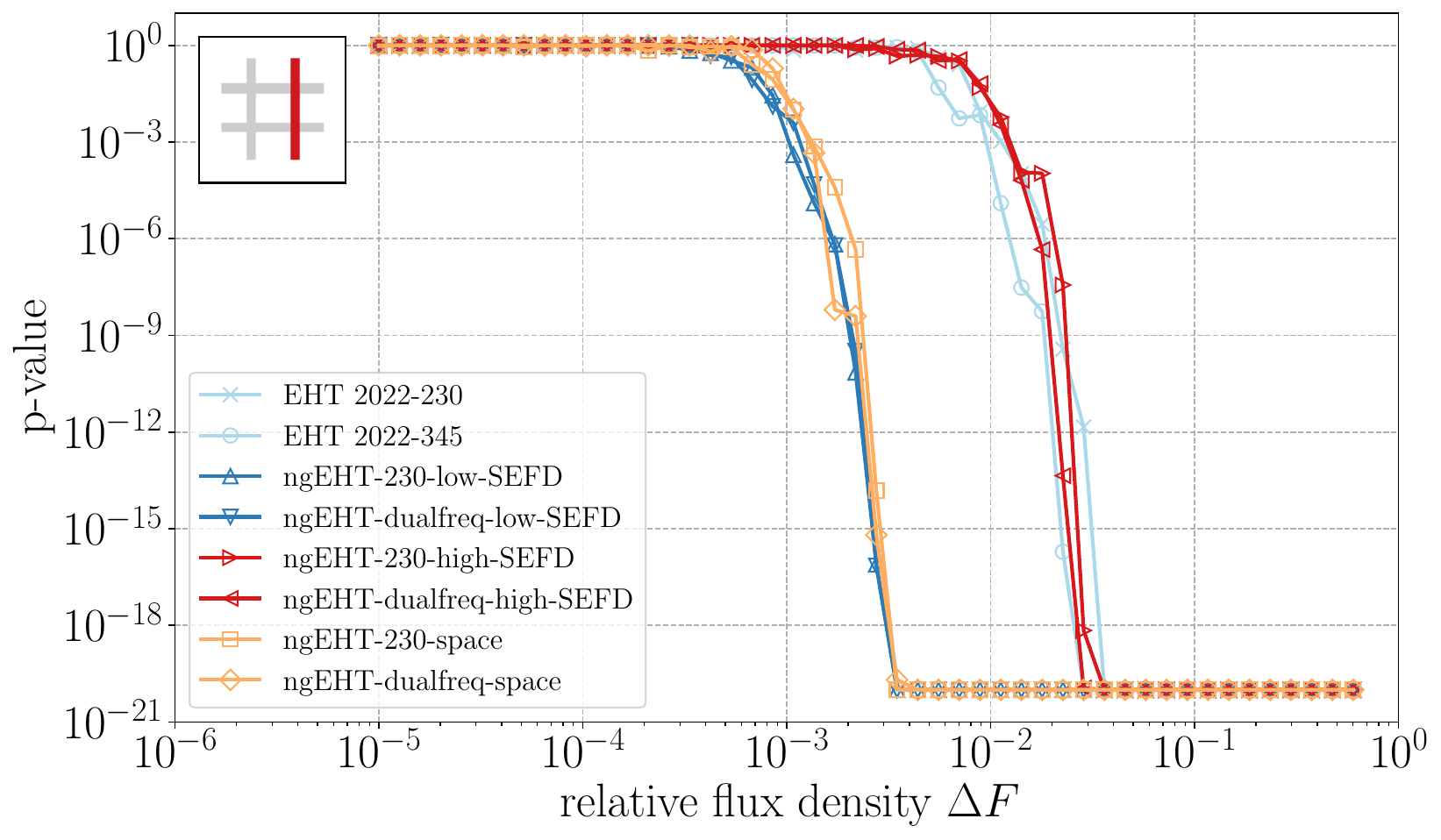}
    \caption{We show the 2-ring detectability (according to the $p$-value test, cf. main text) projected onto the four rays in the $(s,\,\Delta F)$ plane, cf.~Fig.~\ref{fig:s-DeltaF-plane}. A strong constraint on the widths $\omega_{1,2} \leq 2\,\mu$as has been added in the fits.  
     \label{fig:p-value-test_s-DeltaF-plane_tighter_width}
    }
\end{figure*}

To fully demonstrate the power of super-resolution techniques in our simplified setting, we impose a prior  of  $\omega_{1,2} \leq 2\,\mu$as in the fits. This brings the detection threshold for the separation between the two rings to below $2\,\mu \rm as$ for the better-performing arrays and $3-4\,\mu\rm as$ for the worse-performing arrays, cf.~Fig.~\ref{fig:p-value-test_s-DeltaF-plane_tighter_width}. This highlights that there is a nontrivial interplay between i) the scale imposed by the super-resolution constraint, ii) the nominal resolution, and iii) the sensitivity.

Overall, these results suggest that even current Earth-based arrays can distinguish a 2-ring fit from a 1-ring fit at values of $s$ and $\Delta F$ that are relevant to existing new-physics cases, if super-resolution techniques are used. We stress that the super-resolution technique used is a prior on the ring width that follows the expected properties of photon rings in GR and many theories beyond GR. Within the class of theories that produce photon rings which are thin, super-resolution techniques can distinguish between one and two rings, or, in other words, show that the presence of a second ring is a better fit to the data than just a single ring at sufficiently large separation. 
We caution that this result is not sufficient to say whether new-physics cases with these parameters can indeed be ruled out, because we only compare the performances of a 1-ring and a 2-ring model fit and do not perform a more general fit to our simulated data. It is however a result that motivates an in-depth future study that systematically simulates images of such new-physics spacetimes and then systematically analyzes that simulated data with a larger class of fits and/or more general image-reconstruction and data-analysis models that go beyond comparing two flat fitting profiles.

\section{Closure quantities} \label{sec:closure}

\subsection{Motivation and definition of closure quantities}

Thermal noise and systematic uncertainties impact visibility amplitudes, but can (partially) be removed in closure quantities. These are therefore important EHT data products~\cite{Chael:2018oym, 2017isra.book.....T, Lockhart:2021xel, EventHorizonTelescope:2022wok, EventHorizonTelescope:2019ths}. We review these quantities, and discuss how the 1-ring and 2-ring flux density profiles discussed above look like in terms of these variables.

For each pair of telescopes, labelled by indices $i,j$, there is an idealized visibility amplitude $V_{i,j}$ and its actually measured counterpart $\hat{V}_{i,j}$. The measurement is affected by complex gains $g_i$ and thermal noise $\epsilon_{i,j}$, such that
\begin{equation}\label{eq:vijdef}
\hat{V}_{ij}=g_i g_j^*V_{ij}+\epsilon_{ij}.
\end{equation}
Here,
$\epsilon_{ij}$ is a circularly-symmetric (invariant under rotations in the complex plane~\cite{lapidoth2017foundation}) complex Gaussian random variable with zero mean and variance $\sigma_{ij}$ (determined by the radiometer equation~\cite{2017isra.book.....T,2018ApJ...857...23C}) describing thermal noise. $g_i$ are station-based effects including limitations imposed by constituent interferometer elements and atmospheric turbulence. The complex gains $g_i$ describe the dominant noise components.

In contrast to thermal noise, systematic noise is difficult to calibrate. Closure quantities are constructed to remove station gains from the data as far as possible. Closure phases were first defined in \cite{1958MNRAS.118..276J} and closure amplitudes in~\cite{1960Obs....80..153T}. They were first applied in~\cite{1974ApJ...193..293R,1980Natur.285..137R}, later in VLBI in~\cite{Doeleman:2001nr,Fish:2016zqt,Chael:2018oym}, and have been generalized in~\cite{1991InvPr...7..261L,2020ApJ...904..126B,Samuel:2021vfv,2022PhRvD.105d3019T}.

We split $\hat{V}_{ij}$ into its amplitude and its phase. Closure phases vanish identically for symmetric sources (e.g.,~\cite{2003EAS.....6..213M}), and are therefore not useful for the analysis of the idealized synthetic flux density profiles in this paper, thus we focus on closure amplitudes. 

The variable $|\hat{V}_{ij}|$ is distributed according to a Rice distribution, $\mbox{Rice}\left(|g_i||g_j||V_{ij}|,\sigma_{ij}\right)$, as a direct consequence of Eq.~\eqref{eq:vijdef}. Its expectation value is given by 
\be
\langle|\hat{V}_{ij}|\rangle=|g_i| \cdot |g_j| \cdot |V_{ij}|\cdot \left[1+\mathcal{O}(\sigma_{ij}^2)\right]. 
\ee
Due to the $\mathcal{O}(\sigma_{ij}^2)$ term, $\langle|\hat{V}_{ij}|\rangle$ is a biased estimator of the parameter $|g_i| \cdot |g_j| \cdot |V_{ij}|$ in the Rice distribution. Thus, one introduces an unbiased estimator
\begin{equation}
A_{ij}=\sqrt{|\hat{V}_{ij}|^2-\sigma_{ij}^2}.
\end{equation}
Because the expectation values of both $|\hat{V}_{ij}|$ and $A_{ij}$ are proportional to quadratic combinations of gain factors, they are undesirably sensitive to uncertainties in these factors. This sensitivity is removed by defining closure amplitudes. Closure amplitudes can be defined for subsets of 4 stations $\{i,j,k,l\}$. The quantities
\begin{equation}\label{eq:cloamp1}
Z_{ijkl}^{(1)}=\frac{A_{ij}A_{kl}}{A_{ik}A_{jl}},\qquad Z^{(2)}_{ijkl}=\frac{A_{ik}A_{jl}}{A_{il}A_{jk}}
\end{equation}
are independent of gain factors $g_i$ in the absence of thermal noise \footnote{We can also define $Z^{(3)}_{ijkl}=A_{il}A_{jk}/(A_{ij}A_{kl})$, which does not add new information due to the constraint $Z^{(1)}_{ijkl}Z^{(2)}_{ijkl}Z^{(3)}_{ijkl}=1$~\cite{2017isra.book.....T}.}. $\langle Z_{ijkl}^{(1)}\rangle$ and $\langle Z_{ijkl}^{(2)}\rangle$ are only affected by gain factors at subleading order. This is a marked improvement over the visibility amplitude, and the main reason behind the use of these variables. As a final step, we take the logarithm of closure amplitudes, which simplifies the propagation of thermal errors~\cite{2020ApJ...894...31B,Broderick:2020wda}. 

The disadvantage of closure amplitudes is that due to their dependence on four stations, they are naturally represented in a five-dimensional space, making their interpretation more tricky than, e.g., the visibility amplitude.

\subsection{Closure quantities for 2-ring models}
To prepare for the interpretation of synthetic data from an (ng)EHT array, we first analyze an idealized setting with a very large array. We take it to be a square with $N_{\rm st}=M_{\rm st}\times M_{\rm st}$ stations, so that there are two adjustable parameters: the length of the baseline between adjacent corners 
$L_{\rm max}$ (the maximum baseline is $k_{\rm max}=\sqrt{2}L_{\rm max}$), and the number of stations on each side $M_{\rm st}$, which determines the density of stations. For a general array, there are $N_{\rm st}(N_{\rm st}-3)/2$ independent closure amplitudes~\cite{2017isra.book.....T}.  To select an independent set, we follow the algorithm in~\cite{2020ApJ...894...31B}, which takes one of the two independent expressions in Eq.~\eqref{eq:cloamp1} and evaluates it on $N_{\rm st}(N_{\rm st}-3)/2$ independent quadrangles. Using $V(k)$ defined in Eq.~\eqref{eq:visibility_2_rings}, and $k_{ij} = 2 \pi \sqrt{u_{ij}^2 + v_{ij}^2}/\lambda$, we have
\begin{equation}\label{eq:cloamp1}
\ln \left(Z_{ijkl}^{(1)}\right)=\ln \left(\frac{V(k_{ij})V(k_{kl})}{V(k_{ik})V(k_{jl})} \right).
\end{equation}
This expression can be evaluated numerically once $L_{\rm max}$ and $M_{\rm st}$ are fixed.

\begin{figure}[t]
    \centering
    \includegraphics[width=0.65\linewidth]{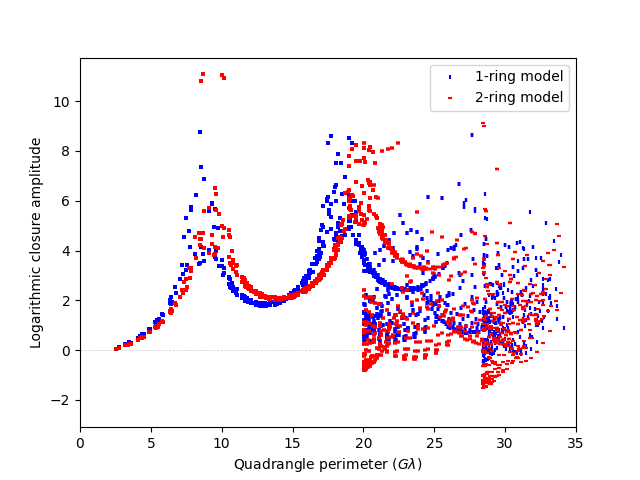}
    \caption{Logarithmic closure amplitudes as a function of the perimeter of the independent quadrangles of a square array with $M_{\rm st}=20$ stations on each side ($N_{\rm st}=400$) and a size $L_{\rm max}=10\mbox{ G}\lambda$. Independent quadrangles are selected following the algorithm in~\cite{2020ApJ...894...31B}, and perimeters up to $35\mbox{ G}\lambda$ are represented. Closure amplitudes are evaluated with the analytical expressions valid for the 1-ring crescent model with $d_1=42\ \mu\mbox{as}$, $\omega_1=2\ \mu\mbox{as}$, and a 2-ring model with an additional ring characterized by $s=5\ \mu\mbox{as}$, $\omega_2=0.5\ \mu\mbox{as}$ and $\Delta F=0.5$.}
    \label{fig:lca_projection}
\end{figure}

To display the 5-dimensional information in logarithmic closure amplitudes, we first represent it as a function of quadrangle perimeter in the $uv$-plane, discarding all other information on the distribution of the stations in the array, see Fig.~\ref{fig:lca_projection}. This type of representation of closure quantities has been used before in e.g.,~\cite{EventHorizonTelescope:2019ggy,2022A&A...663A..35I}.
The only obvious pattern in these figures is the existence of periodic structures, which are, however, somewhat obscured by the representation in terms of quadrangle perimeters. The analytical expressions in Eqs.~\eqref{eq:visibility_2_rings} and~\eqref{eq:cloamp1} indicate that an oscillatory pattern of divergences is expected due to the presence of Bessel functions inside a logarithm. The location of these divergences are controlled by the parameters of the crescent model. By isolating these features, we can therefore identify these parameters. The complete set of logarithmic closure amplitudes evaluated on the $N_{\rm st}(N_{\rm st}-3)/2$ independent quadrangles contains information about the whole image. As discussed next, it is possible to identify subsets of quadrangles containing information about this oscillatory pattern of divergences. 

\subsection{Isolating features of 2-ring models on idealized arrays}

Let us discuss a possible algorithm to identify subsets of quadrangles providing an alternative representation that isolates ring-like features. This representation is based on slicing the space of closure amplitudes by fixing 3 of the stations and forming quadrangles with the remaining station. For the moment, we introduce this ``peak slicing'' procedure on purely theoretical grounds, based on the structure of logarithmic closure quantities. We will discuss practical implementations of this slicing in the next section. 

\begin{figure}[t]
    \centering
    \includegraphics[width=0.5\linewidth]{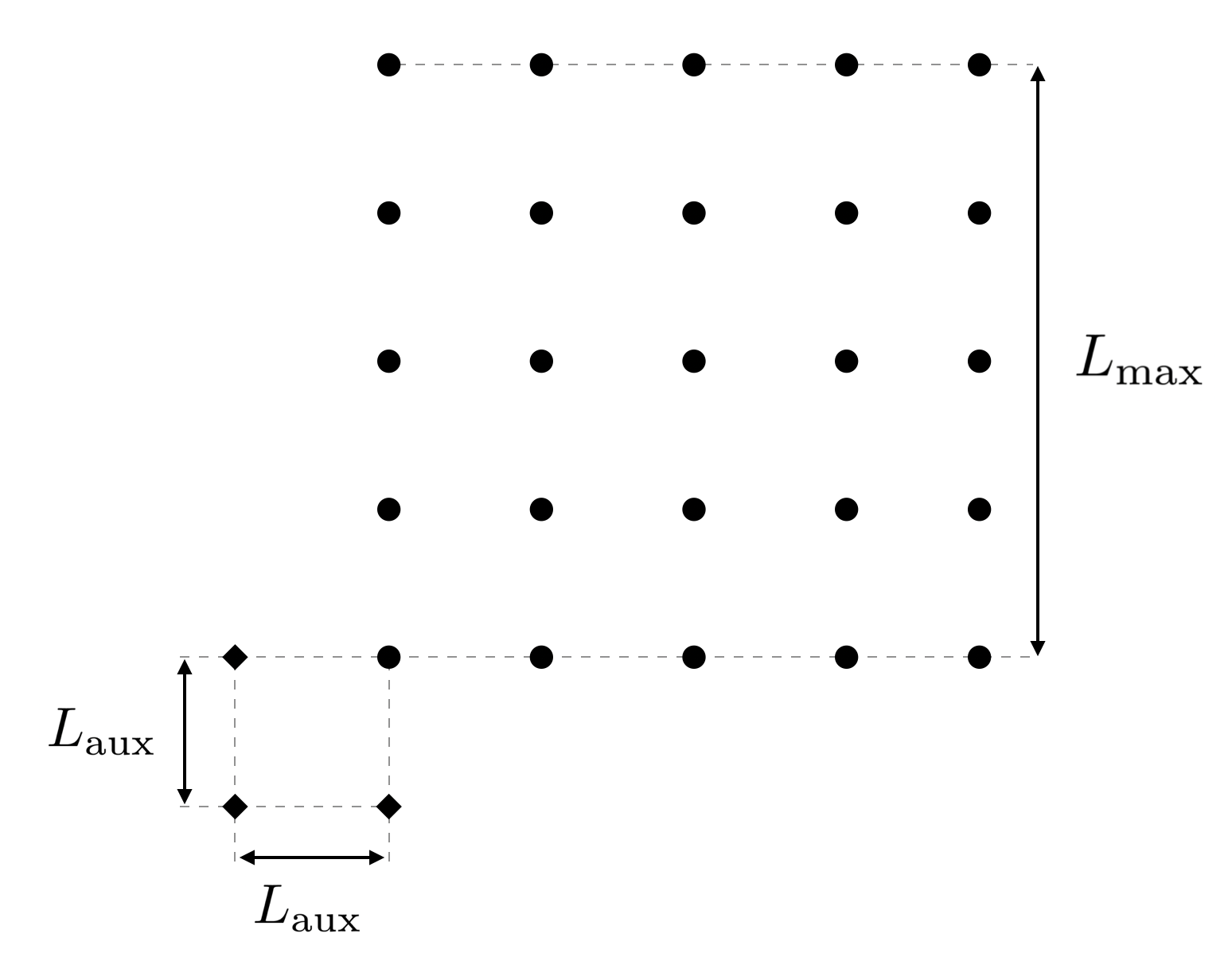}
    \caption{Schematic representation of the array used: a square array with $M_{\rm st}=5$ stations on each side (circles) and maximum baseline between adjacent corners $L_{\rm max}$, and 3 auxiliary stations (diamonds) with relative separation $L_{\rm aux}$, which we take as $L_{\rm aux}=L_{\rm max}/M_{\rm st}$.}
    \label{fig:aux_diag}
\end{figure}

\begin{figure}[t]
    \centering
    \includegraphics[width=0.65\linewidth]{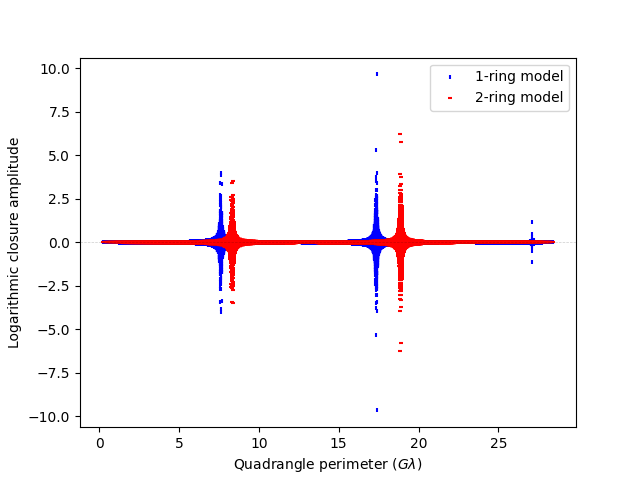}
    \caption{Logarithmic closure amplitudes as a function of the quadrangle perimeter for a square array with $M_{\rm st}=200$ stations on each side ($N_{\rm st}=40000$) and a size $L_{\rm max}=10\mbox{ G}\lambda$, and with 3 auxiliary stations with relative baseline $L_{\rm aux}=L_{\rm max}/M_{\rm st}$. Quadrangles are formed holding the 3 auxiliary stations fixed and choosing each of the stations one by one in the main array. Closure amplitudes are evaluated with the analytical expressions for the 1-ring crescent model with $d_1=42\ \mu\mbox{as}$, $\omega_1=2\ \mu\mbox{as}$, and a 2-ring model with an additional ring characterized by $s=5\ \mu\mbox{as}$, $\omega_2=0.5\ \mu\mbox{as}$ and $\Delta F=0.5$. The 1- and 2-ring models become more distinguishable for larger quadrangle perimeters (baselines), as expected from the fact that larger baselines allow for the detection of smaller features.}
    \label{fig:lca_slicing}
\end{figure}

\begin{figure}[t]
    \centering
    \includegraphics[width=0.65\linewidth]{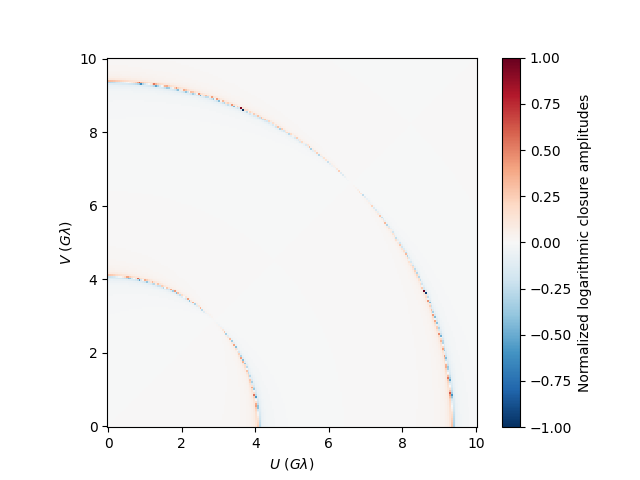}\\
    \includegraphics[width=0.65\linewidth]{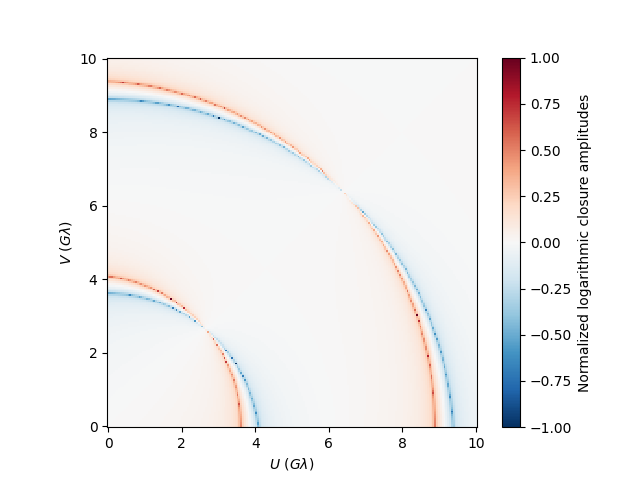}
    \caption{Normalized logarithmic closure amplitudes in the $uv$-plane for a square array with $M_{\rm st}=200$ stations on each side ($N_{\rm st}=40000$) and a size $L_{\rm max}=10\mbox{ G}\lambda$, with 3 auxiliary stations with relative baselines $L_{\rm aux}=L_{\rm max}/M_{\rm st}$ (top panel) and $L_{\rm aux}=10\times L_{\rm max}/M_{\rm st}$ (bottom panel). Closure amplitudes are evaluated for the 2-ring crescent model with $d_1=42\ \mu\mbox{as}$, $\omega_1=2\ \mu\mbox{as}$, $s=5\ \mu\mbox{as}$, $\omega_2=0.5\ \mu\mbox{as}$ and $\Delta F=0.5$. Logarithmic closure amplitudes are positive and formally divergent within the regions marked as dark red, and negative and formally divergent within the regions marked as dark blue. Information about the model parameters is encoded in the location of these divergences, and not the maximum values reached which depend on the parameters of the array, thus we are normalizing the logarithmic closure amplitudes. The larger relative distance between auxiliary stations in the bottom panel allows for a better differentiation of the two types of divergent behavior.
    }
    \label{fig:lca_uv}
\end{figure}

For the purpose of a clear graphical representation, we introduce 3 ``auxiliary stations'' outside the previous square array, see Fig.~\ref{fig:aux_diag} for an illustration of their placement. This slicing contains less information than the full set of closure amplitudes, but provides a cleaner representation of these periodic structures for comparable data densities, as illustrated in Fig.~\ref{fig:lca_slicing}. The figure shows a clear interference pattern, with logarithmic closure amplitudes becoming large (formally, divergent) for specific values of the quadrangle perimeter.
The introduction of auxiliary stations also allows us to represent logarithmic closure amplitudes as functions of the location in the $uv$-plane instead of the quadrangle perimeter, as the remaining stations span a square subset of the $uv$-plane.
In this representation, we associate the value obtained by forming quadrangles with the 3 auxiliary stations and 1 station of the main array to the latter. The result is presented in Fig.~\ref{fig:lca_uv}. These figures show the same interference pattern, and also illustrate the existence of both positive and negative divergences of the logarithmic quantities. Going back to the analytical expressions in Eqs.~\eqref{eq:visibility_2_rings} and~\eqref{eq:cloamp1} allows for a clear interpretation of these features. As 3 of the stations forming quadrangles remain fixed, there are only 2 baselines that change as different stations in the main array are chosen. Depending on the position of the latter station, the Fourier transform along these baselines, Eq.~\eqref{eq:visibility_2_rings}, can be (close to) zero. 
One of these baselines contributes with the Fourier transform to the numerator in the argument of the logarithm, and the other in the denominator, cf.~Eq.~\eqref{eq:cloamp1}; if we pick for instance the index $l$ for the non-fixed station, $V(k_{kl})$ vanishing leads to negative divergences, while $V(k_{jl})$ leads to positive divergences. While the specific location of these divergences depend on the underlying model being used, the existence of these divergences is model-independent and is based on robust interferometric features.
We can then understand Fig.~\ref{fig:lca_projection} as a convolution of these interference patterns and a choice of independent quadrangles that partially obscures these features. This full set of closure quantities contains information about the whole image and is therefore the best possible choice for image reconstruction using fitting procedures. However, some features are easier to see and understand in the sparse representation in Fig.~\ref{fig:lca_projection}.

\begin{figure}[t]
    \centering
    \includegraphics[width=0.5\linewidth]{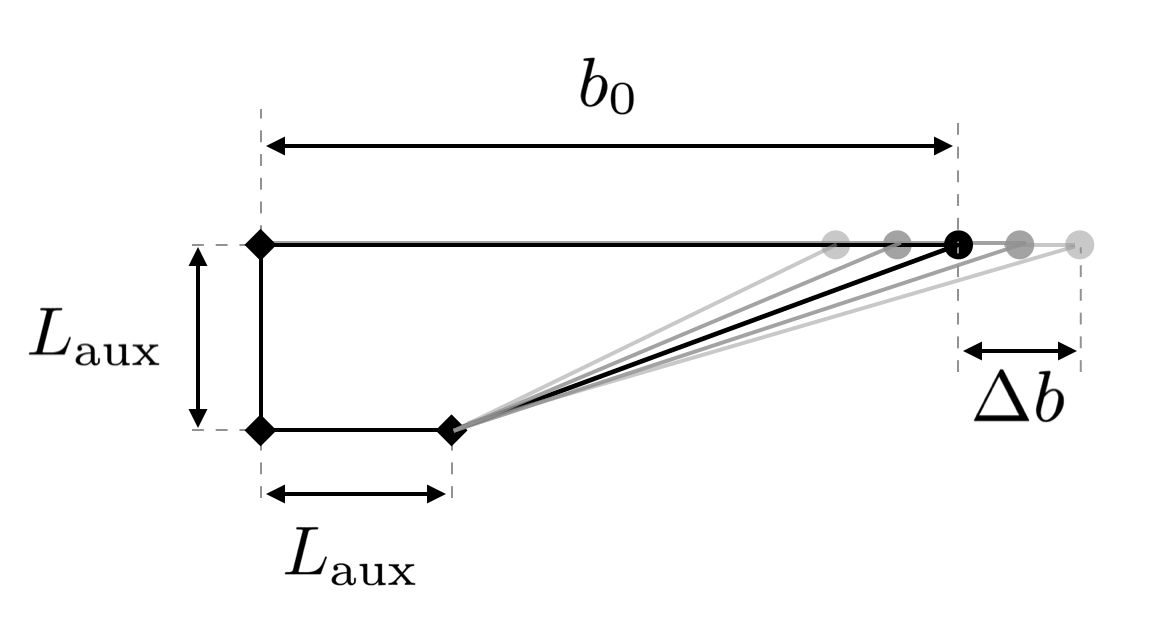}
    \caption{Quadrangles with 3 fixed stations (diamonds) and one movable station (circles). Values for the movable horizontal baseline are given by the set $\{b_0+j\Delta b\}_{j=-J}^J$, which in realistic settings would be naturally provided by Earth's rotation.}
    \label{fig:quad_diag}
\end{figure}

\begin{figure}[h!]
    \centering
    \includegraphics[width=0.55\linewidth]{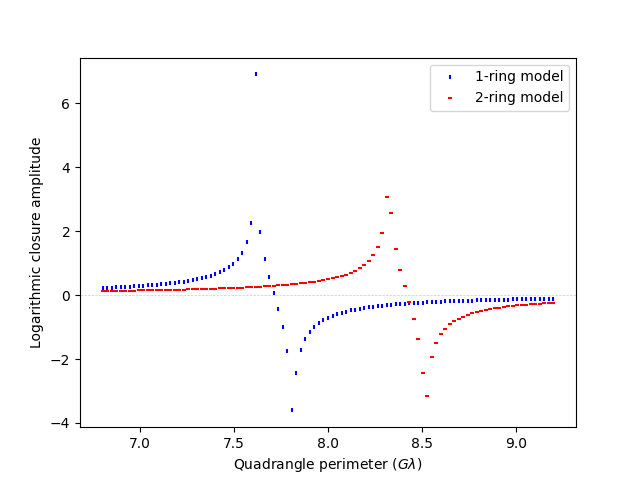}
     \includegraphics[width=0.55\linewidth]{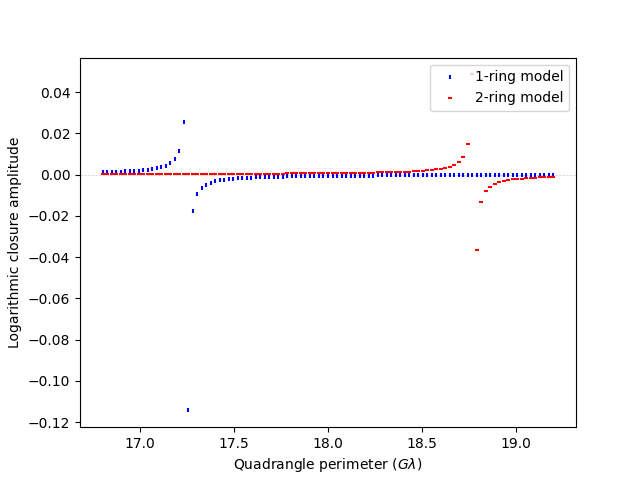}
     \includegraphics[width=0.55\linewidth]{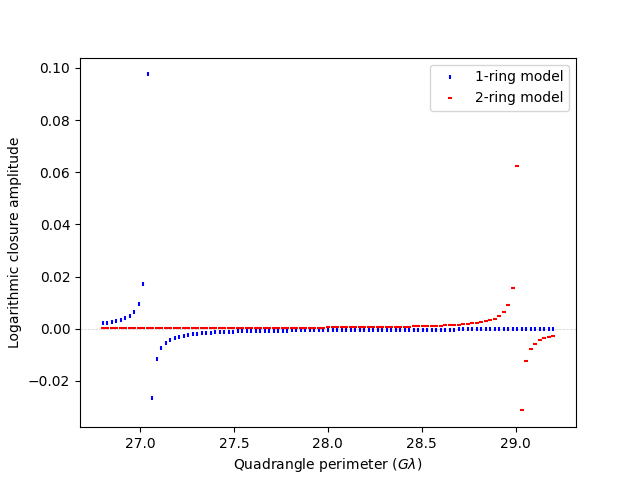}
    \caption{Logarithmic closure amplitudes as a function of the perimeter of the quadrangles depicted in Fig.~\ref{fig:quad_diag}, for the 1-ring crescent model with $d_1=42\ \mu\mbox{as}$, $\omega_1=2\ \mu\mbox{as}$, and a 2-ring model with an additional ring characterized by $s=5\ \mu\mbox{as}$, $\omega_2=0.5\ \mu\mbox{as}$ and $\Delta F=0.5$. In all cases, $\Delta b=0.6\mbox{ G}\lambda$ and $J=50$, while $b_0=3.85\mbox{ G}\lambda$ for the top panel, $b_0=9.0\mbox{ G}\lambda$ for the middle panel, and $b_0=14.0\mbox{ G}\lambda$ for the bottom panel.}
    \label{fig:lca_feature_isolation}
\end{figure}

For instance, we can determine the baseline distances that are needed to distinguish between 0-, 1- and 2-ring crescent models, using the form of the Bessel function $J_1(x)$. The zeros of the Fourier transform of the 0-ring model (i.e. a disk model with radius $R_{\rm outer}$) are controlled by the single parameter $R_{\rm outer}$, and therefore we would need to be able to resolve at least 2 zeros to falsify this model. The second zero of $J_1(x)$ is located at $x\simeq 7$~\cite{1965hmfw.book.....A}, which yields the requirement
\begin{equation}
k_{\rm max}\gtrsim \frac{7}{2\pi}\frac{\lambda}{R_{\rm outer}}.
\end{equation}
For the 1-ring crescent model, we would need to determine the location of at least 3 of these divergences to falsify the model and therefore be able to conclude that the 2-ring model could provide a better fit. We can visually check in Fig.~\ref{fig:lca_uv} that this requirement is satisfied by the square array with $L_{\rm max}=10\mbox{ G}\lambda$. The location of these divergences can be calculated more precisely from numerical values of $J_1(x)$ for the crescent model, or by visually inspecting the behavior of logarithmic closure amplitudes.  This can be performed in detail focusing on quadrangles formed with the three auxiliary stations and one of the remaining stations (see Fig.~\ref{fig:quad_diag}). This procedure yields Fig.~\ref{fig:lca_feature_isolation}, which shows that probing the three first peaks for 1-ring and 2 ring-models characterized by $d_1=42\ \mu\mbox{as}$, $\omega_1=2\ \mu\mbox{as}$, $s=5\ \mu\mbox{as}$, $\omega_2=0.5\ \mu\mbox{as}$ and $\Delta F=0.5$, it is necessary to have quadrangles with 3 close auxiliary stations and one of the remaining baselines of about $b_0=3.9\mbox{ G}\lambda$, $b_0=9.0\mbox{ G}\lambda$ and $b_0=14.0\mbox{ G}\lambda$.

\begin{figure}[t]
    \centering
    \includegraphics[width=0.65\linewidth]{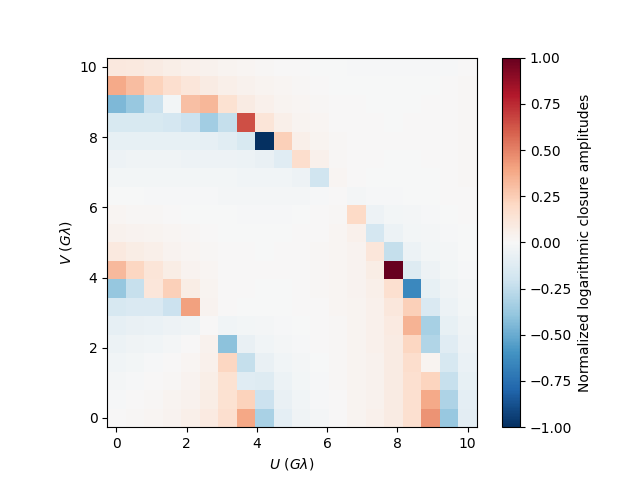}
    \caption{Equivalent of the top panel of Fig.~\ref{fig:lca_uv} but for $M_{\rm st}=20$ (one order of magnitude lower) stations on each side ($N_{\rm st}=400$). The lower density of stations leads to a less precise localization of the divergences of logarithmic closure amplitudes in the $uv$-plane.}
    \label{fig:lca_uv_ld}
\end{figure}

The maximum baseline $k_{\rm max}$ is not the only relevant parameter when assessing the observability of these features. While a minimum value of the latter is a necessary condition to distinguish 0-, 1- and 2-ring models, it is also necessary to have enough density of data points in the space of logarithmic closure amplitudes to be able to determine with confidence the location of the divergences. The higher the density, the better constrained the location of these divergences will be. This is illustrated in Fig.~\ref{fig:lca_uv_ld}.
In the idealized situation we are describing, with a fixed square array with respect to the source, the data density is only related to the density of stations in the array. However, (ng)EHT observations take place during an extended period of time in which the position of the array relative to the source changes due to Earth's rotation, so that the data density is a combination of the density of stations with time resolution. Both factors thus contribute to a better differentiation between 1-ring and 2-ring models.

In order to use the information about the location of the peaks to distinguish between 1-ring and 2-ring models, it is necessary to know the location of the first three first peaks. Knowing the location of the first two peaks allows falsifying a 0-ring model and determining whether a 1-ring model provides a better fit, while the location of the first three peaks allows falsifying the 1-ring model and determining whether a 2-ring model provides a better fit. This motivates the seach for a practical implementation of the peak-slicing procedure that can target the second and third peaks, which is discussed next.

\subsection{Isolating features of 2-ring models with space-based telescopes}

The discussion above is based on an idealized array with a regular geometry, an arbitrarily large number of stations and a particular slicing.

In this section, we propose an implementation of this slicing using a single space-based telescope together with 3 stations on Earth's surface (ALMA, APEX and the planned site LLAMA), and confirm that this implementation works as intended using the \texttt{ehtimaging} toolkit~\cite{Chael:2018oym}. Space-based VLBI has been discussed previously in \cite{Gurvits:2019ioq, Fish:2019epg,Haworth:2019urs,Roelofs:2019nmh,palumbo2019metrics}.

Probing the first peak structure in Fig.~\ref{fig:lca_feature_isolation} requires a baseline of around $8\mbox{ G}\lambda$ that is attainable with stations on Earth.
\begin{figure}[]
    \centering
    \includegraphics[width=0.85\linewidth]{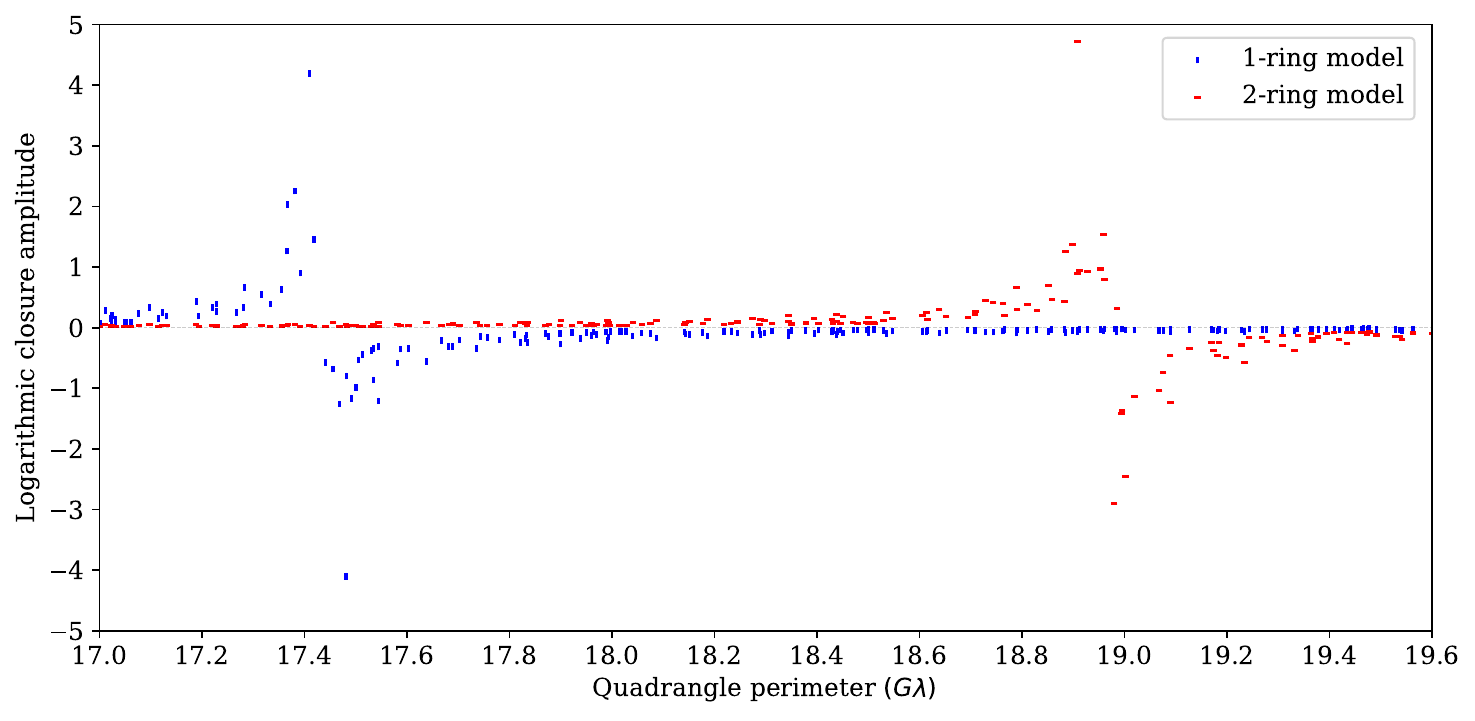}
     \includegraphics[width=0.85\linewidth]{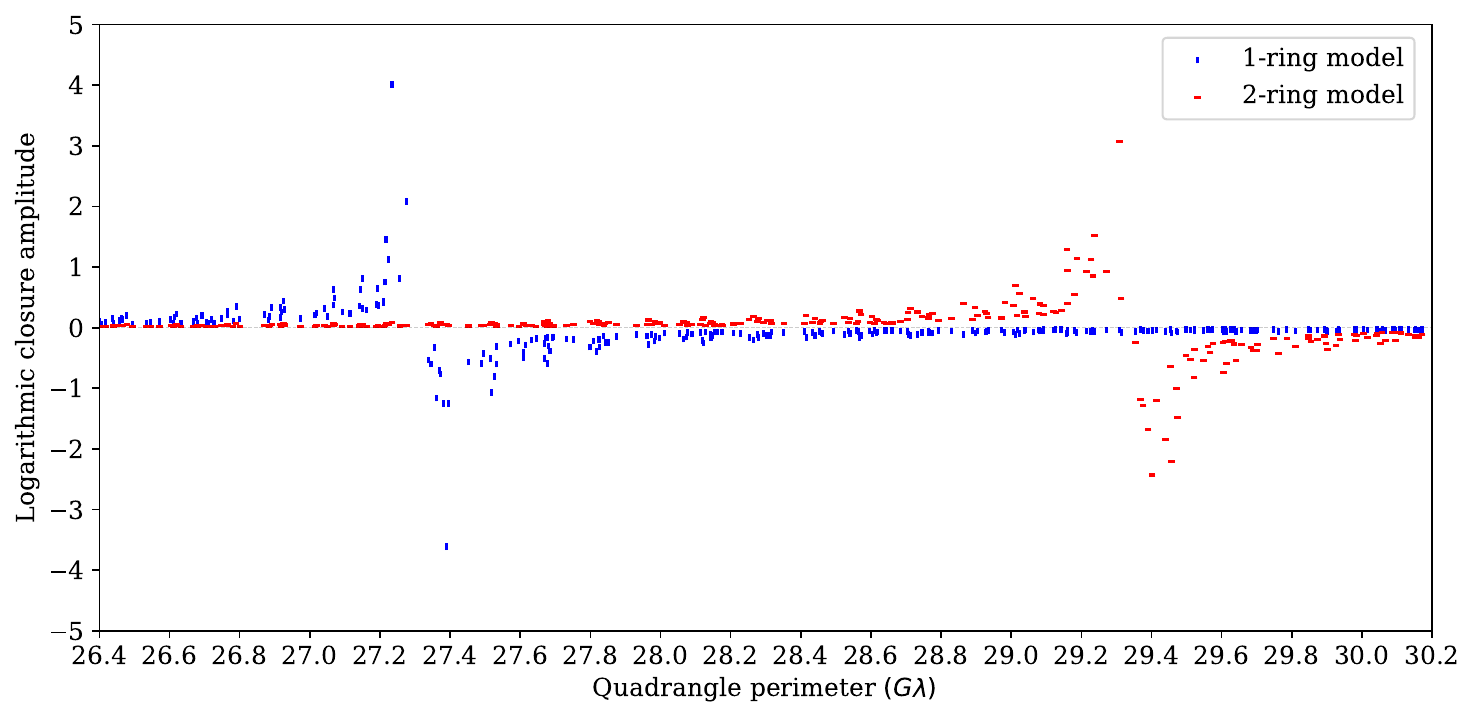}
    \caption{Logarithmic closure amplitudes as a function of the perimeter of the quadrangles formed by the Earth-based stations ALMA, APEX, LLAMA and the space-based station (see Tab.~\ref{tab:array_ngEHT_230_345_and_space_optimistic}), for the 1-ring crescent model with $d_1=42\ \mu\mbox{as}$, $\omega_1=2\ \mu\mbox{as}$, and a 2-ring model with the second ring characterized by $s=5\ \mu\mbox{as}$ and $\omega_2=0.5\ \mu\mbox{as}$. In all cases, the altitude of the space-based station varies between $500$ kms (i.e. $0.38\, {\rm G}\lambda$  at 230 GHz) and $8000$ kms (i.e. $6.15\, {\rm G}\lambda$) above Odense (Denmark), by steps of $200$ kms (i.e. $0.15\, {\rm G}\lambda$).}
    \label{fig:subset_eht_2nd_and_3rd_peaks}
\end{figure}
For the second and third peaks in Fig.~\ref{fig:lca_feature_isolation}, we require longer baselines. Here, we are not concerned with a realistic placement of the space-based telescope. We assume that a placement can be found so that the projected baseline can vary over large distances over the course of an observation night. We mimick this by running simulated observations in which the space-based station is moved by hand. We place it above Odense (Denmark) at distances ranging from 500 kms to 8000 kms over Earth's surface, together with the three Earth-based stations ALMA, APEX and LLAMA (see Tab.~\ref{tab:array_ngEHT_230_345_and_space_optimistic} for the characteristics of all stations). The  set of possible altitudes considered for the space-based station spans a large range of quadrangle perimeters, but only a few of those are actually necessary to target the second and third peaks in logarithmic closure amplitudes with the slicing method. As Earth rotates during an observation period, the projection of the baselines between the Earth-based and space-based stations onto the line of sight of the source changes -- as implemented in the \texttt{ehtimaging} toolkit. This effectively sweeps out a limited range of quadrangle perimeters which, provided that the location and altitude of the space-based station are chosen appropriately, can probe the second or third peak, as shown in Fig.~\ref{fig:subset_eht_2nd_and_3rd_peaks}.
 
\section{Conclusions and outlook}
\label{sec:Conclusions and outlook}

\begin{figure*}[t]
    \centering
    \includegraphics[width=\linewidth]{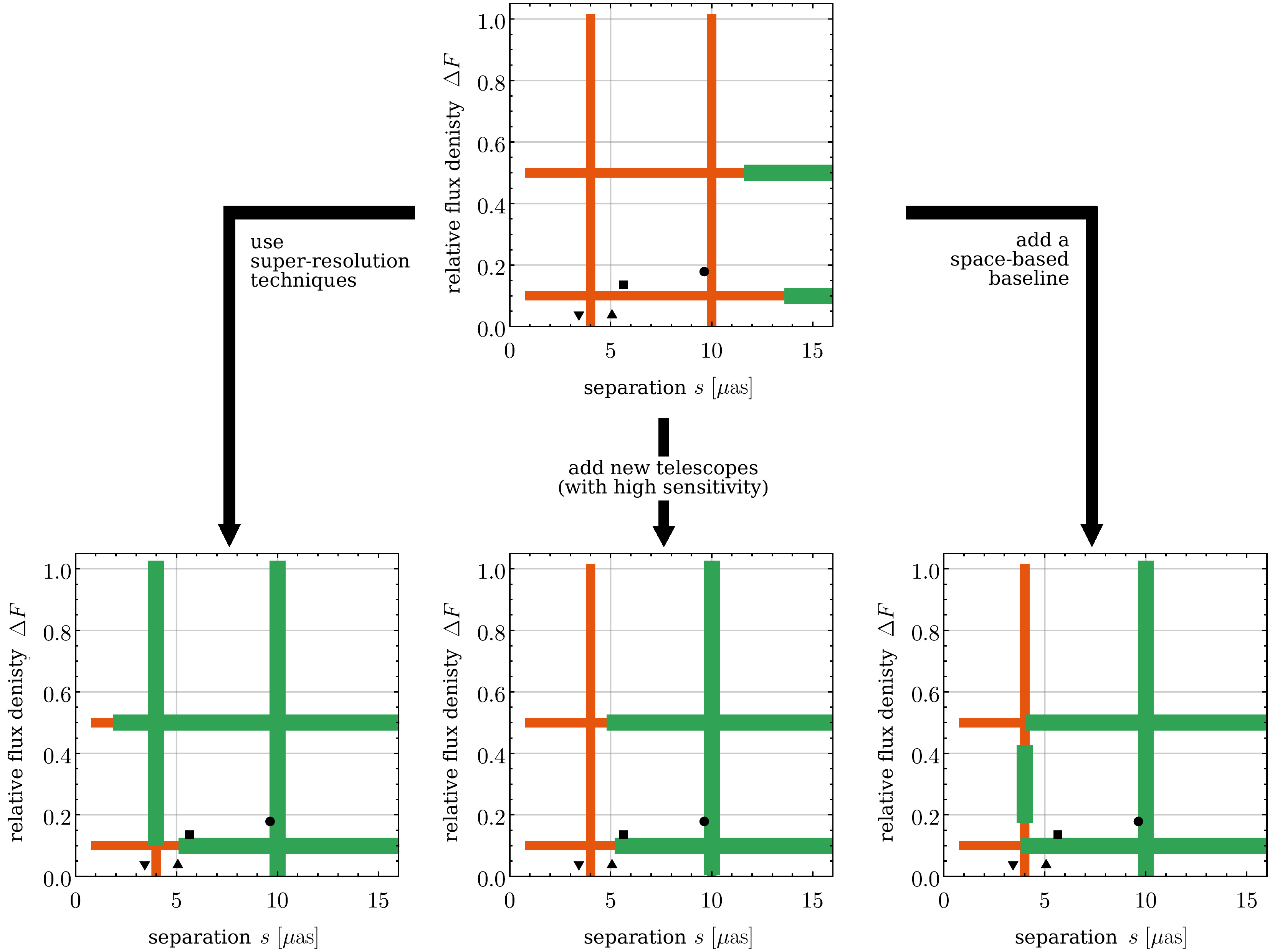}
    \caption{Summary of three tentative pathways to improve detectability, as suggested by the statistical analysis in Fig.~\ref{fig:p-value-test_s-DeltaF-plane_no_constraint} and \ref{fig:p-value-test_s-DeltaF-plane_tighter_width}. In all panels, the detectable (not detectable) parameter ranges are marked as thicker green (thinner red/orange) lines. (Here, detectability refers to a $p$-value of $10^{-5}$.)
	The top panel shows results for the EHT 2022 array and without super-resolution constraints. The three bottom panels show different ways of improving detectability: the left panel shows results for the EHT 2022 array but with a super-resolution constraint; the middle panel shows results for the ngEHT array, assuming optimistic, i.e., low SEFD values; the right panel shows the same ngEHT array with an additional single space-based telescope. All arrays shown here observe at 230 GHz. For details, cf.~notation and figures in Sec.~\ref{sec:Fourier plane-analysis}.
     \label{fig:conclusion-figure}
    }
\end{figure*}

The EHT cannot detect photon rings with the features expected in GR without using super-resolution techniques~\cite{Himwich:2020msm,Broderick:2022tfu}. However, beyond GR, photon rings can be much more widely separated from each other and also be significantly brighter. One example is in horizonless spacetimes with a photon sphere, where both an inner and an outer set of photon rings exist, and the $n=1$ photon ring can be at several $\mu \rm as$ distance\footnote{This assumes a mass of and distance to the source roughly similar to M87*. } from the inner photon rings.\\
This motivates our study, in which we work in a simplified setting to investigate detection capabilities of the EHT and potential future upgrades~\cite{2023Galax..11..107D,Johnson:2023ynn,Ayzenberg:2023hfw}. Our workflow is as follows:
first, we generate synthetic data that contains either one or two thin rings, parameterized by ring separation, relative flux density and widths of the two rings. We put these through a simulated observation and reconstruction, using \texttt{ehtimaging}~\cite{chael_2023_8408352, Chael:2018oym} to obtain the Fourier data of a simulated observation for a given array configuration. To this simulated data, we fit the visibility amplitude of both 1-ring and 2-ring crescent flux density profiles and compare the fit quality. We thereby obtain the detection threshold as a  boundary in parameter space.\\
We also explore the impact of super-resolution techniques by imposing priors on the fits. The relevant prior in our case, in which the simulated data only features thin (compared to the diameter) rings, is a prior on the reconstructed width of the rings. \\

First, we find that for our three new-physics cases which motivate our study, simulated data from the EHT 2022 configuration does not allow to infer the presence of a second ring, see uppermost panel in Fig.~\ref{fig:conclusion-figure}. Thus, VLBI arrays need to be improved. Two properties of VLBI arrays are critical to lower the detection threshold towards rings with lower separation from each other and also a second ring with low relative flux density: first, a high sensitivity (thus low SEFD) and second, a high resolution (thus high frequency and/or much larger baselines).

We find that the following setups bring the detection threshold to what is needed to rule out some of the new-physics cases that motivate our study, see Fig.~\ref{fig:conclusion-figure}: first, an ng-EHT array in which 8 stations are added to the EHT 2022 array and all these stations have low SEFD values corresponding to those of the ALMA station; and where 230 and 345 GHz frequencies are simultaneously used. This provides both the sensitivity and resolution needed. Second, a space-based array in which an additional space-based station is added to the previous ngEHT-configuration, which results in larger baselines and thus lower detection threshold in terms of separation of the rings. Third, the EHT 2022 array combined with super-resolution techniques. These provide a substitute for the resolution that is needed.
\\
The first two options are clearly more expensive arrays and may therefore be more difficult to realize. However, even the existing array, if combined with super-resolution techniques, can approach the region in parameter space where our new-physics examples are located. Using super-resolution techniques implies that any statement about ruling out/detecting signatures of new physics can only be valid within the class of spacetimes that generates photon rings that are thin compared to their diameter. To the best of our knowledge, this is the generic case and no counterexamples are known.

We caution that all our conclusions are to be understood within our simplified setting in which the image consists of two thin rings and a broad image feature corresponding to foreground emission is not accounted for.

The visibility amplitude of a (simulated) observation is subject to systematic uncertainties, some of which can be removed by considering closure quantities, which are based on ratios of visibility amplitudes. Rings generate zeros in visibility amplitudes and accordingly divergences in closure quantities, such as the logarithmic closure amplitude which depends on four stations in the array. We investigate an idealized setting in which three stations are held fixed and a fourth one is moved in a controlled way. In that setting, the divergences of the logarithmic closure amplitude of the 1-ring and 2-ring models can be separated from each other. In a realistic setting, the (projected) baselines between all four stations change over the course of an observation, because of Earth's rotation. Earth's rotation thus paves the way for a practical implementation of this idealized setting, which can be obtained when choosing three auxiliary stations as close as possible and the fourth station placed so that it can target higher-order zeros for specific sources. Using \texttt{ehtimaging}, we have shown that such a setting enables one to probe the first divergence of the logarithmic closure quantity by an Earth-based array, and the second and third by choosing a space-based station as the fourth station. We argue that knowing the locations of the first three divergences is sufficient to distinguish the 1-ring from the 2-ring  model. This provides further motivation for arrays featuring space-based stations.

In summary, our study shows that in an idealized setting, simulated observations with VLBI arrays can distinguish between one and two rings at parameter values motivated by new physics beyond GR, if the array is sensitive enough and dual-frequency capabilities are assumed. Existing arrays can distinguish between one and two rings, if super-resolution techniques are used.

This motivates future upgrades of our investigation along the following lines:
\begin{itemize}
\item First, our simulated data does not come from ray-tracing in a given spacetime geometry, but consists of an ad-hoc geometric model which we used to perform a first parameter study. Given the largely positive outcome of this study, a more extensive study starting from given spacetime geometries is warranted. In such a study, also the following additional points should be addressed, that correspond to simplifying assumptions of our first analysis.
\item Second, our simulated data consists of images with either one or two rings, but no diffuse ($n=0$) emission is included. Investigating the detection capabilities in the presence of a (broad) feature from the diffuse emission is one important future direction. We expect that the detection threshold is shifted towards higher separations and relative flux densities, once foreground emission is accounted for. The extent of the shift depends on the properties of the foreground emission: for a sufficiently broad image feature with approximately constant flux density, from which two rings stand out in peak flux density, our conclusions will likely not be altered much.
\item Third, in such a study, accounting for uncertainties from the imperfectly understood astrophysics of the accretion disk is important; i.e., such a study must fit not only the parameters of the spacetime, but also parameters of an accretion disk model to determine whether two rings can be distinguished from one ring.
\end{itemize}

\noindent
\acknowledgments
We thank Andrew Chael for discussions and for answering questions on \texttt{ehtimaging}, and Maciek Wielgus for discussions. This work is supported by a grant from VILLUM Fonden, no.~29405. 
A.~Held acknowledges support from the PRIME programme of the German Academic Exchange Service (DAAD) with funds from the German Federal Ministry of Education and Research (BMBF). A.~Held and A.~Eichhorn also acknowledge support by the Deutsche Forschungsgemeinschaft (DFG) under Grant No 406116891 within the Research Training Group RTG 2522/1.
\\
H.~Delaporte would like to thank the Perimeter Institute for Theoretical Physics for hospitality during the final phase of this project. Research at Perimeter Institute is supported in part by the Government of Canada through the Department of Innovation, Science and Economic Development and by the Province of Ontario through the Ministry of Colleges and Universities.

\appendix
\section{Gaussian profile}
\label{app:GaussianProfile}
In order to check that, in the thin ring regime, the type of flux density profiles has no impact on the detectability of a second ring (see Fig.~\ref{fig:p-value-test_s-DeltaF-plane_width_profile}), we use another profile, Gaussian, based on the auxiliary function 
\begin{equation}
\nu(r;d,\omega)=\frac{1}{N}e^{-(d-2r)^2/2 \omega^2},
\end{equation}
normalized such that
\be
\int_0^{\infty}dr\,{ 2\pi} r\int_0^{2\pi}d\theta\, \nu(r;d,\omega) = 1,
\label{eq:NormalisationGaussianAuxiliaryFunc}
\ee 
which leads to the normalization factor
\begin{equation}
N=\frac{\pi}{4}\left\{\sqrt{2\pi}d\omega\left[\mbox{erf}\left(d/\sqrt{2}\omega\right)+1 \right] +2\omega^2e^{-d^2/2 \omega^2}\right\},
\end{equation}
where $\mbox{erf}(x)$ is the error function.

The resulting total Gaussian flux density profile combines two Gaussian rings and is then given by
\bea
&{}&F_{\rm Gaussian}[r]= \frac{F_{\rm tot}}{1+\Delta F}\nu(r;d_1,\omega_1)+\frac{F_{\rm tot}}{1+1/\Delta F} \nu(r;d_1-2s,\omega_2).\nonumber
\label{eq:GaussianProfile}
\eea
The total flux density is $F_{\rm tot}$, and the total flux densities for the two rings correspond to the prefactors in Eq.~\eqref{eq:GaussianProfile}, i.e.
\be
F_1 = \frac{F_{\rm tot}}{1+\Delta F},\quad F_2 = \frac{F_{\rm tot}}{1+1/\Delta F}.
\ee

\section{Comparison of two profiles in a quantitative test of detectability}
\label{app:ComparisonProfiles}
\begin{figure}[h!]
    \centering
    \includegraphics[width=0.65\linewidth]{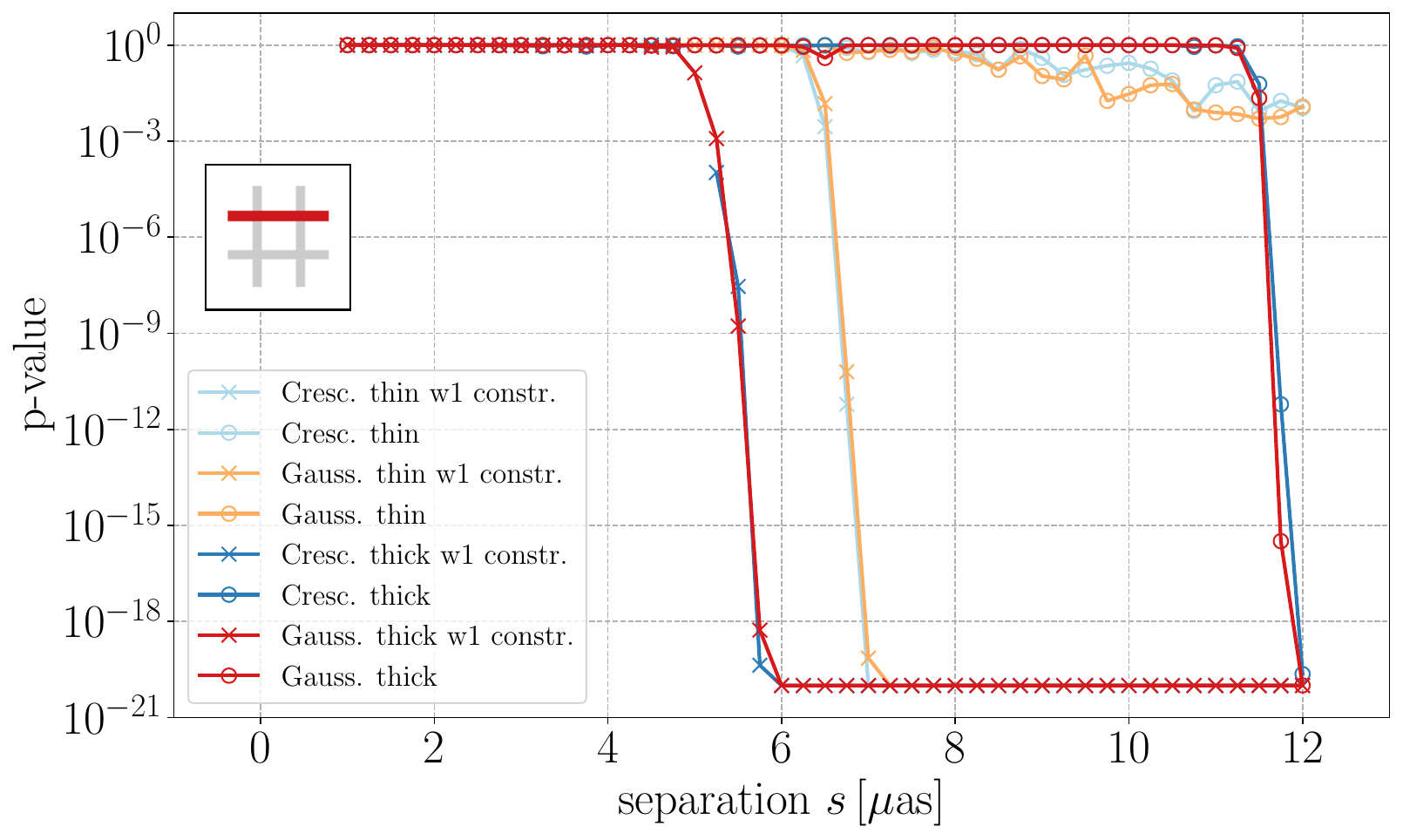}
    \caption{We show the 2-ring detectability (according to the $p$-value test, cf. main text) projected onto the ray $\Delta F = 0.5$, cf.~Fig.~\ref{fig:s-DeltaF-plane}, for the EHT 2022 array at 230 GHz. We vary the profile (either crescent or Gaussian), the width of the outer ring (either $\omega_1 = 2\,\mu {\rm as}$ or $8\, \mu {\rm as}$) and the constraint on the width of the outer ring in the fits (either none or $\omega_1 \leq 10.5\, \mu$as). The remaining 2-ring parameter is chosen as $\omega_2 = 1\, \mu$as.}
    \label{fig:p-value-test_s-DeltaF-plane_width_profile}
\end{figure}

We still focus on synthetic data in the limit of relatively thin rings and use the same fitting profile as in Sec.~\ref{sec:Fourier plane-analysis}. However, we construct synthetic data with two different profiles, namely the crescent profile, Eq.~\eqref{eq:CrescentProfile}, and the Gaussian profile, Eq.~\eqref{eq:GaussianProfile}, with or without a loose constraint on the width of the outer ring $\omega_1 \leq 10.5\, \mu$as in the fits, and for $\omega_1 = 2\, \mu$as (thin) or $\omega_1 = 8\, \mu$as (relatively thin). The detectability test for the 2022 EHT array at 230 GHz as a function of the separation is shown in Fig.~\ref{fig:p-value-test_s-DeltaF-plane_width_profile}. As curves for the Gaussian and crescent profiles are superposed for equal values of the parameters, we deduce that the type of profile does not matter within the thin ring assumption. Note that this assumption remains valid for a relatively thin outer ring, i.e. when $\omega_1 = 8\, \mu$as.

\section{Array specifications}
\label{app:arrays}

We provide details (see \cref{tab:array_EHT2022,tab:array_ngEHT_230_345_optimistic,tab:array_ngEHT_230_345_pessimistic,tab:array_ngEHT_230_345_and_space_optimistic}) on the stations in the VLBI arrays considered in this work and summarized in ~\cref{tab:arrays}.

\begin{table*}[h!]
\resizebox{\linewidth}{!}{%
\begin{threeparttable}
\begin{tabular}{|c|c|c|c|c|c|c|}
\hline
\multicolumn{1}{|c|}{Facility} & Location     & X\tnote{a}\, (m)        & Y\tnote{a}\, (m)         & Z\tnote{a}\, (m)        & SEFD\tnote{b}\,\, at 230 GHz (Jy) &  SEFD\tnote{c}\,\, at 345 GHz (Jy)\\ \hline
\multicolumn{1}{|c|}{ALMA}     & Chile        & 2225061.164  & -5440057.37   & -2481681.15  & 74      & 250             \\ \hline
APEX                           & Chile        & 2225039.53   & -5441197.63   & -2479303.36  & 4700     & 8880            \\ \hline
GLT     &  Greenland & 541547.0     &   -1387978.6   &    6180982.0  &     5000 & 14390 \\ \hline
JCMT                           & Hawaii, USA  & -5464584.68  & -2493001.17   & 2150653.98   & 10500     & 5780            \\ \hline
KPNO & Arizona, US &  -1995954.4   &    -5037389.4   &    3357044.3      & 13000 & 44970 \\ \hline
LMT                            & Mexico       & -768713.9637 & -5988541.7982 & 2063275.9472 & 4500      & 2040            \\ \hline
NOEMA   & France & 4524000.4    &    468042.1   &     4460309.8     & 700 & 1410 \\ \hline
PV 30 m                        & Spain        & 5088967.9000 & -301681.6000  & 3825015.8000 & 1900    & 3850              \\ \hline
SMA                            & Hawaii, USA  & -5464523.400 & -2493147.080  & 2150611.750  & 6200   & 5730              \\ \hline
SMT                            & Arizona, USA & -1828796.200 & -5054406.800  & 3427865.200  & 17100  & 17190               \\ \hline
SPT                            & Antarctica   & 0.01         & 0.01          & -6359609.7   & 19300      & 25440           \\ \hline

\end{tabular}
\begin{tablenotes}
       \item[a] Geocentric coordinates with $X$ pointing to the Greenwich meridian, $Y$ pointing $90\degree$ away in the equatorial plane (eastern longitudes have positive Y), and positive $Z$
pointing in the direction of the North Pole.
		\item[b] SEFD values at 230 GHz from \cite{EventHorizonTelescope:2019uob}
		\item[c] SEFD values at 345 GHz from \cite{Broderick:2021aeg}
     \end{tablenotes}
\end{threeparttable}%
}
\caption{Array specifications for EHT 2022 at 230 and 345 GHz}
\label{tab:array_EHT2022}
\end{table*}

\begin{table*}[h!]
\resizebox{\linewidth}{!}{%
\begin{threeparttable}
\begin{tabular}{|c|c|c|c|c|c|c|}
\hline
\multicolumn{1}{|c|}{Facility} & Location                        & X\tnote{a}\, (m)        & Y\tnote{a}\, (m)         & Z\tnote{a}\, (m)        & SEFD\tnote{b}\,\, at 230 GHz (Jy) &  SEFD\tnote{c}\,\, at 345 GHz (Jy)\\ \hline
\multicolumn{1}{|c|}{ALMA}     & Chile                           & 2225061.164  & -5440057.37   & -2481681.15  & 74  &     250            \\ \hline
APEX                           & Chile                           & 2225039.53   & -5441197.63   & -2479303.36  & 4700   &  8880            \\ \hline
BAJA                           & Baja California, Mexico         & -2352576.0   & -4940331.0    & 3271508.0    & 74  &     250         \\ \hline
CNI                            & La Palma, Canary Islands, Spain & 5311000.0    & -1725000.0    & 3075000.0    & 74    &    250        \\ \hline
GAM                            & Gamsberg, Namibia               & 5627890.0    & 1637767.0     & -2512493.0   & 74  &   250           \\ \hline
GLT                            & Greenland                       & 541647.0     & -1388536.0    & 6180829.0    & 5000  &   14390           \\ \hline
HAY                            & Masachussetts, USA              & 1521000.0     & -4417000.0    & 4327000.0    & 74   &   250           \\ \hline
JCMT                           & Hawaii, USA                     & -5464584.68  & -2493001.17   & 2150653.98   & 10500   &  5780            \\ \hline
KP 12 m                        & Arizona, USA                    & -1994314.0   & -5037909.0    & 3357619.0    & 13000   &  44970            \\ \hline
KVN-YS & Korea  & -3042280.9137  &   4045902.7164  &  3867374.3544  &  74  & 250   \\ \hline
LAS  & Chile &  1818163.826  &  -5280331.162   &  -3074870.820  &   74  & 250   \\ \hline
LLAMA &  Argentina   & 2325327.209   &  -5341469.111  &  -2599682.209   &  74 & 250  \\ \hline
LMT                            & Mexico                          & -768713.9637 & -5988541.7982 & 2063275.9472 & 4500    &   2040           \\ \hline
NOEMA                          & France                          & 4523998.40   & 468045.240    & 4460309.760  & 700    &   1410           \\ \hline
OVRO                           & California, USA                 & -2409598.0   & -4478348.0    & 3838607.0    & 74   &     250        \\ \hline
PV 30 m                        & Spain                           & 5088967.9000 & -301681.6000  & 3825015.8000 & 1900   &    3850           \\ \hline
SMA                            & Hawaii, USA                     & -5464523.400 & -2493147.080  & 2150611.750  & 6200    &    5730         \\ \hline
SMT                            & Arizona, USA                    & -1828796.200 & -5054406.800  & 3427865.200  & 17100    &  17190           \\ \hline
SPT                            & Antarctica                      & 0.01         & 0.01          & -6359609.7   & 19300   &   25440           \\ \hline
\end{tabular}%
\begin{tablenotes}
       \item[a] Geocentric coordinates with $X$ pointing to the Greenwich meridian, $Y$ pointing $90\degree$ away in the equatorial plane (eastern longitudes have positive Y), and positive $Z$
pointing in the direction of the North Pole.
		\item[b] SEFD values at 230 GHz from \cite{EventHorizonTelescope:2019uob} for EHT 2022 sites, and taken as ALMA's (lowest) SEFD value for ngEHT planned and phase-1 sites from \cite{Johnson:2023ynn} 
		\item[c] SEFD values at 345 GHz from \cite{Broderick:2021aeg} for EHT 2022 sites, and taken as ALMA's (lowest) SEFD value for ngEHT planned and phase-1 sites from \cite{Johnson:2023ynn}
     \end{tablenotes}
\end{threeparttable}
}
\caption{Array specifications for ngEHT-low (low SEFD values) at 230 and 345 GHz.}
\label{tab:array_ngEHT_230_345_optimistic}
\end{table*}

\begin{table*}[h!]
\resizebox{\linewidth}{!}{%
\begin{threeparttable}
\begin{tabular}{|c|c|c|c|c|c|c|}
\hline
\multicolumn{1}{|c|}{Facility} & Location                        & X\tnote{a}\, (m)        & Y\tnote{a}\, (m)         & Z\tnote{a}\, (m)        & SEFD\tnote{b}\,\, at 230 GHz (Jy) &  SEFD\tnote{b}\,\, at 345 GHz (Jy)\\ \hline
\multicolumn{1}{|c|}{ALMA}     & Chile                           & 2225061.164  & -5440057.37   & -2481681.15  & 74  &  250            \\ \hline
APEX                           & Chile                           & 2225039.53   & -5441197.63   & -2479303.36  & 4700   &  8880            \\ \hline
BAJA                           & Baja California, Mexico         & -2352576.0   & -4940331.0    & 3271508.0    & 19300  &   44970    \\ \hline
CNI                            & La Palma, Canary Islands, Spain & 5311000.0    & -1725000.0    & 3075000.0    & 19300    &   44970   \\ \hline
GAM                            & Gamsberg, Namibia               & 5627890.0    & 1637767.0     & -2512493.0   & 19300  &   44970    \\ \hline
GLT                            & Greenland                       & 541647.0     & -1388536.0    & 6180829.0    & 5000  &   14390           \\ \hline
HAY                            & Masachussetts, USA              & 1521000.0     & -4417000.0    & 4327000.0    & 19300   &  44970           \\ \hline
JCMT                           & Hawaii, USA                     & -5464584.68  & -2493001.17   & 2150653.98   & 10500   &  5780            \\ \hline
KP 12 m                        & Arizona, USA                    & -1994314.0   & -5037909.0    & 3357619.0    & 13000   &  44970            \\ \hline
KVN-YS & Korea  & -3042280.9137  &   4045902.7164  &  3867374.3544  &  19300  &  44970  \\ \hline
LAS  & Chile &  1818163.826  &  -5280331.162   &  -3074870.820  &  19300  & 44970   \\ \hline
LLAMA &  Argentina   & 2325327.209   &  -5341469.111  &  -2599682.209   &  19300 &  44970  \\ \hline
LMT                            & Mexico                          & -768713.9637 & -5988541.7982 & 2063275.9472 & 4500    &   2040           \\ \hline
NOEMA                          & France                          & 4523998.40   & 468045.240    & 4460309.760  & 700    &   1410           \\ \hline
OVRO                           & California, USA                 & -2409598.0   & -4478348.0    & 3838607.0    & 19300  &   44970        \\ \hline
PV 30 m                        & Spain                           & 5088967.9000 & -301681.6000  & 3825015.8000 & 1900   &    3850           \\ \hline
SMA                            & Hawaii, USA                     & -5464523.400 & -2493147.080  & 2150611.750  & 6200    &    5730         \\ \hline
SMT                            & Arizona, USA                    & -1828796.200 & -5054406.800  & 3427865.200  & 17100    &  17190           \\ \hline
SPT                            & Antarctica                      & 0.01         & 0.01          & -6359609.7   & 19300   &   25440           \\ \hline
\end{tabular}%
\begin{tablenotes}
       \item[a] Geocentric coordinates with $X$ pointing to the Greenwich meridian, $Y$ pointing $90\degree$ away in the equatorial plane (eastern longitudes have positive Y), and positive $Z$
pointing in the direction of the North Pole.
		\item[b] SEFD values at 230 GHz from \cite{EventHorizonTelescope:2019uob} for EHT 2022 sites, and taken as SPT's (highest) SEFD value for ngEHT planned and phase-1 sites from \cite{Johnson:2023ynn} 
		\item[c] SEFD values at 345 GHz from \cite{Broderick:2021aeg} for EHT 2022 sites, and taken as KP's (highest) SEFD value for ngEHT planned and phase-1 sites from \cite{Johnson:2023ynn}
     \end{tablenotes}
\end{threeparttable}
}
\caption{Array specifications for ngEHT-high (high SEFD values) at 230 and 345 GHz.}
\label{tab:array_ngEHT_230_345_pessimistic}
\end{table*}

\begin{table*}[h!]
\resizebox{\linewidth}{!}{%
\begin{threeparttable}
\begin{tabular}{|c|c|c|c|c|c|c|}
\hline
\multicolumn{1}{|c|}{Facility} & Location                        & X\tnote{a}\, (m)        & Y\tnote{a}\, (m)         & Z\tnote{a}\, (m)        & SEFD\tnote{b}\,\, at 230 GHz (Jy) &  SEFD\tnote{b}\,\, at 345 GHz (Jy)\\ \hline
\multicolumn{1}{|c|}{ALMA}     & Chile                           & 2225061.164  & -5440057.37   & -2481681.15  & 74  &     250            \\ \hline
APEX                           & Chile                           & 2225039.53   & -5441197.63   & -2479303.36  & 4700   &  8880            \\ \hline
BAJA                           & Baja California, Mexico         & -2352576.0   & -4940331.0    & 3271508.0    & 74  &     250         \\ \hline
CNI                            & La Palma, Canary Islands, Spain & 5311000.0    & -1725000.0    & 3075000.0    & 74    &    250        \\ \hline
GAM                            & Gamsberg, Namibia               & 5627890.0    & 1637767.0     & -2512493.0   & 74  &   250           \\ \hline
GLT                            & Greenland                       & 541647.0     & -1388536.0    & 6180829.0    & 5000  &   14390           \\ \hline
HAY                            & Masachussetts, USA              & 1521000.0     & -4417000.0    & 4327000.0    & 74   &   250           \\ \hline
JCMT                           & Hawaii, USA                     & -5464584.68  & -2493001.17   & 2150653.98   & 10500   &  5780            \\ \hline
KP 12 m                        & Arizona, USA                    & -1994314.0   & -5037909.0    & 3357619.0    & 13000   &  44970            \\ \hline
KVN-YS & Korea  & -3042280.9137  &   4045902.7164  &  3867374.3544  &  74  & 250   \\ \hline
LAS  & Chile &  1818163.826  &  -5280331.162   &  -3074870.820  &   74  & 250   \\ \hline
LLAMA &  Argentina   & 2325327.209   &  -5341469.111  &  -2599682.209   &  74 & 250  \\ \hline
LMT                            & Mexico                          & -768713.9637 & -5988541.7982 & 2063275.9472 & 4500    &   2040           \\ \hline
NOEMA                          & France                          & 4523998.40   & 468045.240    & 4460309.760  & 700    &   1410           \\ \hline
OVRO                           & California, USA                 & -2409598.0   & -4478348.0    & 3838607.0    & 74   &     250        \\ \hline
PV 30 m                        & Spain                           & 5088967.9000 & -301681.6000  & 3825015.8000 & 1900   &    3850           \\ \hline
SMA                            & Hawaii, USA                     & -5464523.400 & -2493147.080  & 2150611.750  & 6200    &    5730         \\ \hline
SMT                            & Arizona, USA                    & -1828796.200 & -5054406.800  & 3427865.200  & 17100    &  17190           \\ \hline
SPT                            & Antarctica                      & 0.01         & 0.01          & -6359609.7   & 19300   &   25440           \\ \hline
space-based\tnote{d}  &  Above Odense (Denmark) at 35786 km   &  23560747.282  &  4319365.430   &   34681814.518  &  36600  &  56000 \\ \hline
\end{tabular}%
\begin{tablenotes}
       \item[a] Geocentric coordinates with $X$ pointing to the Greenwich meridian, $Y$ pointing $90\degree$ away in the equatorial plane (eastern longitudes have positive Y), and positive $Z$
pointing in the direction of the North Pole.
		\item[b] SEFD values at 230 GHz from \cite{EventHorizonTelescope:2019uob} for EHT 2022 sites, and taken as ALMA's (lowest) SEFD value for ngEHT planned and phase-1 sites from \cite{Johnson:2023ynn} 
		\item[c] SEFD values at 345 GHz from \cite{Broderick:2021aeg} for EHT 2022 sites, and taken as ALMA's (lowest) SEFD value for ngEHT planned and phase-1 sites from \cite{Johnson:2023ynn}
		\item[d] SEFD values for the space-based telescope at 230 and 345 GHz estimated from Tab. 1 in \cite{Roelofs:2019nmh}
     \end{tablenotes}
\end{threeparttable}
}
\caption{Array specifications for ngEHT-space (low SEFD values) at 230 and 345 GHz.}
\label{tab:array_ngEHT_230_345_and_space_optimistic}
\end{table*}

\clearpage
\bibliographystyle{unsrt}
\bibliography{References}

\begin{thebibliography}{100}

\bibitem{Chapline:2000en}
G.~Chapline, E.~Hohlfeld, R.~B. Laughlin, and D.~I. Santiago.
\newblock {Quantum phase transitions and the breakdown of classical general
  relativity}.
\newblock {\em Int. J. Mod. Phys. A}, 18:3587--3590, 2003.

\bibitem{Mazur:2001fv}
Pawel~O. Mazur and Emil Mottola.
\newblock {Gravitational Condensate Stars: An Alternative to Black Holes}.
\newblock {\em Universe}, 9(2):88, 2023.

\bibitem{Mathur:2005zp}
Samir~D. Mathur.
\newblock {The Fuzzball proposal for black holes: An Elementary review}.
\newblock {\em Fortsch. Phys.}, 53:793--827, 2005.

\bibitem{Almheiri:2012rt}
Ahmed Almheiri, Donald Marolf, Joseph Polchinski, and James Sully.
\newblock {Black Holes: Complementarity or Firewalls?}
\newblock {\em JHEP}, 02:062, 2013.

\bibitem{Barcelo:2014npa}
Carlos Barcel\'o, Ra\'ul Carballo-Rubio, and Luis~J. Garay.
\newblock {Mutiny at the white-hole district}.
\newblock {\em Int. J. Mod. Phys. D}, 23(12):1442022, 2014.

\bibitem{Rovelli:2014cta}
Carlo Rovelli and Francesca Vidotto.
\newblock {Planck stars}.
\newblock {\em Int. J. Mod. Phys. D}, 23(12):1442026, 2014.

\bibitem{Haggard:2014rza}
Hal~M. Haggard and Carlo Rovelli.
\newblock {Quantum-gravity effects outside the horizon spark black to white
  hole tunneling}.
\newblock {\em Phys. Rev. D}, 92(10):104020, 2015.

\bibitem{Barcelo:2014cla}
Carlos Barcel\'o, Ra\'ul Carballo-Rubio, Luis~J. Garay, and Gil Jannes.
\newblock {The lifetime problem of evaporating black holes: mutiny or
  resignation}.
\newblock {\em Class. Quant. Grav.}, 32(3):035012, 2015.

\bibitem{Barcelo:2015noa}
Carlos Barcel\'o, Ra\'ul Carballo-Rubio, and Luis~J. Garay.
\newblock {Where does the physics of extreme gravitational collapse reside?}
\newblock {\em Universe}, 2(2):7, 2016.

\bibitem{Haggard:2016ibp}
Hal~M. Haggard and Carlo Rovelli.
\newblock {Quantum Gravity Effects around Sagittarius A*}.
\newblock {\em Int. J. Mod. Phys. D}, 25(12):1644021, 2016.

\bibitem{Bacchini:2021fig}
Fabio Bacchini, Daniel~R. Mayerson, Bart Ripperda, Jordy Davelaar, H\'ector
  Olivares, Thomas Hertog, and Bert Vercnocke.
\newblock {Fuzzball Shadows: Emergent Horizons from Microstructure}.
\newblock {\em Phys. Rev. Lett.}, 127(17):171601, 2021.

\bibitem{Eichhorn:2022bbn}
Astrid Eichhorn and Aaron Held.
\newblock {Quantum gravity lights up spinning black holes}.
\newblock {\em JCAP}, 01:032, 2023.

\bibitem{Bena:2022rna}
Iosif Bena, Emil~J. Martinec, Samir~D. Mathur, and Nicholas~P. Warner.
\newblock {Fuzzballs and Microstate Geometries: Black-Hole Structure in String
  Theory}.
\newblock 4 2022.

\bibitem{Eichhorn:2023xxx}
Astrid Eichhorn, Pedro G.~S. Fernandes, Aaron Held, and Hector~O.Z Silva.
\newblock {Breaking black-hole uniqueness at supermassive scales}.
\newblock 2023.

\bibitem{LIGOScientific:2016aoc}
B.~P. Abbott et~al.
\newblock {Observation of Gravitational Waves from a Binary Black Hole Merger}.
\newblock {\em Phys. Rev. Lett.}, 116(6):061102, 2016.

\bibitem{EventHorizonTelescope:2019dse}
Kazunori Akiyama et~al.
\newblock {First M87 Event Horizon Telescope Results. I. The Shadow of the
  Supermassive Black Hole}.
\newblock {\em Astrophys. J. Lett.}, 875:L1, 2019.

\bibitem{NANOGrav:2023gor}
Gabriella Agazie et~al.
\newblock {The NANOGrav 15 yr Data Set: Evidence for a Gravitational-wave
  Background}.
\newblock {\em Astrophys. J. Lett.}, 951(1):L8, 2023.

\bibitem{Roelofs:2022lux}
Freek Roelofs et~al.
\newblock {The ngEHT Analysis Challenges}.
\newblock {\em Galaxies}, 11(1):12, 2023.

\bibitem{Johnson:2023ynn}
Michael~D. Johnson et~al.
\newblock {Key Science Goals for the Next-Generation Event Horizon Telescope}.
\newblock {\em Galaxies}, 11:61, 2023.

\bibitem{Ayzenberg:2023hfw}
D.~Ayzenberg et~al.
\newblock {Fundamental Physics Opportunities with the Next-Generation Event
  Horizon Telescope}.
\newblock 12 2023.

\bibitem{EventHorizonTelescope:2019uob}
Kazunori Akiyama et~al.
\newblock {First M87 Event Horizon Telescope Results. II. Array and
  Instrumentation}.
\newblock {\em Astrophys. J. Lett.}, 875(1):L2, 2019.

\bibitem{EventHorizonTelescope:2019jan}
Kazunori Akiyama et~al.
\newblock {First M87 Event Horizon Telescope Results. III. Data Processing and
  Calibration}.
\newblock {\em Astrophys. J. Lett.}, 875(1):L3, 2019.

\bibitem{EventHorizonTelescope:2019ths}
Kazunori Akiyama et~al.
\newblock {First M87 Event Horizon Telescope Results. IV. Imaging the Central
  Supermassive Black Hole}.
\newblock {\em Astrophys. J. Lett.}, 875(1):L4, 2019.

\bibitem{EventHorizonTelescope:2019pgp}
Kazunori Akiyama et~al.
\newblock {First M87 Event Horizon Telescope Results. V. Physical Origin of the
  Asymmetric Ring}.
\newblock {\em Astrophys. J. Lett.}, 875(1):L5, 2019.

\bibitem{EventHorizonTelescope:2019ggy}
Kazunori Akiyama et~al.
\newblock {First M87 Event Horizon Telescope Results. VI. The Shadow and Mass
  of the Central Black Hole}.
\newblock {\em Astrophys. J. Lett.}, 875(1):L6, 2019.

\bibitem{EventHorizonTelescope:2021bee}
Kazunori Akiyama et~al.
\newblock {First M87 Event Horizon Telescope Results. VII. Polarization of the
  Ring}.
\newblock {\em Astrophys. J. Lett.}, 910(1):L12, 2021.

\bibitem{EventHorizonTelescope:2021srq}
Kazunori Akiyama et~al.
\newblock {First M87 Event Horizon Telescope Results. VIII. Magnetic Field
  Structure near The Event Horizon}.
\newblock {\em Astrophys. J. Lett.}, 910(1):L13, 2021.

\bibitem{EventHorizonTelescope:2022wkp}
Kazunori Akiyama et~al.
\newblock {First Sagittarius A* Event Horizon Telescope Results. I. The Shadow
  of the Supermassive Black Hole in the Center of the Milky Way}.
\newblock {\em Astrophys. J. Lett.}, 930(2):L12, 2022.

\bibitem{EventHorizonTelescope:2022apq}
Kazunori Akiyama et~al.
\newblock {First Sagittarius A* Event Horizon Telescope Results. II. EHT and
  Multiwavelength Observations, Data Processing, and Calibration}.
\newblock {\em Astrophys. J. Lett.}, 930(2):L13, 2022.

\bibitem{EventHorizonTelescope:2022wok}
Kazunori Akiyama et~al.
\newblock {First Sagittarius A* Event Horizon Telescope Results. III. Imaging
  of the Galactic Center Supermassive Black Hole}.
\newblock {\em Astrophys. J. Lett.}, 930(2):L14, 2022.

\bibitem{EventHorizonTelescope:2022exc}
Kazunori Akiyama et~al.
\newblock {First Sagittarius A* Event Horizon Telescope Results. IV.
  Variability, Morphology, and Black Hole Mass}.
\newblock {\em Astrophys. J. Lett.}, 930(2):L15, 2022.

\bibitem{EventHorizonTelescope:2022urf}
Kazunori Akiyama et~al.
\newblock {First Sagittarius A* Event Horizon Telescope Results. V. Testing
  Astrophysical Models of the Galactic Center Black Hole}.
\newblock {\em Astrophys. J. Lett.}, 930(2):L16, 2022.

\bibitem{EventHorizonTelescope:2022xqj}
Kazunori Akiyama et~al.
\newblock {First Sagittarius A* Event Horizon Telescope Results. VI. Testing
  the Black Hole Metric}.
\newblock {\em Astrophys. J. Lett.}, 930(2):L17, 2022.

\bibitem{Lamy:2018zvj}
Fr\'ed\'eric Lamy, Eric Gourgoulhon, Thibaut Paumard, and Fr\'ed\'eric~H.
  Vincent.
\newblock {Imaging a non-singular rotating black hole at the center of the
  Galaxy}.
\newblock {\em Class. Quant. Grav.}, 35(11):115009, 2018.

\bibitem{Vincent:2020dij}
F.~H. Vincent, M.~Wielgus, M.~A. Abramowicz, E.~Gourgoulhon, J.~P. Lasota,
  T.~Paumard, and G.~Perrin.
\newblock {Geometric modeling of M87* as a Kerr black hole or a non-Kerr
  compact object}.
\newblock {\em Astron. Astrophys.}, 646:A37, 2021.

\bibitem{Kumar:2020ltt}
Rahul Kumar and Sushant~G. Ghosh.
\newblock {Photon ring structure of rotating regular black holes and no-horizon
  spacetimes}.
\newblock {\em Class. Quant. Grav.}, 38(8):8, 2021.

\bibitem{Eichhorn:2021etc}
Astrid Eichhorn and Aaron Held.
\newblock {Image features of spinning regular black holes based on a locality
  principle}.
\newblock {\em Eur. Phys. J. C}, 81(10):933, 2021.

\bibitem{Eichhorn:2021iwq}
Astrid Eichhorn and Aaron Held.
\newblock {From a locality-principle for new physics to image features of
  regular spinning black holes with disks}.
\newblock {\em JCAP}, 05:073, 2021.

\bibitem{Eichhorn:2022oma}
Astrid Eichhorn, Aaron Held, and Philipp-Vincent Johannsen.
\newblock {Universal signatures of singularity-resolving physics in photon
  rings of black holes and horizonless objects}.
\newblock {\em JCAP}, 01:043, 2023.

\bibitem{KumarWalia:2022ddq}
Rahul Kumar~Walia.
\newblock {Observational predictions of LQG motivated polymerized black holes
  and constraints from Sgr A* and M87*}.
\newblock {\em JCAP}, 03:029, 2023.

\bibitem{Ling:2022vrv}
Yi~Ling and Meng-He Wu.
\newblock {The Shadows of Regular Black Holes with Asymptotic Minkowski Cores}.
\newblock {\em Symmetry}, 14(11):2415, 2022.

\bibitem{Islam:2022wck}
Shafqat~Ul Islam, Jitendra Kumar, Rahul Kumar~Walia, and Sushant~G. Ghosh.
\newblock {Investigating Loop Quantum Gravity with Event Horizon Telescope
  Observations of the Effects of Rotating Black Holes}.
\newblock {\em Astrophys. J.}, 943(1):22, 2023.

\bibitem{Cunha:2017wao}
Pedro V.~P. Cunha, Jos\'e~A. Font, Carlos Herdeiro, Eugen Radu, Nicolas
  Sanchis-Gual, and Miguel Zilh\~ao.
\newblock {Lensing and dynamics of ultracompact bosonic stars}.
\newblock {\em Phys. Rev. D}, 96(10):104040, 2017.

\bibitem{Eichhorn:2022fcl}
Astrid Eichhorn, Roman Gold, and Aaron Held.
\newblock {Horizonless Spacetimes As Seen by Present and Next-generation Event
  Horizon Telescope Arrays}.
\newblock {\em Astrophys. J.}, 950(2):117, 2023.

\bibitem{Guerrero:2022msp}
Merce Guerrero, Gonzalo~J. Olmo, Diego Rubiera-Garcia, and Diego
  S\'aez-Chill\'on~G\'omez.
\newblock {Multiring images of thin accretion disk of a regular naked compact
  object}.
\newblock {\em Phys. Rev. D}, 106(4):044070, 2022.

\bibitem{Carballo-Rubio:2022bgh}
Ra\'ul Carballo-Rubio, Vitor Cardoso, and Ziri Younsi.
\newblock {Toward very large baseline interferometry observations of black hole
  structure}.
\newblock {\em Phys. Rev. D}, 106(8):084038, 2022.

\bibitem{Guerrero:2022qkh}
Merce Guerrero, Gonzalo~J. Olmo, Diego Rubiera-Garcia, and Diego G\'omez
  S\'aez-Chill\'on.
\newblock {Light ring images of double photon spheres in black hole and
  wormhole spacetimes}.
\newblock {\em Phys. Rev. D}, 105(8):084057, 2022.

\bibitem{Olmo:2023lil}
Gonzalo~J. Olmo, Joao~Luis Rosa, Diego Rubiera-Garcia, and Diego
  Saez-Chillon~Gomez.
\newblock {Shadows and photon rings of regular black holes and geonic
  horizonless compact objects}.
\newblock {\em Class. Quant. Grav.}, 40(17):174002, 2023.

\bibitem{Neto:2022pmu}
M\'ario~Raia Neto, Daniela P\'erez, and Joaqu\'\i{}n Pelle.
\newblock {The shadow of charged traversable wormholes}.
\newblock {\em Int. J. Mod. Phys. D}, 32(02):2250137, 2023.

\bibitem{Gyulchev:2021dvt}
Galin Gyulchev, Petya Nedkova, Tsvetan Vetsov, and Stoytcho Yazadjiev.
\newblock {Image of the thin accretion disk around compact objects in the
  Einstein\textendash{}Gauss\textendash{}Bonnet gravity}.
\newblock {\em Eur. Phys. J. C}, 81(10):885, 2021.

\bibitem{Sengo:2022jif}
Ivo Sengo, Pedro V.~P. Cunha, Carlos A.~R. Herdeiro, and Eugen Radu.
\newblock {Kerr black holes with synchronised Proca hair: lensing, shadows and
  EHT constraints}.
\newblock {\em JCAP}, 01:047, 2023.

\bibitem{Lara:2021zth}
Guillermo Lara, Sebastian~H. V\"olkel, and Enrico Barausse.
\newblock {Separating astrophysics and geometry in black hole images}.
\newblock {\em Phys. Rev. D}, 104(12):124041, 2021.

\bibitem{Kocherlakota:2022jnz}
Prashant Kocherlakota and Luciano Rezzolla.
\newblock {Distinguishing gravitational and emission physics in black hole
  imaging: spherical symmetry}.
\newblock {\em Mon. Not. Roy. Astron. Soc.}, 513(1):1229--1243, 2022.

\bibitem{Younsi:2021dxe}
Ziri Younsi, Dimitrios Psaltis, and Feryal \"Ozel.
\newblock {Black Hole Images as Tests of General Relativity: Effects of
  Spacetime Geometry}.
\newblock {\em Astrophys. J.}, 942(1):47, 2023.

\bibitem{darwin1959gravity}
Charles~Galton Darwin.
\newblock The gravity field of a particle.
\newblock {\em Proceedings of the Royal Society of London. Series A.
  Mathematical and Physical Sciences}, 249(1257):180--194, 1959.

\bibitem{Bardeen:1972fi}
James~M. Bardeen, William~H. Press, and Saul~A Teukolsky.
\newblock {Rotating black holes: Locally nonrotating frames, energy extraction,
  and scalar synchrotron radiation}.
\newblock {\em Astrophys. J.}, 178:347, 1972.

\bibitem{Luminet:1979nyg}
J.~P. Luminet.
\newblock {Image of a spherical black hole with thin accretion disk}.
\newblock {\em Astron. Astrophys.}, 75:228--235, 1979.

\bibitem{Gralla:2019xty}
Samuel~E. Gralla, Daniel~E. Holz, and Robert~M. Wald.
\newblock {Black Hole Shadows, Photon Rings, and Lensing Rings}.
\newblock {\em Phys. Rev. D}, 100(2):024018, 2019.

\bibitem{Johnson:2019ljv}
Michael~D. Johnson et~al.
\newblock {Universal interferometric signatures of a black
  hole\textquoteright{}s photon ring}.
\newblock {\em Sci. Adv.}, 6(12):eaaz1310, 2020.

\bibitem{Cardenas-Avendano:2023dzo}
Alejandro C\'ardenas-Avenda\~no and Alexandru Lupsasca.
\newblock {Prediction for the interferometric shape of the first black hole
  photon ring}.
\newblock {\em Phys. Rev. D}, 108(6):064043, 2023.

\bibitem{bisnovatyi1974accretion}
GS~Bisnovatyi-Kogan and AA~Ruzmaikin.
\newblock The accretion of matter by a collapsing star in the presence of a
  magnetic field.
\newblock {\em Astrophysics and Space Science}, 28:45--59, 1974.

\bibitem{Igumenshchev:2003rt}
Igor~V. Igumenshchev, Ramesh Narayan, and Marek~A. Abramowicz.
\newblock {Three-dimensional mhd simulations of radiatively inefficient
  accretion flows}.
\newblock {\em Astrophys. J.}, 592:1042--1059, 2003.

\bibitem{Narayan:2003by}
Ramesh Narayan, Igor~V. Igumenshchev, and Marek~A. Abramowicz.
\newblock {Magnetically arrested disk: an energetically efficient accretion
  flow}.
\newblock {\em Publ. Astron. Soc. Jap.}, 55:L69, 2003.

\bibitem{DeVilliers:2003gr}
Jean-Pierre De~Villiers, John~F. Hawley, and Julian~H. Krolik.
\newblock {Magnetically driven accretion flows in the kerr metric I: models and
  overall structure}.
\newblock {\em Astrophys. J.}, 599:1238, 2003.

\bibitem{Gammie:2003rj}
Charles~F. Gammie, Jonathan~C. McKinney, and Gabor Toth.
\newblock {HARM: A Numerical scheme for general relativistic
  magnetohydrodynamics}.
\newblock {\em Astrophys. J.}, 589:444--457, 2003.

\bibitem{Narayan:2012yp}
Ramesh Narayan, Aleksander Sadowski, Robert~F. Penna, and Akshay~K. Kulkarni.
\newblock {GRMHD Simulations of Magnetized Advection-Dominated Accretion on a
  Non-Spinning Black Hole: Role of Outflows}.
\newblock {\em Mon. Not. Roy. Astron. Soc.}, 426:3241, 2012.

\bibitem{Broderick:2021ohx}
Avery~E. Broderick, Paul Tiede, Dominic~W. Pesce, and Roman Gold.
\newblock {Measuring Spin from Relative Photon-ring Sizes}.
\newblock {\em Astrophys. J.}, 927(1):6, 2022.

\bibitem{LIGOScientific:2018dkp}
B.~P. Abbott et~al.
\newblock {Tests of General Relativity with GW170817}.
\newblock {\em Phys. Rev. Lett.}, 123(1):011102, 2019.

\bibitem{LIGOScientific:2019fpa}
B.~P. Abbott et~al.
\newblock {Tests of General Relativity with the Binary Black Hole Signals from
  the LIGO-Virgo Catalog GWTC-1}.
\newblock {\em Phys. Rev. D}, 100(10):104036, 2019.

\bibitem{LIGOScientific:2020tif}
R.~Abbott et~al.
\newblock {Tests of general relativity with binary black holes from the second
  LIGO-Virgo gravitational-wave transient catalog}.
\newblock {\em Phys. Rev. D}, 103(12):122002, 2021.

\bibitem{LIGOScientific:2021sio}
R.~Abbott et~al.
\newblock {Tests of General Relativity with GWTC-3}.
\newblock 12 2021.

\bibitem{Jiang:2023kpx}
Nan Jiang.
\newblock {\em {Testing General Relativity With Gravitational Waves From
  Compact Binaries}}.
\newblock PhD thesis, Virginia U., 2023.

\bibitem{Lu:2015cqa}
H.~Lu, A.~Perkins, C.~N. Pope, and K.~S. Stelle.
\newblock {Black Holes in Higher-Derivative Gravity}.
\newblock {\em Phys. Rev. Lett.}, 114(17):171601, 2015.

\bibitem{Fernandes:2023vux}
Pedro G.~S. Fernandes.
\newblock {Rotating black holes in semiclassical gravity}.
\newblock {\em Phys. Rev. D}, 108(6):L061502, 2023.

\bibitem{Doneva:2017bvd}
Daniela~D. Doneva and Stoytcho~S. Yazadjiev.
\newblock {New Gauss-Bonnet Black Holes with Curvature-Induced Scalarization in
  Extended Scalar-Tensor Theories}.
\newblock {\em Phys. Rev. Lett.}, 120(13):131103, 2018.

\bibitem{Dima:2020yac}
Alexandru Dima, Enrico Barausse, Nicola Franchini, and Thomas~P. Sotiriou.
\newblock {Spin-induced black hole spontaneous scalarization}.
\newblock {\em Phys. Rev. Lett.}, 125(23):231101, 2020.

\bibitem{Doneva:2022ewd}
Daniela~D. Doneva, Fethi~M. Ramazano\u{g}lu, Hector~O. Silva, Thomas~P.
  Sotiriou, and Stoytcho~S. Yazadjiev.
\newblock {Scalarization}.
\newblock 11 2022.

\bibitem{Witek:2020uzz}
Helvi Witek, Leonardo Gualtieri, and Paolo Pani.
\newblock {Towards numerical relativity in scalar Gauss-Bonnet gravity: $3+1$
  decomposition beyond the small-coupling limit}.
\newblock {\em Phys. Rev. D}, 101(12):124055, 2020.

\bibitem{Silva:2020omi}
Hector~O. Silva, Helvi Witek, Matthew Elley, and Nicol\'as Yunes.
\newblock {Dynamical Descalarization in Binary Black Hole Mergers}.
\newblock {\em Phys. Rev. Lett.}, 127(3):031101, 2021.

\bibitem{Corman:2022xqg}
Maxence Corman, Justin~L. Ripley, and William~E. East.
\newblock {Nonlinear studies of binary black hole mergers in
  Einstein-scalar-Gauss-Bonnet gravity}.
\newblock {\em Phys. Rev. D}, 107(2):024014, 2023.

\bibitem{Held:2023aap}
Aaron Held and Hyun Lim.
\newblock {Nonlinear evolution of quadratic gravity in 3+1 dimensions}.
\newblock {\em Phys. Rev. D}, 108(10):104025, 2023.

\bibitem{Pretorius:2005gq}
Frans Pretorius.
\newblock {Evolution of binary black hole spacetimes}.
\newblock {\em Phys. Rev. Lett.}, 95:121101, 2005.

\bibitem{Bishop:2016lgv}
Nigel~T. Bishop and Luciano Rezzolla.
\newblock {Extraction of Gravitational Waves in Numerical Relativity}.
\newblock {\em Living Rev. Rel.}, 19:2, 2016.

\bibitem{Broderick:2022tfu}
Avery~E. Broderick et~al.
\newblock {The Photon Ring in M87*}.
\newblock {\em Astrophys. J.}, 935:61, 2022.

\bibitem{Vincent:2022fwj}
Frederic~H. Vincent, Samuel~E. Gralla, Alexandru Lupsasca, and Maciek Wielgus.
\newblock {Images and photon ring signatures of thick disks around black
  holes}.
\newblock {\em Astron. Astrophys.}, 667:A170, 2022.

\bibitem{Himwich:2020msm}
Elizabeth Himwich, Michael~D. Johnson, Alexandru Lupsasca, and Andrew
  Strominger.
\newblock {Universal polarimetric signatures of the black hole photon ring}.
\newblock {\em Phys. Rev. D}, 101(8):084020, 2020.

\bibitem{Paugnat:2022qzy}
Hadrien Paugnat, Alexandru Lupsasca, Fr\'ed\'eric Vincent, and Maciek Wielgus.
\newblock {Photon ring test of the Kerr hypothesis: Variation in the ring
  shape}.
\newblock {\em Astron. Astrophys.}, 668:A11, 2022.

\bibitem{Tiede:2022grp}
Paul Tiede, Michael~D. Johnson, Dominic~W. Pesce, Daniel C.~M. Palumbo,
  Dominic~O. Chang, and Peter Galison.
\newblock {Measuring Photon Rings with the ngEHT}.
\newblock {\em Galaxies}, 10(6):111, 2022.

\bibitem{Lockhart:2022rui}
Will Lockhart and Samuel~E. Gralla.
\newblock {How narrow is the M87* ring \textendash{} II. A new geometric
  model}.
\newblock {\em Mon. Not. Roy. Astron. Soc.}, 517(2):2462--2470, 2022.

\bibitem{Wielgus:2021peu}
Maciek Wielgus.
\newblock {Photon rings of spherically symmetric black holes and robust tests
  of non-Kerr metrics}.
\newblock {\em Phys. Rev. D}, 104(12):124058, 2021.

\bibitem{Staelens:2023jgr}
Seppe Staelens, Daniel~R. Mayerson, Fabio Bacchini, Bart Ripperda, and Lorenzo
  K\"uchler.
\newblock {Black hole photon rings beyond general relativity}.
\newblock {\em Phys. Rev. D}, 107(12):124026, 2023.

\bibitem{Ayzenberg:2022twz}
Dimitry Ayzenberg.
\newblock {Testing gravity with black hole shadow subrings}.
\newblock {\em Class. Quant. Grav.}, 39(10):105009, 2022.

\bibitem{Cardenas-Avendano:2023obg}
Alejandro C\'ardenas-Avenda\~no and Aaron Held.
\newblock {A Lensing-Band Approach to Spacetime Constraints}.
\newblock 12 2023.

\bibitem{Bonanno:2000ep}
Alfio Bonanno and Martin Reuter.
\newblock {Renormalization group improved black hole space-times}.
\newblock {\em Phys. Rev. D}, 62:043008, 2000.

\bibitem{Falls:2010he}
Kevin Falls, Daniel~F. Litim, and Aarti Raghuraman.
\newblock {Black Holes and Asymptotically Safe Gravity}.
\newblock {\em Int. J. Mod. Phys. A}, 27:1250019, 2012.

\bibitem{Platania:2019kyx}
Alessia Platania.
\newblock {Dynamical renormalization of black-hole spacetimes}.
\newblock {\em Eur. Phys. J. C}, 79(6):470, 2019.

\bibitem{Eichhorn:2022bgu}
Astrid Eichhorn and Aaron Held.
\newblock {Black holes in asymptotically safe gravity and beyond}.
\newblock 12 2022.

\bibitem{Platania:2023srt}
Alessia Platania.
\newblock {Black Holes in Asymptotically Safe Gravity}.
\newblock 2 2023.

\bibitem{Ashtekar:2005qt}
Abhay Ashtekar and Martin Bojowald.
\newblock {Quantum geometry and the Schwarzschild singularity}.
\newblock {\em Class. Quant. Grav.}, 23:391--411, 2006.

\bibitem{Modesto:2005zm}
Leonardo Modesto.
\newblock {Loop quantum black hole}.
\newblock {\em Class. Quant. Grav.}, 23:5587--5602, 2006.

\bibitem{Campiglia:2007pb}
Miguel Campiglia, Rodolfo Gambini, and Jorge Pullin.
\newblock {Loop quantization of spherically symmetric midi-superspaces : The
  Interior problem}.
\newblock {\em AIP Conf. Proc.}, 977(1):52--63, 2008.

\bibitem{Ashtekar:2023cod}
Abhay Ashtekar, Javier Olmedo, and Parampreet Singh.
\newblock {Regular black holes from Loop Quantum Gravity}.
\newblock 1 2023.

\bibitem{Ayon-Beato:1998hmi}
Eloy Ayon-Beato and Alberto Garcia.
\newblock {Regular black hole in general relativity coupled to nonlinear
  electrodynamics}.
\newblock {\em Phys. Rev. Lett.}, 80:5056--5059, 1998.

\bibitem{Ayon-Beato:1999kuh}
Eloy Ayon-Beato and Alberto Garcia.
\newblock {New regular black hole solution from nonlinear electrodynamics}.
\newblock {\em Phys. Lett. B}, 464:25, 1999.

\bibitem{Bronnikov:2000vy}
Kirill~A. Bronnikov.
\newblock {Regular magnetic black holes and monopoles from nonlinear
  electrodynamics}.
\newblock {\em Phys. Rev. D}, 63:044005, 2001.

\bibitem{Bronnikov:2022ofk}
Kirill~A. Bronnikov.
\newblock {Regular black holes sourced by nonlinear electrodynamics}.
\newblock 11 2022.

\bibitem{Babichev:2020qpr}
Eugeny Babichev, Christos Charmousis, Adolfo Cisterna, and Mokhtar Hassaine.
\newblock {Regular black holes via the Kerr-Schild construction in DHOST
  theories}.
\newblock {\em JCAP}, 06:049, 2020.

\bibitem{Baake:2021jzv}
Olaf Baake, Christos Charmousis, Mokhtar Hassaine, and Miguel San~Juan.
\newblock {Regular black holes and gravitational particle-like solutions in
  generic DHOST theories}.
\newblock {\em JCAP}, 06:021, 2021.

\bibitem{Hayward:2005gi}
Sean~A. Hayward.
\newblock {Formation and evaporation of regular black holes}.
\newblock {\em Phys. Rev. Lett.}, 96:031103, 2006.

\bibitem{Simpson:2018tsi}
Alex Simpson and Matt Visser.
\newblock {Black-bounce to traversable wormhole}.
\newblock {\em JCAP}, 02:042, 2019.

\bibitem{Carballo-Rubio:2019fnb}
Ra\'ul Carballo-Rubio, Francesco Di~Filippo, Stefano Liberati, and Matt Visser.
\newblock {Geodesically complete black holes}.
\newblock {\em Phys. Rev. D}, 101:084047, 2020.

\bibitem{Mazza:2021rgq}
Jacopo Mazza, Edgardo Franzin, and Stefano Liberati.
\newblock {A novel family of rotating black hole mimickers}.
\newblock {\em JCAP}, 04:082, 2021.

\bibitem{Simpson:2021dyo}
Alex Simpson and Matt Visser.
\newblock {The eye of the storm: a regular Kerr black hole}.
\newblock {\em JCAP}, 03(03):011, 2022.

\bibitem{Carballo-Rubio:2023mvr}
Ra\'ul Carballo-Rubio, Francesco Di~Filippo, Stefano Liberati, and Matt Visser.
\newblock {Singularity-free gravitational collapse: From regular black holes to
  horizonless objects}.
\newblock 1 2023.

\bibitem{Bambibook}
C.~Bambi, editor.
\newblock {\em Regular black holes: Towards a new paradigm of gravitational
  collapse}.
\newblock Springer, 2023.

\bibitem{Carballo-Rubio:2018pmi}
Ra\'ul Carballo-Rubio, Francesco Di~Filippo, Stefano Liberati, Costantino
  Pacilio, and Matt Visser.
\newblock {On the viability of regular black holes}.
\newblock {\em JHEP}, 07:023, 2018.

\bibitem{Carballo-Rubio:2022kad}
Ra\'ul Carballo-Rubio, Francesco Di~Filippo, Stefano Liberati, Costantino
  Pacilio, and Matt Visser.
\newblock {Regular black holes without mass inflation instability}.
\newblock {\em JHEP}, 09:118, 2022.

\bibitem{Franzin:2022wai}
Edgardo Franzin, Stefano Liberati, Jacopo Mazza, and Vania Vellucci.
\newblock {Stable rotating regular black holes}.
\newblock {\em Phys. Rev. D}, 106(10):104060, 2022.

\bibitem{Carballo-Rubio:2021bpr}
Ra\'ul Carballo-Rubio, Francesco Di~Filippo, Stefano Liberati, Costantino
  Pacilio, and Matt Visser.
\newblock {Inner horizon instability and the unstable cores of regular black
  holes}.
\newblock {\em JHEP}, 05:132, 2021.

\bibitem{Bonanno:2020fgp}
Alfio Bonanno, Amir-Pouyan Khosravi, and Frank Saueressig.
\newblock {Regular black holes with stable cores}.
\newblock {\em Phys. Rev. D}, 103(12):124027, 2021.

\bibitem{Barcelo:2022gii}
Carlos Barcel\'o, Valentin Boyanov, Ra\'ul Carballo-Rubio, and Luis~J. Garay.
\newblock {Classical mass inflation versus semiclassical inner horizon
  inflation}.
\newblock {\em Phys. Rev. D}, 106(12):124006, 2022.

\bibitem{DiFilippo:2022qkl}
Francesco Di~Filippo, Ra\'ul Carballo-Rubio, Stefano Liberati, Costantino
  Pacilio, and Matt Visser.
\newblock {On the Inner Horizon Instability of Non-Singular Black Holes}.
\newblock {\em Universe}, 8(4):204, 2022.

\bibitem{Bonanno:2022jjp}
Alfio Bonanno, Amir-Pouyan Khosravi, and Frank Saueressig.
\newblock {Regular evaporating black holes with stable cores}.
\newblock {\em Phys. Rev. D}, 107(2):024005, 2023.

\bibitem{Carballo-Rubio:2022twq}
Ra\'ul Carballo-Rubio, Francesco Di~Filippo, Stefano Liberati, Costantino
  Pacilio, and Matt Visser.
\newblock {Comment on ''Stability properties of Regular Black Holes''}.
\newblock 12 2022.

\bibitem{Delaporte:2022acp}
H\'elo\"\i{}se Delaporte, Astrid Eichhorn, and Aaron Held.
\newblock {Parameterizations of black-hole spacetimes beyond circularity}.
\newblock {\em Class. Quant. Grav.}, 39(13):134002, 2022.

\bibitem{Papapetrou:1966zz}
Achille Papapetrou.
\newblock {Champs gravitationnels stationnaires a symetrie axiale}.
\newblock {\em Ann. Inst. H. Poincare Phys. Theor.}, 4:83--105, 1966.

\bibitem{Weyl:1917gp}
H.~Weyl.
\newblock {The theory of gravitation}.
\newblock {\em Annalen Phys.}, 54:117--145, 1917.

\bibitem{Lewis:1932zz}
T.~Lewis.
\newblock {Some speical solutions to the equations of axially symmetric
  gravitational fields}.
\newblock {\em Proc. Roy. Soc. Lond. A}, 136:179--192, 1932.

\bibitem{Konoplya:2016jvv}
Roman Konoplya, Luciano Rezzolla, and Alexander Zhidenko.
\newblock {General parametrization of axisymmetric black holes in metric
  theories of gravity}.
\newblock {\em Phys. Rev. D}, 93(6):064015, 2016.

\bibitem{Benenti:1979erw}
S.~Benenti and M.~Francaviglia.
\newblock {Remarks on certain separability structures and their applications to
  general relativity}.
\newblock {\em Gen. Rel. Grav.}, 10(1):79--92, 1979.

\bibitem{Johannsen:2013szh}
Tim Johannsen.
\newblock {Regular Black Hole Metric with Three Constants of Motion}.
\newblock {\em Phys. Rev. D}, 88(4):044002, 2013.

\bibitem{Vigeland:2011ji}
Sarah Vigeland, Nicolas Yunes, and Leo Stein.
\newblock {Bumpy Black Holes in Alternate Theories of Gravity}.
\newblock {\em Phys. Rev. D}, 83:104027, 2011.

\bibitem{1993tegp.book.....W}
Clifford~M. {Will}.
\newblock {\em {Theory and Experiment in Gravitational Physics}}.
\newblock 1993.

\bibitem{Carballo-Rubio:2017tlh}
Ra\'ul Carballo-Rubio.
\newblock {Stellar equilibrium in semiclassical gravity}.
\newblock {\em Phys. Rev. Lett.}, 120(6):061102, 2018.

\bibitem{Arrechea:2021xkp}
Julio Arrechea, Carlos Barcel\'o, Ra\'ul Carballo-Rubio, and Luis~J. Garay.
\newblock {Semiclassical relativistic stars}.
\newblock {\em Sci. Rep.}, 12(1):15958, 2022.

\bibitem{Carballo-Rubio:2022nuj}
Ra\'ul Carballo-Rubio, Francesco Di~Filippo, Stefano Liberati, and Matt Visser.
\newblock {A connection between regular black holes and horizonless
  ultracompact stars}.
\newblock {\em JHEP}, 08:046, 2023.

\bibitem{Bonanno:2019ilz}
Alfio Bonanno, Roberto Casadio, and Alessia Platania.
\newblock {Gravitational antiscreening in stellar interiors}.
\newblock {\em JCAP}, 01:022, 2020.

\bibitem{Carballo-Rubio:2022aed}
Ra\'ul Carballo-Rubio, Vitor Cardoso, and Ziri Younsi.
\newblock {Toward very large baseline interferometry observations of black hole
  structure}.
\newblock {\em Phys. Rev. D}, 106(8):084038, 2022.

\bibitem{Shaikh:2018lcc}
Rajibul Shaikh, Prashant Kocherlakota, Ramesh Narayan, and Pankaj~S. Joshi.
\newblock {Shadows of spherically symmetric black holes and naked
  singularities}.
\newblock {\em Mon. Not. Roy. Astron. Soc.}, 482(1):52--64, 2019.

\bibitem{Gyulchev:2020cvo}
Galin Gyulchev, Jutta Kunz, Petya Nedkova, Tsvetan Vetsov, and Stoytcho
  Yazadjiev.
\newblock {Observational signatures of strongly naked singularities: image of
  the thin accretion disk}.
\newblock {\em Eur. Phys. J. C}, 80(11):1017, 2020.

\bibitem{Kamruddin:2013iea}
Ayman~Bin Kamruddin and Jason Dexter.
\newblock {A geometric crescent model for black hole images}.
\newblock {\em Mon. Not. Roy. Astron. Soc.}, 434:765, 2013.

\bibitem{chael_2023_8408352}
Andrew Chael.
\newblock eht-imaging, October 2023.

\bibitem{Chael:2018oym}
Andrew~A. Chael, Michael~D. Johnson, Katherine~L. Bouman, Lindy~L. Blackburn,
  Kazunori Akiyama, and Ramesh Narayan.
\newblock {Interferometric Imaging Directly with Closure Phases and Closure
  Amplitudes}.
\newblock {\em Astrophys. J.}, 857(1):23, 2018.

\bibitem{Roelofs:2019nmh}
Freek Roelofs et~al.
\newblock {Simulations of imaging the event horizon of Sagittarius A* from
  space}.
\newblock {\em Astron. Astrophys.}, 625:A124, 2019.

\bibitem{2023Galax..11..107D}
Sheperd~S. {Doeleman}, John {Barrett}, Lindy {Blackburn}, Katherine~L.
  {Bouman}, Avery~E. {Broderick}, Ryan {Chaves}, Vincent~L. {Fish}, Garret
  {Fitzpatrick}, Mark {Freeman}, Antonio {Fuentes}, Jos{\'e}~L. {G{\'o}mez},
  Kari {Haworth}, Janice {Houston}, Sara {Issaoun}, Michael~D. {Johnson}, Mark
  {Kettenis}, Laurent {Loinard}, Neil {Nagar}, Gopal {Narayanan}, Aaron
  {Oppenheimer}, Daniel C.~M. {Palumbo}, Nimesh {Patel}, Dominic~W. {Pesce},
  Alexander~W. {Raymond}, Freek {Roelofs}, Ranjani {Srinivasan}, Paul {Tiede},
  Jonathan {Weintroub}, and Maciek {Wielgus}.
\newblock {Reference Array and Design Consideration for the Next-Generation
  Event Horizon Telescope}.
\newblock {\em Galaxies}, 11(5):107, October 2023.

\bibitem{Broderick:2021aeg}
Avery Broderick.
\newblock ngehtexplorer.

\bibitem{newville_matthew_2014_11813}
Matthew Newville, Till Stensitzki, Daniel~B. Allen, and Antonino Ingargiola.
\newblock {LMFIT: Non-Linear Least-Square Minimization and Curve-Fitting for
  Python}, September 2014.

\bibitem{2010arXiv1012.3754A}
Rene {Andrae}, Tim {Schulze-Hartung}, and Peter {Melchior}.
\newblock {Dos and don'ts of reduced chi-squared}.
\newblock {\em arXiv e-prints}, page arXiv:1012.3754, December 2010.

\bibitem{2017isra.book.....T}
A.~Richard {Thompson}, James~M. {Moran}, and Jr. {Swenson}, George~W.
\newblock {\em {Interferometry and Synthesis in Radio Astronomy, 3rd Edition}}.
\newblock 2017.

\bibitem{Lockhart:2021xel}
Will Lockhart and Samuel~E. Gralla.
\newblock {How narrow is the M87* ring? I. The choice of closure likelihood
  function}.
\newblock {\em Mon. Not. Roy. Astron. Soc.}, 509(3):3643--3659, 2021.

\bibitem{lapidoth2017foundation}
A.~Lapidoth.
\newblock {\em A Foundation in Digital Communication}.
\newblock Cambridge University Press, 2017.

\bibitem{2018ApJ...857...23C}
Andrew~A. {Chael}, Michael~D. {Johnson}, Katherine~L. {Bouman}, Lindy~L.
  {Blackburn}, Kazunori {Akiyama}, and Ramesh {Narayan}.
\newblock {Interferometric Imaging Directly with Closure Phases and Closure
  Amplitudes}.
\newblock {\em ApJ}, 857(1):23, April 2018.

\bibitem{1958MNRAS.118..276J}
R.~C. {Jennison}.
\newblock {A phase sensitive interferometer technique for the measurement of
  the Fourier transforms of spatial brightness distributions of small angular
  extent}.
\newblock {\em Mon. Not. Roy. Astron. Soc.}, 118:276, January 1958.

\bibitem{1960Obs....80..153T}
R.~Q. {Twiss}, A.~W.~L. {Carter}, and A.~G. {Little}.
\newblock {Brightness distribution over some strong radio sources at 1427
  Mc/s}.
\newblock {\em The Observatory}, 80:153--159, August 1960.

\bibitem{1974ApJ...193..293R}
A.~E.~E. {Rogers}, H.~F. {Hinteregger}, A.~R. {Whitney}, C.~C. {Counselman},
  I.~I. {Shapiro}, J.~J. {Wittels}, W.~K. {Klemperer}, W.~W. {Warnock}, T.~A.
  {Clark}, L.~K. {Hutton}, G.~E. {Marandino}, B.~O. {Ronnang}, O.~E.~H.
  {Rydbeck}, and A.~E. {Niell}.
\newblock {The structure of radio sources 3C 273B and 3C 84 deduced from the
  ``closure'' phases and visibility amplitudes observed with three-element
  interferometers.}
\newblock {\em Astrophys. J.}, 193:293--301, October 1974.

\bibitem{1980Natur.285..137R}
A.~C.~S. {Readhead}, R.~C. {Walker}, T.~J. {Pearson}, and M.~H. {Cohen}.
\newblock {Mapping radio sources with uncalibrated visibility data}.
\newblock {\em Nature}, 285(5761):137--140, May 1980.

\bibitem{Doeleman:2001nr}
S.~S. Doeleman et~al.
\newblock {Structure of sagittarius a* at 86 ghz using vlbi closure
  quantities}.
\newblock {\em Astron. J.}, 121:2610, 2001.

\bibitem{Fish:2016zqt}
Vincent~L. Fish et~al.
\newblock {Persistent Asymmetric Structure of Sagittarius A* on Event Horizon
  Scales}.
\newblock {\em Astrophys. J.}, 820(2):90, 2016.

\bibitem{1991InvPr...7..261L}
A.~{Lannes}.
\newblock {Phase and amplitude calibration in aperture synthesis. Algebraic
  structures}.
\newblock {\em Inverse Problems}, 7(2):261--298, April 1991.

\bibitem{2020ApJ...904..126B}
Avery~E. {Broderick} and Dominic~W. {Pesce}.
\newblock {Closure Traces: Novel Calibration-insensitive Quantities for Radio
  Astronomy}.
\newblock {\em Astrophys. J.}, 904(2):126, December 2020.

\bibitem{Samuel:2021vfv}
Joseph Samuel, Rajaram Nityananda, and Nithyanandan Thyagarajan.
\newblock {Invariants in Polarimetric Interferometry: A Non-Abelian Gauge
  Theory}.
\newblock {\em Phys. Rev. Lett.}, 128(9):091101, 2022.

\bibitem{2022PhRvD.105d3019T}
Nithyanandan {Thyagarajan}, Rajaram {Nityananda}, and Joseph {Samuel}.
\newblock {Invariants in copolar interferometry: An Abelian gauge theory}.
\newblock {\em Phys. Rev. D}, 105(4):043019, February 2022.

\bibitem{2003EAS.....6..213M}
J.~D. {Monnier}.
\newblock {Astrophysics with Closure Phases}.
\newblock In G.~{Perrin} and F.~{Malbet}, editors, {\em EAS Publications
  Series}, volume~6 of {\em EAS Publications Series}, page 213, January 2003.

\bibitem{2020ApJ...894...31B}
Lindy {Blackburn}, Dominic~W. {Pesce}, Michael~D. {Johnson}, Maciek {Wielgus},
  Andrew~A. {Chael}, Pierre {Christian}, and Sheperd~S. {Doeleman}.
\newblock {Closure Statistics in Interferometric Data}.
\newblock {\em Astrophys. J.}, 894(1):31, May 2020.

\bibitem{Broderick:2020wda}
Avery~E. Broderick, Dominic~W. Pesce, Paul Tiede, Hung-Yi Pu, and Roman Gold.
\newblock {Hybrid Very Long Baseline Interferometry Imaging and Modeling with
  themis}.
\newblock {\em Astrophys. J.}, 898(1):9, 2020.

\bibitem{2022A&A...663A..35I}
J.~W. {Isbell} et~al.
\newblock {The dusty heart of Circinus. I. Imaging the circumnuclear dust in
  N-band}.
\newblock {\em Astron. Astrophys.}, 663:A35, July 2022.

\bibitem{1965hmfw.book.....A}
Milton {Abramowitz} and Irene~A. {Stegun}.
\newblock {\em {Handbook of mathematical functions with formulas, graphs, and
  mathematical tables}}.
\newblock 1965.

\bibitem{Gurvits:2019ioq}
Leonid~I. Gurvits et~al.
\newblock {THEZA: TeraHertz Exploration and Zooming-in for Astrophysics: An ESA
  Voyage 2050 White Paper}.
\newblock {\em Exper. Astron.}, 51(3):559--594, 2021.

\bibitem{Fish:2019epg}
Vincent~L. Fish, Maura Shea, and Kazunori Akiyama.
\newblock {Imaging black holes and jets with a VLBI array including multiple
  space-based telescopes}.
\newblock {\em Adv. Space Res.}, 65:821--830, 2020.

\bibitem{Haworth:2019urs}
Kari Haworth et~al.
\newblock {Studying black holes on horizon scales with space-VLBI}.
\newblock 9 2019.

\bibitem{palumbo2019metrics}
Daniel~CM Palumbo, Sheperd~S Doeleman, Michael~D Johnson, Katherine~L Bouman,
  and Andrew~A Chael.
\newblock Metrics and motivations for earth--space vlbi: Time-resolving sgr a*
  with the event horizon telescope.
\newblock {\em The Astrophysical Journal}, 881(1):62, 2019.

\end{thebibliography}
\end{document}